\documentclass[12pt]{iopart}

\usepackage{graphicx}
\usepackage{graphics}
\usepackage{epsfig}
\usepackage{color}
\usepackage{hyperref}
\usepackage{helvet}
\usepackage{courier}
\usepackage{exscale}		%zum richtigen Skalieren von Schriftarten über 10pt

\newcommand{\dC}{\,$^{\circ}$C}

\newcommand{\cytc}{cytochrome~\textit{c}\,}
\newcommand{\cytcol}{cytochrome~\textit{c}}

\newcommand{\etalp}{\textit{et~al.~}}

\begin{document}

\title[]{Soft matter in hard confinement: phase transition thermodynamics, structure, texture, diffusion and flow in nanoporous media}

%Phase Transformations, Diffusion and Flow of Molecular Condensates in Nanoporous Media}
% How molecular assemblies arrange, diffuse and flow in nanoporous media
% How molecular assemblies arrange and huddle in nanoporous media
%Phase Transitions, Texture Formation, Thermal Diffusion and Capillary Flow in Nanoporous Media}

\author{Patrick Huber}

\address{Hamburg University of Technology (TUHH), Institute of Materials Physics and Technology, Ei\ss endorfer Str. 42, D-21073 Hamburg-Harburg (Germany)}
\ead{patrick.huber@tuhh.de}
\begin{abstract}
Spatial confinement in nanoporous media affects the structure, thermodynamics and mobility of molecular soft matter often markedly. This article reviews thermodynamic equilibrium phenomena, such as physisorption, capillary condensation, crystallisation, self-diffusion, and structural phase transitions as well as selected aspects of the emerging field of spatially confined, non-equilibrium physics, \textit{i.e.} the rheology of liquids, capillarity-driven flow phenomena, and imbibition front broadening in nanoporous materials. The observations in the nanoscale systems are related to the corresponding bulk phenomenologies. The complexity of the confined molecular species is varied from simple building blocks, like noble gas atoms, normal alkanes and alcohols to liquid crystals, polymers, ionic liquids, proteins and water. Mostly, experiments with mesoporous solids of alumina, carbon, gold, silica, and silicon having pore diameters ranging from a few up to 50 nanometers are presented. The observed peculiarities of nanopore-confined condensed matter are also discussed with regard to applications. A particular emphasis is put on texture formation upon crystallisation in nanoporous media, a topic both of high fundamental interest and of increasing nanotechnological importance, \textit{\textit{e.g.}}, for the synthesis of organic/inorganic hybrid materials by melt infiltration, the usage of nanoporous solids in crystal nucleation or in template-assisted electrochemical deposition of nano structures. \\ \\Pre-print version of the article - for the published article see \color{blue}{\href{http://iopscience.iop.org/0953-8984/27/10/103102/}{Journal of Physics: Condensed Matter 27, 103102 (2015)}}, \href{http://dx.doi.org/10.1088/0953-8984/27/10/103102}{doi: 10.1088/0953-8984/27/10/103102}\color{black} . \\ \\
Online supplementary data available from \color{blue} {\href{http:stacks.iop.org/JPCM/27/103102/mmedia}{stacks.iop.org/JPCM/27/103102/mmedia}}\color{black}.
\end{abstract}

\maketitle
%\begin{counted}
\section{Introduction \& Motivation}
The properties of molecular condensates confined in pores a few nanometers across play a dominant role in phenomena ranging from clay swelling, frost heave, oil recovery and catalysis, to colloidal stability, protein folding and transport across artificial nanostructures, bio-membranes and tissues \cite{Drake1990, Christenson2001, Alba-Simionesco2006, Knorr2008, Binder2008, Perkin2013, Jiang2014}. Nanoporous media are also gaining an increasing relevance in template-assisted (electro-)deposition of nano structures \cite{Huczko2000, Yin2001, Steinhart2002, Sander2003} and in the synthesis of soft-hard hybrid materials \cite{Coakley2003, Ford2005, Hoffmann2006, Sousa2014, Martin2014} by melt-infiltration \cite{Jongh2013}, for example in the field of battery and supercapacitor design \cite{Westover2014} and for the preparation of multifunctional structural materials \cite{Elbert2014, Wang2013}. Moreover, a rapidly increasing number of studies on applications as sensors and actuators, in the design and delivery of drugs, and in protein crystallisation have been reported \cite{Chayen2006}. Therefore the advent of tailorable porous materials, most prominently based on carbon \cite{Holt2006, Presser2011}, silicon \cite{Canham1990, Lehmann1991, Cullis1991, Black2001, Sailor2011, Canham2015}, gold \cite{Erlebacher2001, Qi2013}, silica \cite{Zhao1998, Inayat2013, Kresge2013}, alumina \cite{Keller1953, Masuda1995, Lee2006} and titania \cite{Alberius2002, Macak2005}, has led to a growing interest in the thermodynamic equilibrium and non-equilibrium behaviour of solids and fluids confined in nanoporous media. 

Here, I review this rapidly evolving research field with a particular emphasis on fundamental aspects of structural, thermodynamical and transport properties, and relate them to applications of nanopore-confined soft matter systems. The complexity of the confined solids and fluids is varied from simple building blocks, like noble gas atoms, linear hydrocarbons and neat alcohols to liquid crystals, polymers, ionic liquids, proteins and water. 

Following the notation of the International Union of Pure and Applied Chemistry (IUPAC) \cite{Rouquerol1994} almost solely experimental studies and theoretical backgrounds with regard to ''mesoporous'' media, that are materials with a mean pore diameter between 2 nm and 50 nm, are presented. The very interesting realm of microporous materials with pore diameters smaller than 2 nm (like most zeolites) is governed by new types of collective phenomena and strong confinement effects \cite{Auerbach2003}, e.g. single-file diffusion, and will only marginally be addressed here.

The interesting field of chemical reactions \cite{Moller1998, Wang2014}, in particular catalysis \cite{Taguchi2005} goes beyond the scope of this topical review. Driven both by fundamental and applied biophysical interest with regard to water, ion and macromolecular transport at biomembranes \cite{King2004, Gouaux2005, Zimmermann2011}, there is also a huge number of studies on single-molecular behaviour, most prominently nucleic acids, proteins and water across single nano pores in solid state membranes \cite{Meller2000, Aksimentiev2004, Dekker2007, McMullen2014}. Here I shall focus on the collective behaviour of molecular condensates in bulk nanoporous media and will start in the following section with thermodynamic equilibrium phenomena.

\section{Equilibrium Phenomena}
\subsection{Film Adsorption and Capillary Condensation}
The analogon of the bulk liquid-vapour transition for the confined state is capillary condensation. This phenomenon is not only of fundamental interest \cite{ Knorr2008, Ball1989, Evans1990}. It also provides the basis for one of the most important techniques in order to characterise nanoporous media, that are sorption isotherm measurements \cite{Schueth2002, Gregg1982, Thommes2014}. 

Whereas in the past mostly simple van-der-Waals interacting systems, such as argon and nitrogen, were studied, in the meantime also adsorption and capillary condensation of more complex molecules, such as linear alkanes or liquid crystals have been examined \cite{Sluckin1990a, Kocevar2001, Wolff2010, Huber2013}. 

\begin{figure}[htbp]
\center
\includegraphics[width=0.6\columnwidth]{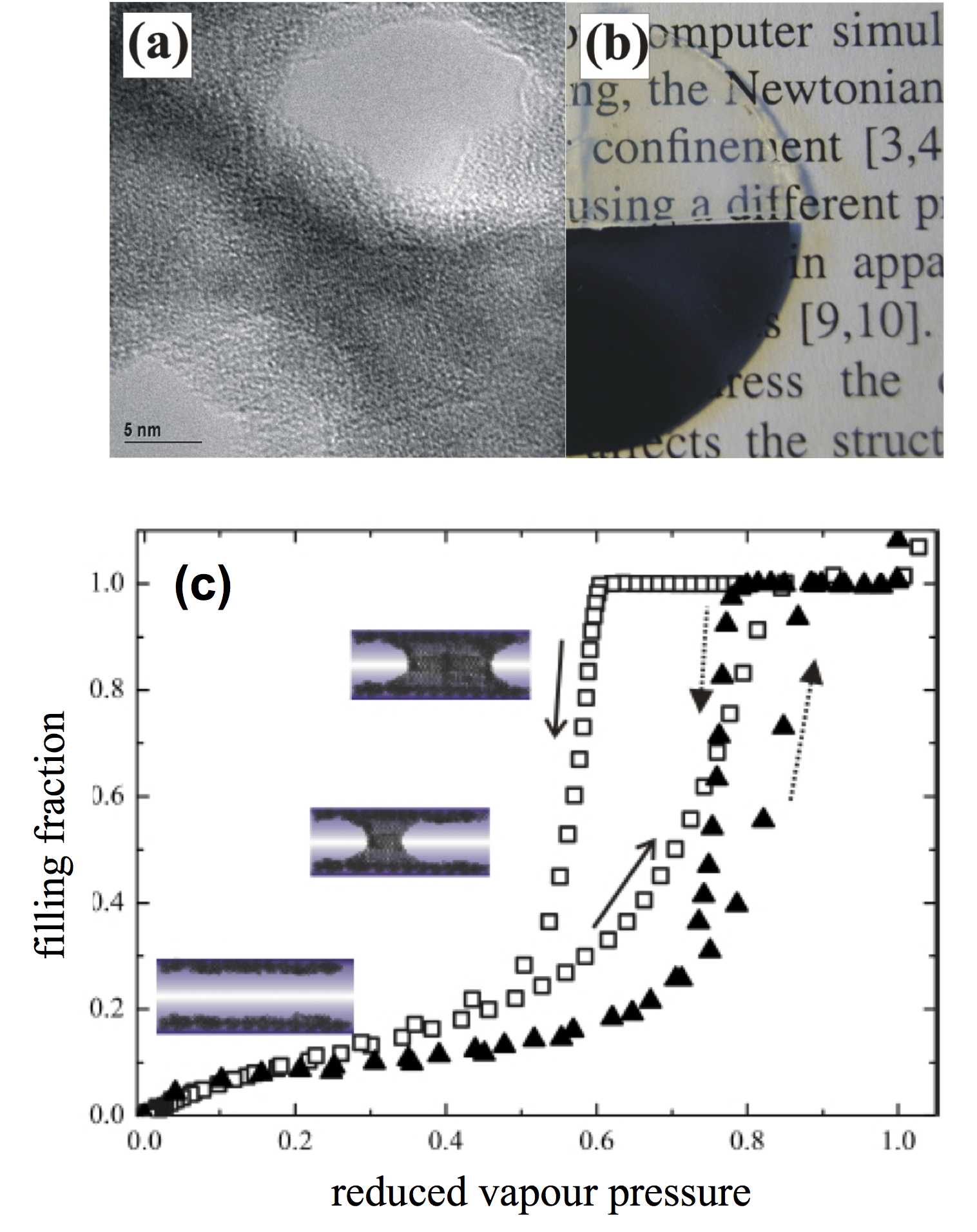}
\caption{(a)~Transmission electron micrograph of pore entrances in porous silicon. (b)~Thermal oxidation at 800\dC~of an as-prepared opaque porous silicon membrane (bottom) results in a monolithic, optically transparent silica membrane (top). (c)~Sorption isotherm of argon (triangles) and n-hexane (squares) condensed in a porous silicon membrane with a mean pore diameter of 10~nm at $T=$86~K and $T=$273~K, respectively. The arrows indicate capillary condensation (in ad- and desorption) of the liquids in the nanoporous membrane.} \label{fig:IsothermArC6Silicon}
\end{figure}

Arguably, the most powerful technique for exploring physisorption and capillary condensation is a sorption isotherm measurement, where the amount of condensate adsorbed in a porous medium is measured as a function of its vapour pressure at a fixed temperature $T$. As a first example of a nanoporous host medium for such an experiment we present mesoporous silicon. The preparation route of this material was rather accidentally found 1956 by Arthur Uhlir Jr. and Ingeborg Uhlir at the Bell Labs \cite{Sailor2011, Canham2015}. They observed the formation of parallel channels a few nanometers across aligned in the $\langle$100$\rangle$ crystalline direction in electropolishing experiments on single-crystalline silicon - see Fig.~\ref{fig:IsothermArC6Silicon}(a,b). The etching depth and accordingly the pore length can be simply controlled via the etching time \cite{Lehmann1991, Lehmann2000}. Moreover, on demand, the porous matrix can be detached from the underlying silicon wafer after finishing the etching process rendering this porous structure particularly versatile for studies of nanopore-condensed matter \cite{Knorr2008, Canham2015, Dolino1996, Faivre1999, Coasne2002a, Kovalev2002, Kunzner2003, Wallacher2004, DeStefano2007, Rumpf2008, Anglin2008, Kumar2008, Khokhlov2007, Naumov2008, DeTommasi2009, Acquaroli2011}.

In Fig.~\ref{fig:IsothermArC6Silicon}c volumetric isotherms of a short chain hydrocarbon (n-hexane) is depicted in comparison with an argon isotherm for vapour condensation in a silicon membrane with 10~nm pore diameter. The adsorbed amount of molecules is plotted versus the vapour pressure of the adsorbed species (here reported as normalised pressure $p$ with respect to the vapour pressure of the bulk liquid at the given temperature $T$). 

Typically, for sorption in nanoporous media one can distinguish two regimes as illustrated in the insets of Fig. \ref{fig:IsothermArC6Silicon}c: the film-condensed regime, where in reversible manner with respect to ad- and desorption the molecules form a film on the pore walls. Beyond a critical vapour pressure $p_c$ in adsorption the second characteristic sorption regime, capillary condensation sets in, that is the formation of liquid bridges with concave menisci in the pore centre. Upon further adding of molecules the pores are completely filled by a movement of the menisci toward the pore ends, in the ideal case at the fixed value $p_c$. As soon as the pore space is completely filled, a plateau is reached (filling fraction $f=1$) and eventually bulk droplets are formed outside of the porous medium ($p=1$). 

The characteristic capillary condensation pressure $p_c$ depends on the curvature of the meniscus formed upon capillary condensation and thus on the pore radius. Vice versa, the pore radius or its variation in pore space can be determined by an analysis of the sorption isotherm shape: For a real system with a sizeable pore diameter variation both capillary condensation and capillary evaporation branches are smeared and not abrupt occurring at fixed values of $p$ and this smearing can be related to the pore size distribution. This can also be seen in the two isotherms of Fig. \ref{fig:IsothermArC6Silicon}c, where for example the capillary condensation of Ar is observable for $0.75<p<0.85$ and for n-hexane at $0.55<p<0.75$.

Sometimes, there is, instead of the plateau at $f=1$, an additional smooth increase of the adsorbed amount of liquid towards $p=1$ observable. This so-called post-filling regime upon approach of bulk liquid-vapour coexistence and after filling of the mesoporous pore spaces originates in gas condensation in the tapered pore mouths, in niches or in thin films between the porous grains, in the case of a powder \cite{Rascon2000, Gang2005, Kityk2008a}. 

\begin{figure}[htbp]
\center
\includegraphics[width=0.7\columnwidth]{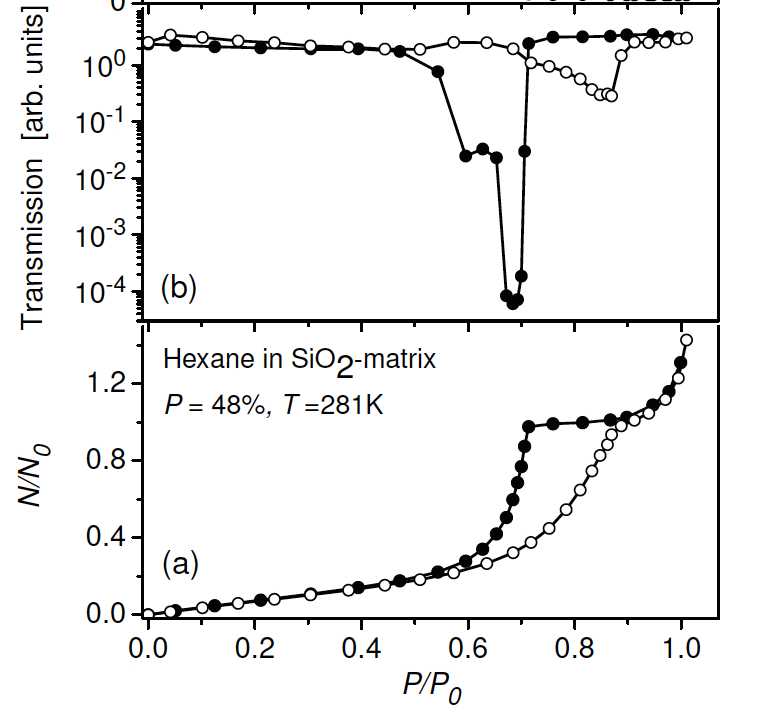}
\caption{(a) Sorption isotherm of n-hexane condensed in a silica membrane at $T=$281~K along with the simultaneously measured optical transmission (b). Open symbols refer to adsorption, solid symbols to desorption. Reprinted (adapted) with permission from Kityk \etalp \cite{Kityk2009}. Copyright
2009 American Physical Society.} \label{fig:OpticsCapillaryCondensationHexanepSilica}
\end{figure}

Note, however, that even for a perfect cylindrical pore the liquid-vapour transition is hysteretic. The capillary evaporation pressure $p_e$ is smaller than the capillary condensation pressure, since thermodynamically metastable liquid layers grow on the pore walls before onset of capillary condensation in adsorption. Cole and Saam treated this phenomenon in a mean-field description for van-der-Waals interacting systems in cylindrical pore geometry \cite{Cole1974, Saam1975}. In many real porous media, for example sponge-like systems like porous Vycor glass \cite{Nordberg1944, Levitz1991} or silica gels, this hysteretic behaviour is additionally affected by the pore connectivity and the pore radius variation. The resulting pore-network effects can dramatically slow down the sorption dynamics and result in a plethora of long-living metastable states \cite{Valiullin2006}. Additionally, the emptying of wide pore segments can be hampered, if they are surrounded by narrow pore segments filled with liquid. This phenomenon is called pore blocking and can result in a coarsening of the pore condensate over a large range in length scales.

For sponge-like, transparent glasses this coarsening of the pore condensate manifests itself by light scattering during liquid condensation, and even more impressively during liquid evaporation. For example, Vycor glass, with its relatively large pore diameter distribution turns completely white upon onset of capillary evaporation. Pore blocking effects and the nucleation of connected vapour-filled pore clusters result in a structuring of the pore condensate on length scales much larger than the nanoscopic pore diameter, \textit{i.e.}, on the length scale of visible light \cite{Kityk2009, Page1993, Page1995,Soprunyuk2003,  Ogawa2013, Huber2013}. 

Interestingly, this phenomenon occurs even in the isolated, tubular channels of a monolithic silica membrane prepared by thermal oxidation of porous silicon - see Fig.~\ref{fig:IsothermArC6Silicon}b. Whereas the matrix stays optically transparent on adsorption, it starts to strongly scatter light and thus turns completely white upon onset of liquid desorption. The optical transmission drops by four orders of magnitude in a Laser light experiment with a nanoporous glass sheet of 0.4 mm thickness \cite{Kityk2009} - see Fig.~\ref{fig:OpticsCapillaryCondensationHexanepSilica}b. Obviously, the pore size variation along with pore surface roughness in the isolated channels result in a similarly large coarsening of the liquid distribution as for sponge-like Vycor glass. This conclusion is corroborated by mean-field calculations for fluid adsorption, which indicate a strong influence of geometric inhomogeneities (and thus quenched random disorder) on the capillary condensation and hysteresis phenomenology even for such a simple pore topology \cite{Naumov2008, Naumov2009}. 

Obviously, the complex interplay of these network effects with the hysteresis intrinsic to the single-pore capillary condensation phenomenology render the pore size analysis from sorption isotherms much more challenging than the simple picture of capillary condensation initially presented. Moreover, for pore sizes comparable to the molecular diameters of the adsorbed species, macroscopic concepts for the description of the condensates (such as meniscus radius, mean density, surface tension) are not properly defined or may substantially deviate from the bulk state. For example, confinement-induced layering in the pore wall proximity can alter the structural and thermodynamic state of the condensate. Then, an appropriate description can solely be achieved by microscopic modelling of the adsorption processes \cite{Evans1990, Gelb1999, Gelb1999b, Ravikovitch2001,  Klapp2002,  Neimark2003,Bohlen2005, Paul2005, Binder2008, Lu2010, Winkler2010, Monson2012, Nguyen2013,  Landers2013, Miyahara2014}.

\begin{figure}[htbp]
\center
\includegraphics[width=0.6\columnwidth]{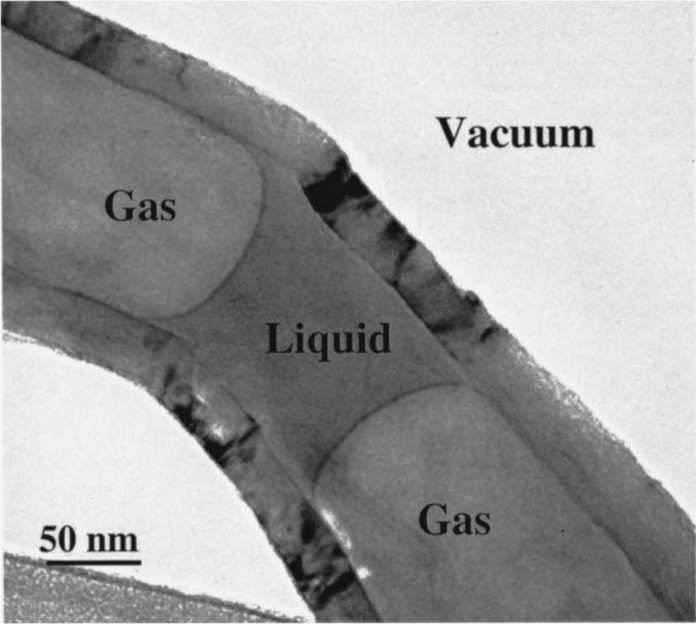}
\caption{Transmission electron micrograph of an elbow portion of a capped carbon nanotube made hydrothermally. The nanotube contains a liquid inclusion constrained between two menisci separating the aqueous liquid from the adjoining gas. The environment outside the nanotube is the high vacuum of the electron microscope column. The volume of the liquid inclusion is of the order of 10$^{-18}$ litres. Reprinted (adapted) with permission from Megaridis \etalp  \cite{Megaridis2002}. Copyright 2002 AIP Publishing LLC.} \label{fig:TEMLiquidCarbon}
\end{figure}

Leaving aside the coarsening of the pore condensate upon desorption, typical geometrical features of liquids condensed in nanoporous media, such as their menisci, are on the nanometer scale. This renders their experimental study particularly challenging. X-ray and neutron tomography methods have not (or not yet) reached such a spatial resolution \cite{Dierolf2010, Wilke2012}. Electron microscopy has the appropriate spatial resolution. However, the high vapour pressures of most liquids make high-resolution studies with this ultra-high vacuum technique difficult. Nevertheless, Megaridis \etalp \cite{Megaridis2002} and Rossi \etalp \cite{Rossi2004} demonstrated in pioneering experimental studies the principal feasibility of liquid menisci shape studies in end-capped carbon nanotubes by transmission electron microscopy with nano- and subnanometer resolution - see Fig.~\ref{fig:TEMLiquidCarbon}. This allowed them to directly document a remarkable robustness of macroscopic concepts with regard to the interface shape and of the mobility of the confined liquid. %Albeit, the sealed channels hampered filling fraction-dependent measurements.

\begin{figure}[htbp]
\center
\includegraphics[width=0.6\columnwidth]{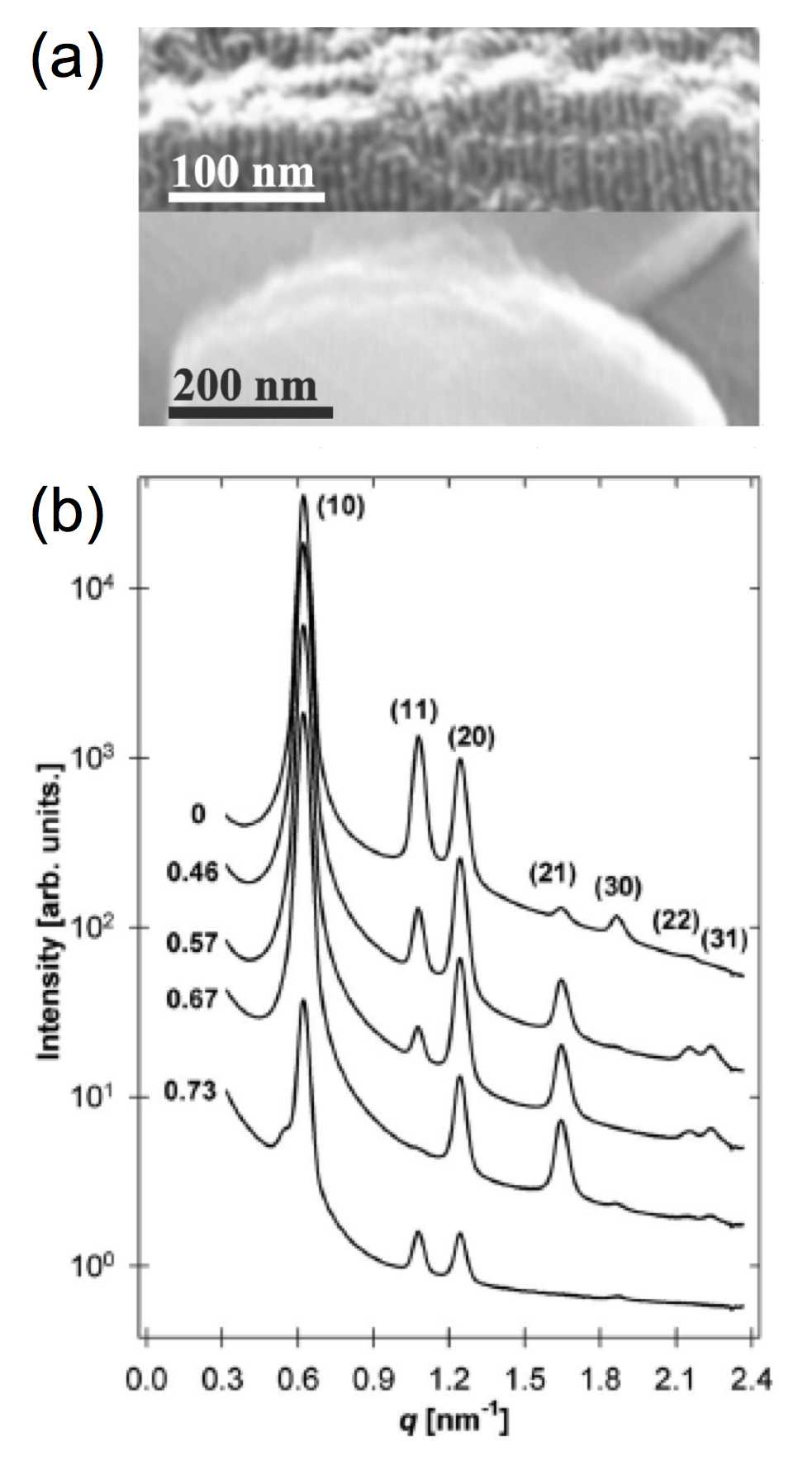}
\caption{(a) Scanning electron micrographs of a SBA-15 grain taken with different spatial resolutions as indicated in the figure \cite{Schaefer2008}. Reprinted (adapted) with permission from Schaefer \etalp  \cite{Schaefer2008}. Copyright 2008 American Physical Society. (b) A typical series of small-angle x-ray diffraction profiles (scattering intensity versus wave vector transfer $q$) of C$_5$F$_{12}$ in SBA-15. For the sake of clarity, the curves are shifted vertically by a factor of 3 with respect to each other. The values given at the left side of the figure refer to relative vapour pressures $p/p_{0}$. The top four profiles were measured at pressures below the pore condensation point and the last profile was measured above the pore condensation point. Reprinted with permission from Zickler \etalp  \cite{Zickler2006}. Copyright 2006 American Physical Society.} \label{fig:SEMSAXSPentanSBA15CapillaryCondensation}
\end{figure}

X-ray and neutron scattering are the most powerful techniques to gain information on the microscopic arrangement of liquids in or at the surface of nanoporous media as a function of pore filling (and thus vapour pressure in the coexisting gas phase) \cite{Hofmann2005, Zickler2006, Alvine2006, Gommes2013, Gang2013}. Hofmann \etalp \cite{Hofmann2005} and Zickler \etalp \cite{Zickler2006} studied the capillary condensation of fluids in the template-grown mesoporous material SBA-15 \cite{Zhao1998}, a powder of grains with hexagonally arranged, parallel nano capillaries by small angle X-ray diffraction - see Fig.~\ref{fig:SEMSAXSPentanSBA15CapillaryCondensation}. The diffraction patterns consist of Bragg peaks typical of the 2D-hexagonal arrangement of the tubular channels. As a function of fluid adsorption the intensities of these Bragg peaks vary in a characteristic manner, dictated by the changes in the formfactor of the single channels as a function of film- and capillary-condensation. Therefore, the authors were able to study \textit{\textit{in-situ}} the fluid sorption and could confirm phenomenological models, in particular the Saam-Cole model \cite{Cole1974, Hofmann2005} with regard to the onset of capillary condensation. Those studies also revealed scattering contributions, which are explainable only by a significant microporosity of the pore walls \cite{Imperor-Clerc2000}. Presently, pore wall corrugations along the cylindrical channels and how they additionally influence fluid adsorption of SBA-15 materials are intensively discussed in the literature \cite{Gommes2009, Gommes2012, Morishige2013}.

%Post-filling regime (Tommy)
\subsection{Capillary Sublimation}
Capillary sublimation, that is the vapour-solid transition in confined spaces has also attracted quite some interest over the years. It exhibits a phenomenology quite analogous to capillary condensation \cite{Christenson2001, Knorr2008, Huber1999}. 

\begin{figure}[htbp]
\center
\includegraphics[width=0.8\columnwidth]{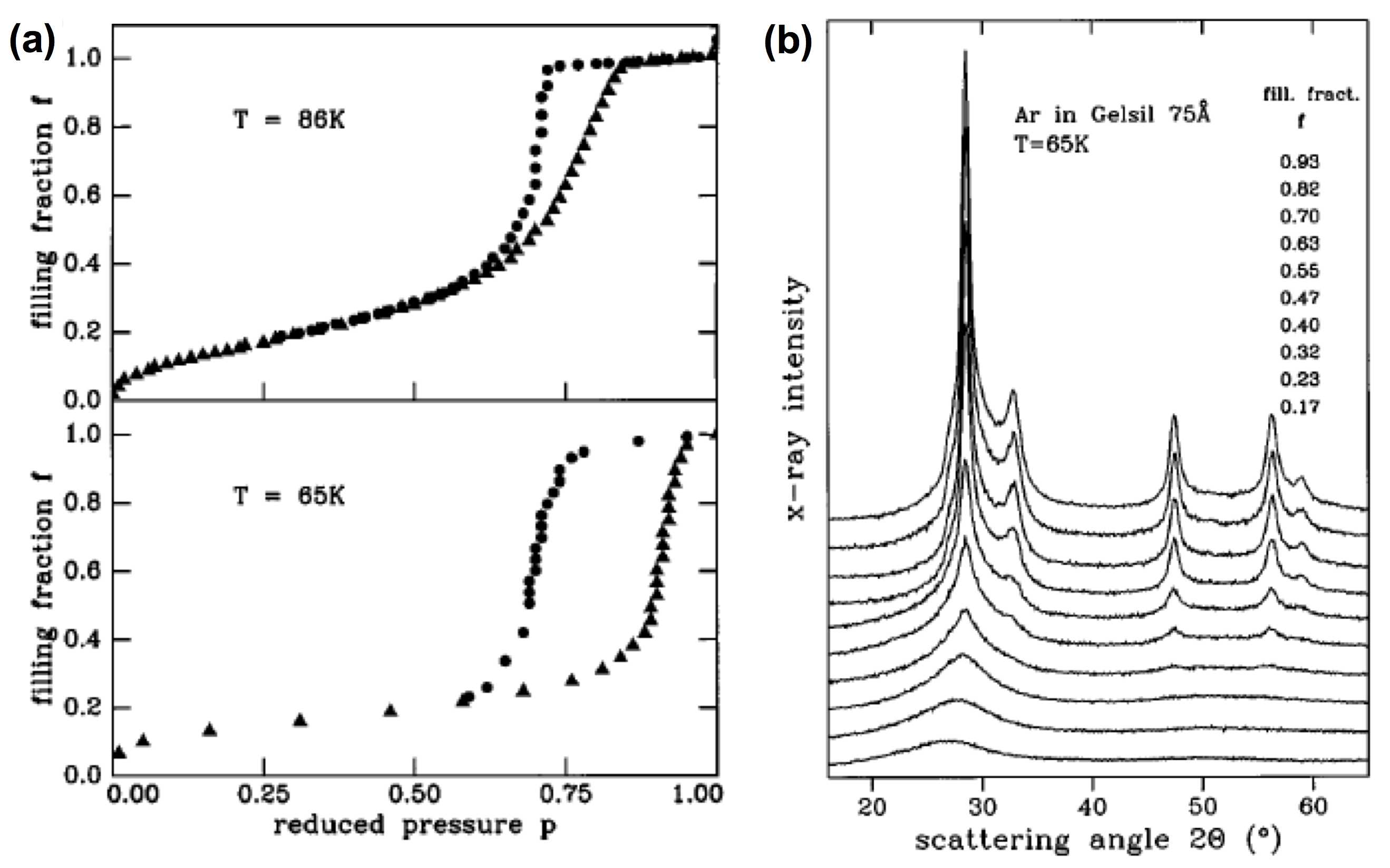}
\caption{(a) Adsorption-desorption isotherms of argon in the liquid state at 86~K (capillary condensation) and in the solid state at 65~K (capillary sublimation) in a controlled pore glass with 7.5~nm mean pore diameter. (b) The evolution of the X-ray diffraction patterns upon capillary sublimation of argon into this matrix. Reprinted (adapted) with permission from Huber and Knorr \cite{Huber1999}. Copyright 1999 American Physical Society.} \label{fig:IsothermXDiffArGelsilSublimation}
\end{figure}

As an example may serve the capillary sublimation of argon in a controlled pore glass, a sponge-like silica-gel with 7.5~nm mean pore diameter \cite{Huber1999}. Plotted in Fig.~\ref{fig:IsothermXDiffArGelsilSublimation} is the filling fraction $f$ as a function of argon vapour pressure upon formation of liquid ($T=86$~K) and solid argon ($T=65$~K) in pore space. A comparison of the two isotherms reveals immediately that for both cases two sorption regimes can be distinguished, a film-condensed and a capillary-condensed or more strictly spoken capillary-sublimated state with hysteresis. 

Parallel to the argon sorption experiment, X-ray diffraction patterns of the pore sublimate were \textit{in-situ} recorded. They allow one to explore the structural state of the pore condensate as a function of reduced vapour pressure, see Fig. \ref{fig:IsothermXDiffArGelsilSublimation}. In the film-condensed state mono- (or multi-) layers grow on the glass surface with short-range hexagonal in-plane arrangements. Starting with capillary sublimation, that is the steep increase in the sorption isotherm upon adsorption, argon crystals with grain sizes larger then the pore diameter form. An analysis of the diffraction patterns yields that solid argon adapts to the confined geometry by the introduction of a sizeable number of stacking faults in comparison to the perfect ABC close-packed stacking sequence of hexagonal layers, typical of the bulk face-center-cubic structure \cite{Huber1999}. The number of these stacking faults is smaller than the one observed for the pore solid prepared by freezing of the liquid - see Ref. \cite{Huber1999} for a detailed discussion on the stacking fault probabilities in pore-confined argon. 

Specific heat measurements \cite{Wallacher2001a, Amanuel2009}, early self-diffusion measurements with nuclear magnetic resonance on pore condensates in porous glasses \cite{Kimmich1996, Stapf1997} and more recent studies of the elasticity of pore-condensed argon \cite{Schappert2008} suggest that the contact layers at the pore walls formed during film condensation are highly mobile, liquid-like over a sizeable temperature range below the freezing temperature of the condensate in the pore center.

It is interesting to note that capillary sublimation, again in analogy to capillary condensation, is also drastically affected by the quenched random disorder intrinsic to most nanoporous media. Detailed experimental studies of the resulting light scattering and a discussion with regard to intimately related rearrangements of the pore condensate upon freezing and melting as a function of filling of the porous medium, can be found in Refs. \cite{Soprunyuk2003, Wallacher2005} for argon in Vycor.
  
\subsection{Protein Adsorption}
In recent years, there has been an increasing interest in the properties of complex molecules, most prominently polymers and proteins in porous media. Especially nanoporous materials and their interaction with biomolecules have been in the focus of intensive research \cite{Anglin2008, TinsleyBown2000, Prestidge2007, Sailor2007, Salonen2008, Chen2011, Fried2013, Boecking2012, Lasave2013, Hartmann2013}, since they combine a high inner surface area with pore sizes large enough to comfortably host most proteins and enzymes \cite{Adiga2009}. Some proteins show an enhanced stability against denaturating conditions (chemical as well as thermal)  and retain or even increase their electro-chemical activity when encapsulated in silica mesopores \cite{Washmon-Kriel2000}. More trivially, since microorganisms like bacteria or fungi are far too large to penetrate mesoporous structures, encapsulated proteins are well protected from biological decomposition. This opens a wide field of biochemical applications that employ the enzymatic activity of proteins under conditions which would otherwise destroy the enzymes.

The analogies between artificial nanopores and biological transport pores in biomembranes \cite{King2004, Gouaux2005, Zimmermann2011, Mahendran2013} has also stimulated fundamental studies on biomolecular adsorption, diffusion, and translocation processes in nanoporous media \cite{Keil2000, Causserand2001, Katiyar2006, Striemer2007, Javidpour2008, Uehara2009, Javidpour2009, Firnkes2010, Lee2012, Plesa2013, Mihovilovic2013}. Moreover nanoporous media promise a vast variety of applications in biochemical technologies, like enzymatic catalysis\cite{Hartmann2005}, protein crystallisation \cite{Chayen2006, Khurshid2014} {and the fractionation of biological fluids like blood into their individual components} \cite{Yang2006}. New means of targeted drug delivery\cite{Adiga2009, Slowing2007, Liu2013, Herranz-Blanco2014}, biosensors \cite{Chen2011, Janshoff1998, Kilian2009, Guan2011} and nanocomposite materials \cite{Boecking2012}, partially inspired by Nature's biomaterials \cite{Fratzl2007,Zlotnikov2014}, can be envisaged. 

Their intermolecular interaction of proteins and the interaction with the nanoporous medium are of course much more complicated than the ones of rare-gases or short-length, normal alkanes discussed above. In the case of proteins, the sorption and molecular mobility in nanoporous media depends sensitively on the charge state and the chemistry of the porous medium. Beyond pure size-dependent steric interactions, van-der-Waals interactions, electrostatic, entropic interactions, and also the interaction directly and indirectly mediated by counter-ions in solvents can be of importance. 

The highest pore loadings, \textit{i.e.}, the amount of bound protein per mass of nanoporous medium, are often found close to the isoelectric point pI, the pH-value where the overall charge of the protein is zero. For example, Vinu \etalp \cite{Vinu2004} studied the pH-dependent adsorption of horse heart \cytc~a globular protein on SBA-15 and observed the highest pore loading at pH $9.6$ which is only slightly below the pI of \cytc. This is often interpreted in terms of a balancing between an attractive protein-wall interaction and protein-protein repulsion \cite{Hartmann2005, Miyahara2006}. The loss of electrostatic repulsion between the molecules at their pI facilitates the observed dense packing of the adsorbing molecules while the attraction to the silica surface is driven by patches of charged amino acid residues on the protein surface \cite{Essa2007, Hartvig2011}. At pH values far from the isoelectric point, the proteins will repel each other and thus cause a less compact packing density on the adsorbing surface.

\begin{figure}[htbp]
\center
\includegraphics[width=0.6\columnwidth]{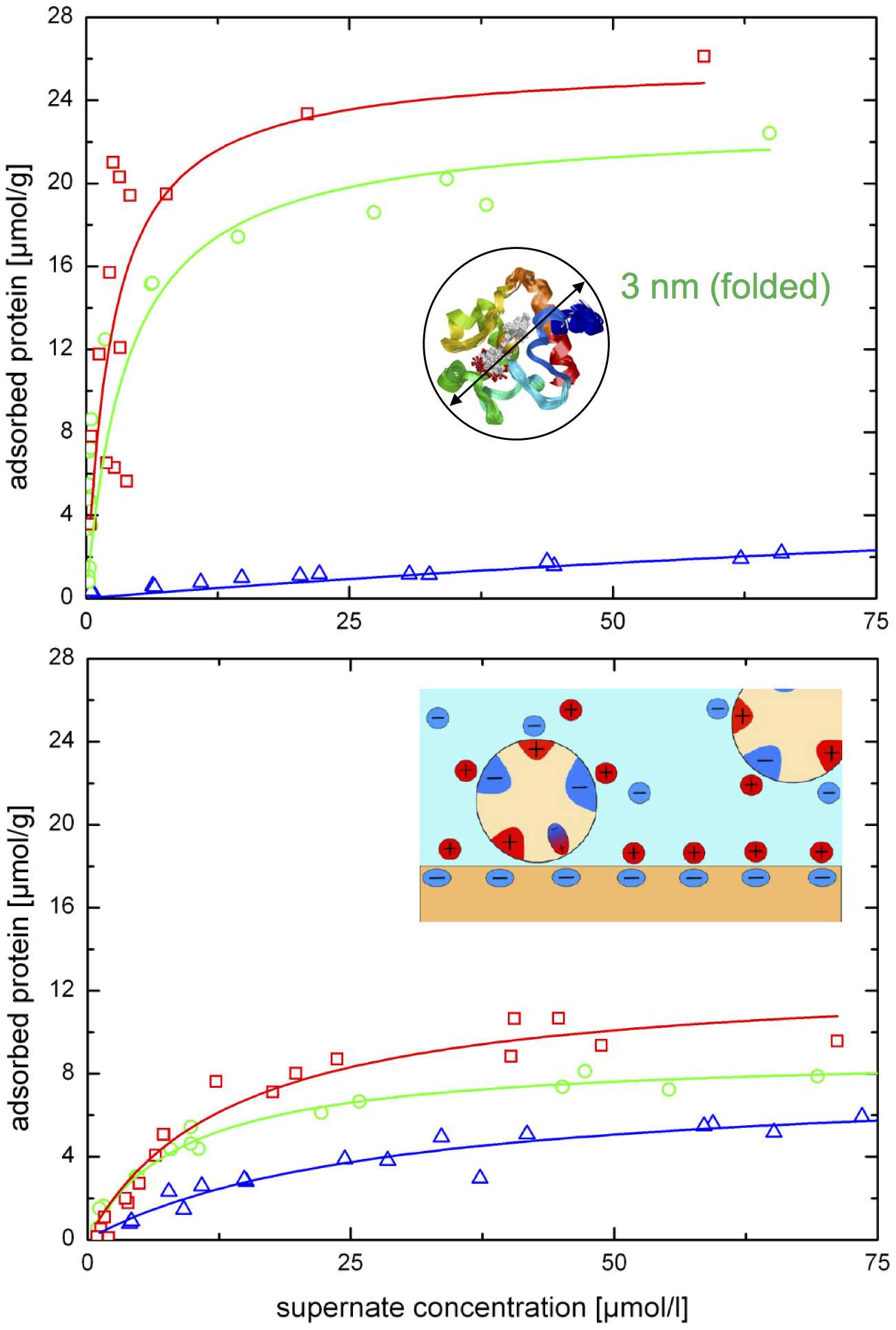}
\caption{Cytochrome \emph{c} adsorption isotherms for the native, folded state (top) and for the unfolded state (bottom). Solid lines are Langmuir-type fits to the experimental data. The isotherms were recorded at pH\,9.7 (red squares), pH\,6.4 (green circles) and pH\,4.4 (blue triangles). The inset in the upper panel depicts the \cytc structure. The inset in the lower panel illustrates the influence of the protein and pore wall charging as well as the counter ions in the solution on the adsorption process. Reprinted (adapted) with permission from Moerz and Huber \cite{Moerz2014}. Copyright 2014 American Chemical Society.} \label{fig:IsothermProteinSBA15}
\end{figure}

A study by Moerz and Huber \cite{Moerz2014} on \cytc adsorption in SBA 15 revealed that also the folding state of the protein can sensitively affect the adsorption behaviour in a nanoporous medium - see Fig.~\ref{fig:IsothermProteinSBA15}. The experiments were done at three different pH values corresponding to three fundamentally disparate electrochemical conditions. The red line and symbols correspond to a pH of 9.7, close to the isoelectric point of the \cytc. The protein is virtually uncharged under these conditions, while the silica exhibits strong negative surface charges. The green line and symbols were measured under near-neutral conditions where both the protein and the surface are charged with opposite signs, resulting in mutual attraction. Finally, the blue lines and symbols represent the measurements near the SBA-15's isoelectric point. The protein has a strong positive overall charge at this pH, while the surface is mostly neutral.

For both folding states of \cytc the pore loadings, \textit{i.e.}, the amount of adsorbed protein, decreases with decreasing pH value. This decrease is quite moderate for unfolded \cytc, but severe for the native, folded type. Unfolded \cytcol, on the other hand, seems less susceptible to changes in the buffer pH. These peculiarities along with an additional study on the influence of the ionic strength of the buffer solution on the pH-dependency revealed the importance of an attractive interaction between protein and pore wall, that is associated with the entropy gained by the release of immobile counter ions around the charged protein and at the charged surface upon protein adsorption at the pore wall (counter-ion release mechanism).

\subsection{Sorption-induced matrix deformation, Young-Laplace and tensile pressures}
Finally, it should be mentioned that Young-Laplace pressures act at the liquid/vapour, liquid/solid, and vapour/solid interfaces in nanoporous media, because of the curved shapes of these interfaces and the interfacial tension. Their magnitude can in first approximation be estimated by the Young-Laplace formula, $\Delta p=\sigma/R$, where $\sigma$ is the interfacial tension and $R$ is the Gauss mean curvature of the interface considered. At the concave menisci of a capillary-condensed liquid bridge in a nanopore, which means a negative mean curvature, the resulting, negative pressure in the capillary condensed liquid can easily reach hundred MPa. By the same token, in equilibrium this tensile pressure has to be balanced by the mechanically hard confining medium.

It was noticed quite early in the study of nanopore-confined liquids that even materials considered infinitely hard compared to the soft sorbents, like silica or silicon macroscopically deform upon uptake of liquids: In the film-condensed regime the solid macroscopically expands, because of the decrease of the solid/vapour tension at the pore wall by the film formation of the adsorbate (Bangham effect, reported in 1928) \cite{Bangham1928}, then there is a capillary contraction regime upon formation of the concave menisci during capillary condensation \cite{Dolino1996, Amberg1952, Dourdain2008, Guenther2008, Sharifi2014, Balzer2014} and eventually the matrix expands upon complete filling \cite{Guenther2008, Findenegg2010}, when the highly curved menisci vanish. 

A proper treatment of this phenomenon considers not only the Young-Laplace pressure (in a macroscopic sense), but also the dispersive solid/fluid interaction potentials which additionally modify the pressure within the confined liquid \cite{Landers2013, Guenther2008}. Gor and Neimark \cite{Gor2010, Gor2011} developed a pertinent phenomenological model which semi-quantitatively describes a series of experimental studies on physisorption-induced deformations of nanoporous media published in recent years \cite{Prass2009, Gor2013}. Moreover, it has been discussed that the pressure acting on the confining solid medium and on the confined pore condensate, if it is in a solid state, cannot be simply identified with the isotropic solvation pressure, by contrast it corresponds to the surface stress at the solid-liquid (or solid-solid interface), which has to be linked to bulk stress and strain in the solid via the generalised capillary equation for solids \cite{Shao2010}. 

Sorption-induced pressures or deformations are of obvious interest for applications of nanoporous media as sensors or mechanical actuators. In the case of soft porous materials, \textit{e.g.}, aerogels \cite{Herman2006}, porous polymers \cite{Weber2008} and biomaterials \cite{Fratzl2007}, the deformations are particularly large and can therefore be used for large-scale, reversible locomotion \cite{Elbaum2007, Zickler2012, Zhao2014} induced via fluid sorption. And this in a remarkably energy-efficient manner, as has been documented in a recent study on water sorption-induced deformation of wood by Bertinetti, Fischer and Fratzl \cite{Bertinetti2013}.

Given the large Young-Laplace pressures achievable in pore condensates, nanoporous hard matrixes can be readily employed to explore the effects of large tensile pressures on condensed matter phases. In fact, we will see in the following sections a couple of instances, where tensile pressures affect the properties of pore condensates markedly. For example, recently Schappert and Pelster \cite{Schappert2014} found experimental hints in ultrasonic measurements that the longitudinal elastic modulus of liquid argon is a few percent, but significantly, smaller under negative Young-Laplace pressure compared to bulk liquid argon under its own vapour pressure. This is maybe not too surprising, since the equilibrium argon-argon distance (without external pressure) is determined by a minimum in the Lennard-Jones potential $\Phi_{LJ}$, which is asymmetric (and more importantly not parabolic), given by a repulsive part scaling with $1/r_{LJ}{12}$ and an attractive one scaling with $1/r_{LJ}^{6}$. Here, $r_{LJ}$ is the interatomar distance.  Hence, for a larger argon-argon molecular distance (induced by the tensile pressure acting on the system), the derivative of $\Phi_{LJ}$ with respect to $r_{LJ}$, which determines the elastic modulus, is smaller than in the equilibrium situation. The ultrasonic measurements of Schappert and Pelster have, however, the shortcoming that they are sensitive to the ratio of elastic modulus and mass density only. Since the mass densities are also affected by the tensile pressure (as has been documented for a series of pore solids in X-ray diffraction experiments \cite{Huber1999a, Morishige2006, Kojda2014}) and the authors had to rely on bulk data sets, a full quantitative understanding of this interesting effect presumably necessitates complementary simulation studies.

\subsection{Liquid-Solid Transition, Crystallisation and Texture Formation}
It has been known for a long time that freezing and melting or more generally spoken the liquid-solid transition in nanoporous media are significantly affected both by pure spatial confinement and by the interaction with the pore walls \cite{Christenson2001, Alba-Simionesco2006, Knorr2008, Jiang2014}. Changes in the relative chemical potentials of the confined solid and liquid phases with respect to the corresponding bulk phases \cite{Huber1999, Duffy1995} along with interfacial melting (starting from the pore walls) \cite{Wallacher2001a} could be successfully related to reduced solidification temperatures compared to the corresponding bulk systems and to peculiar hysteresis phenomena between cooling and heating of the pore condensates \cite{Christenson2001, Alba-Simionesco2006, Knorr2008}.

Even for the liquid-solid transition of a simple system like argon or nitrogen in SBA-15 specific heat measurements as a function of fractional filling of the pores indicate a remarkably complex phenomenology \cite{Schaefer2008, Moerz2012}: While interfacial melting leads to a single melting peak in the specific heat measurements, homogeneous and heterogeneous freezing along with a delayering transition for partial fillings of the pores result in a freezing mechanism explainable only by a consideration of regular adsorption sites (in the cylindrical mesopores) and irregular adsorption sites (in niches of the rough external surfaces of the grains and at points of mutual contact of the powder grains). 

Moreover, the large tensile pressure release upon reaching bulk liquid/vapour coexistence could quantitatively account for an upward shift of the melting/freezing temperature observed while overfilling the mesopores, when the concave menisci of the pore condensate and the corresponding negative Young-Laplace pressures in the pore condensate vanish. Analogous observations for a liquid crystal, water, methanol, and ethanol condensed in nanoporous media \cite{Huber2013, Schreiber2001, Morishige2012, Findenegg2013} document the importance of the large, negative hydrostatic pressures \cite{Tas2003} typical of the capillary-condensed state in nanoscale confinement on the phase transition behaviour.

\begin{figure}[htbp]
\center
\includegraphics[width=0.6\columnwidth]{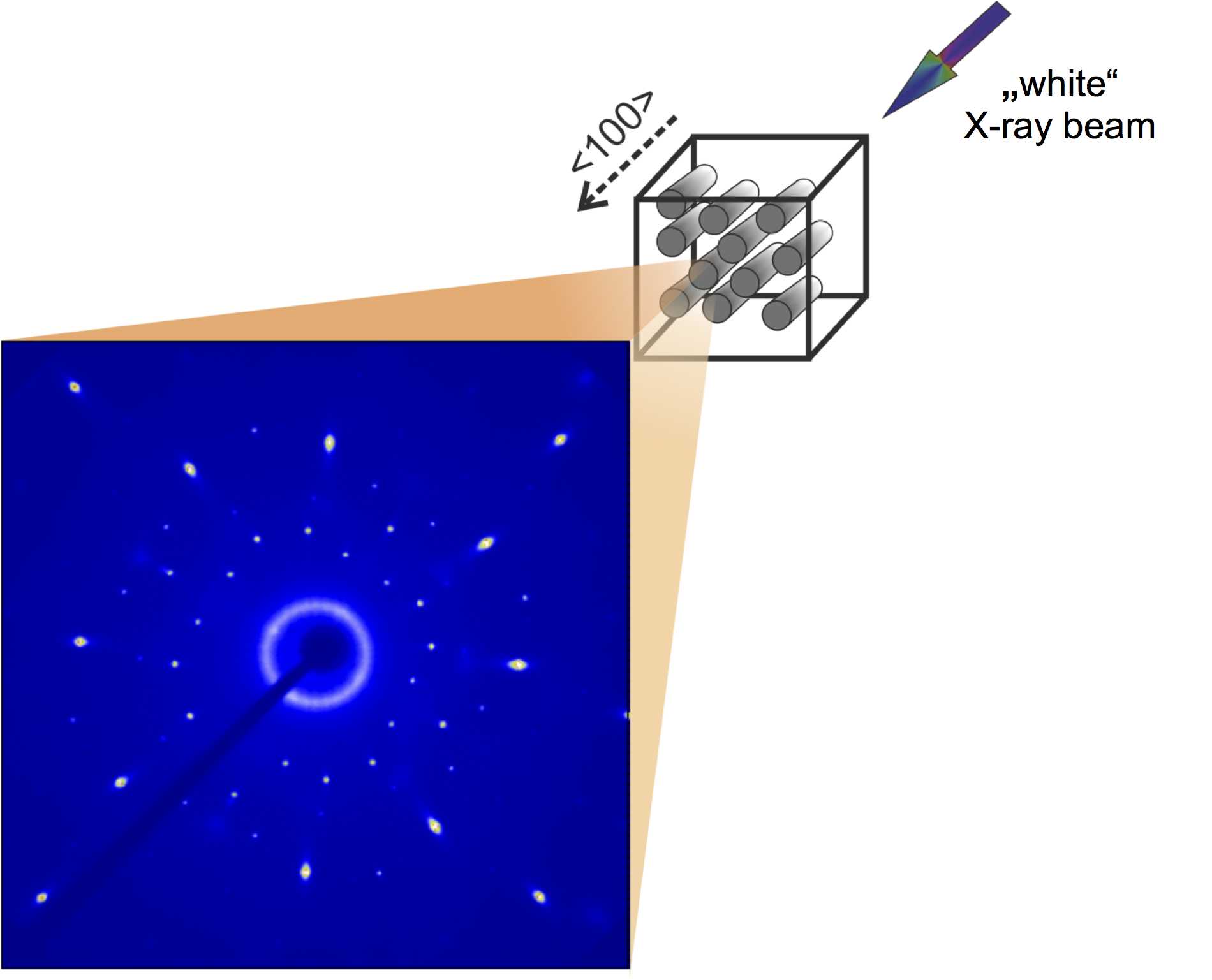}
\caption{Laue X-ray transmission pattern recorded from a nanoporous silicon membrane filled with n-heptadecane at room temperature ($T$=293~K). The incident X-ray beam is parallel to the $<100>$ crystalline direction of the porous sheet and thus parallel to the long axes of the tubular pores and the surface normal of the membrane, as illustrated in the figure. The Bragg peaks in the diffraction pattern result from the single-crystalline matrix, whereas the intensity ring is typical of the first maximum of the structure factor of the nano-confined liquid alkane.} \label{fig:XLaueC24Silicon}
\end{figure}

Recently, the advent of ordered tubular pore arrays in monolithic alumina, silicon and silica, allows for an improved characterisation of the crystalline state in pore space by neutron and X-ray diffraction. Particularly, the orientation of the pore solids can be explored with respect to selected pore directions. Porous silicon with its single crystalline matrix structure allows additionally to relate the orientation of the pore solids to the crystalline directions of the matrix. Moreover, the high crystalline symmetry results in a low scattering background and a well defined scattering pattern of the matrix, which makes the scattering background subtraction of the matrix much easier than for amorphous hosts \cite{Huber1999,Schafer1996, Huber1998a, Huber1999a}. This is demonstrated in Fig.~\ref{fig:XLaueC24Silicon} by a Laue X-ray diffraction image recorded from a mesoporous silicon membrane infiltrated with the melt of a linear alkane n-C$_{17}$H$_{36}$ (C17). The diffraction spots result from the crystalline porous matrix, whereas the central ring represents the first maximum of the structure factor of the nano-confined liquid wax - the chain-chain correlation peak \cite{Henschel2007}. 

\begin{figure}[htbp]
\center
\includegraphics[width=0.8\columnwidth]{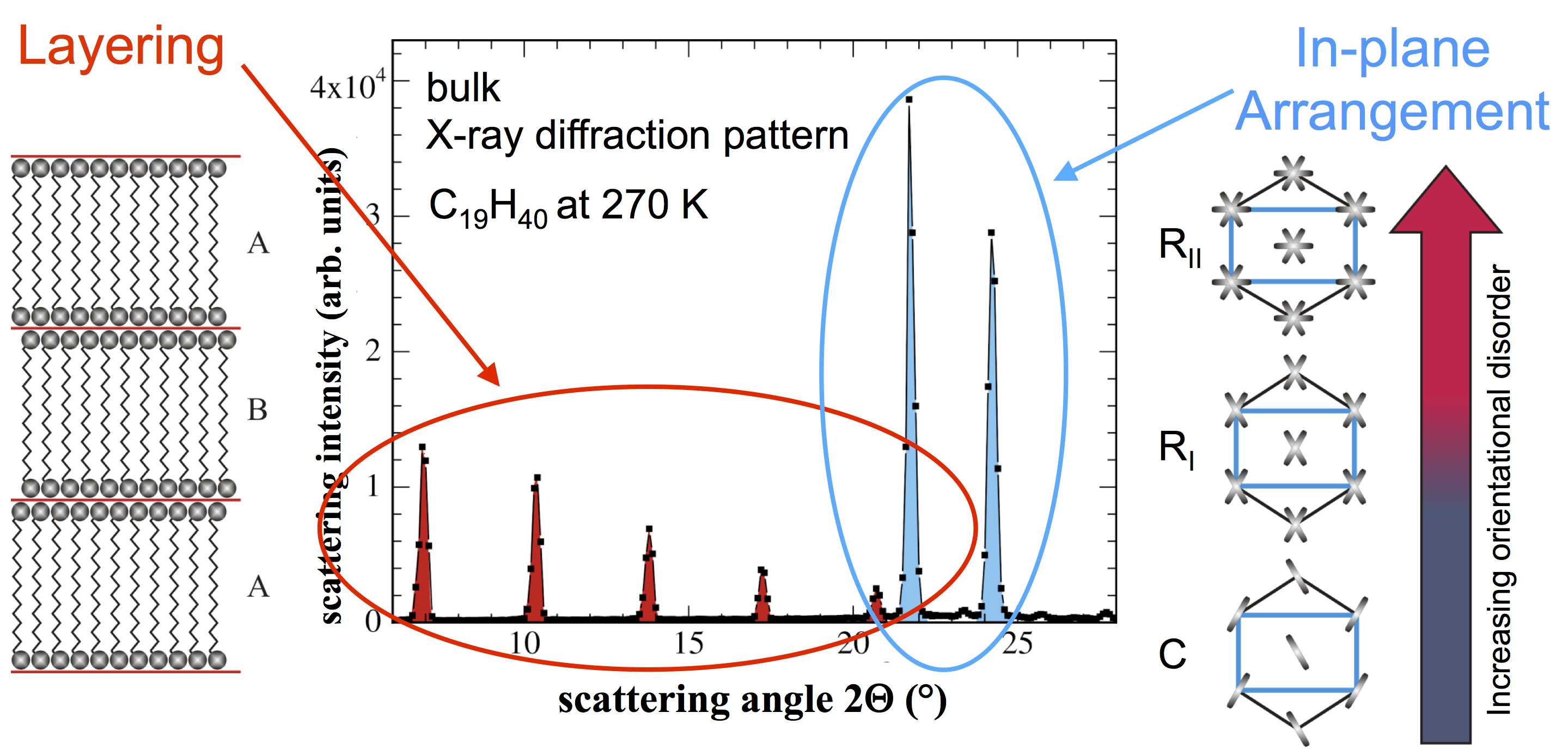}
\caption{X-ray powder diffraction pattern of bulk n-heptadecane at T=291~K. The Bragg peaks typical of the lamellar ordering and of the in-plane order are indicated. The changes in the in-plane backbone ordering upon heating, encompassing a herringbone, a Rotator II and a Rotator I phase, are illustrated in the right panel.} \label{fig:XDiffC19Bulk}
\end{figure}

In the crystalline state of medium-length n-alkanes the C atoms of the zig-zag backbone are all in the trans configuration, so that all of them are located in a plane. Gauche defects which lead to kinks and twists of the -C-C- chain are abundant in longer alkanes, but have little effect on the crystalline structures of medium-length n-alkanes. Thus, the molecular crystals are governed by two architectural principles: Firstly, the molecules form layers. Secondly, within the layers the molecules are 2D close-packed, side by side, in a quasi-hexagonal 2D array. Since the layering periodicity is much larger than the molecule-molecule distance within the layers, these two building principles are nicely separated in diffraction patterns. At low wave vector transfer $q$ (or small scattering angle 2 $\Theta$) a series of layering reflections typical of the lamellar ordering is observable, whereas at somewhat higher $q$ the Bragg peaks typical of the in-plane order can be found - see the powder X-ray diffraction pattern of bulk C17 in Fig.~\ref{fig:XDiffC19Bulk}. 

\begin{figure}[htbp]
\center
\includegraphics[width=0.7\columnwidth]{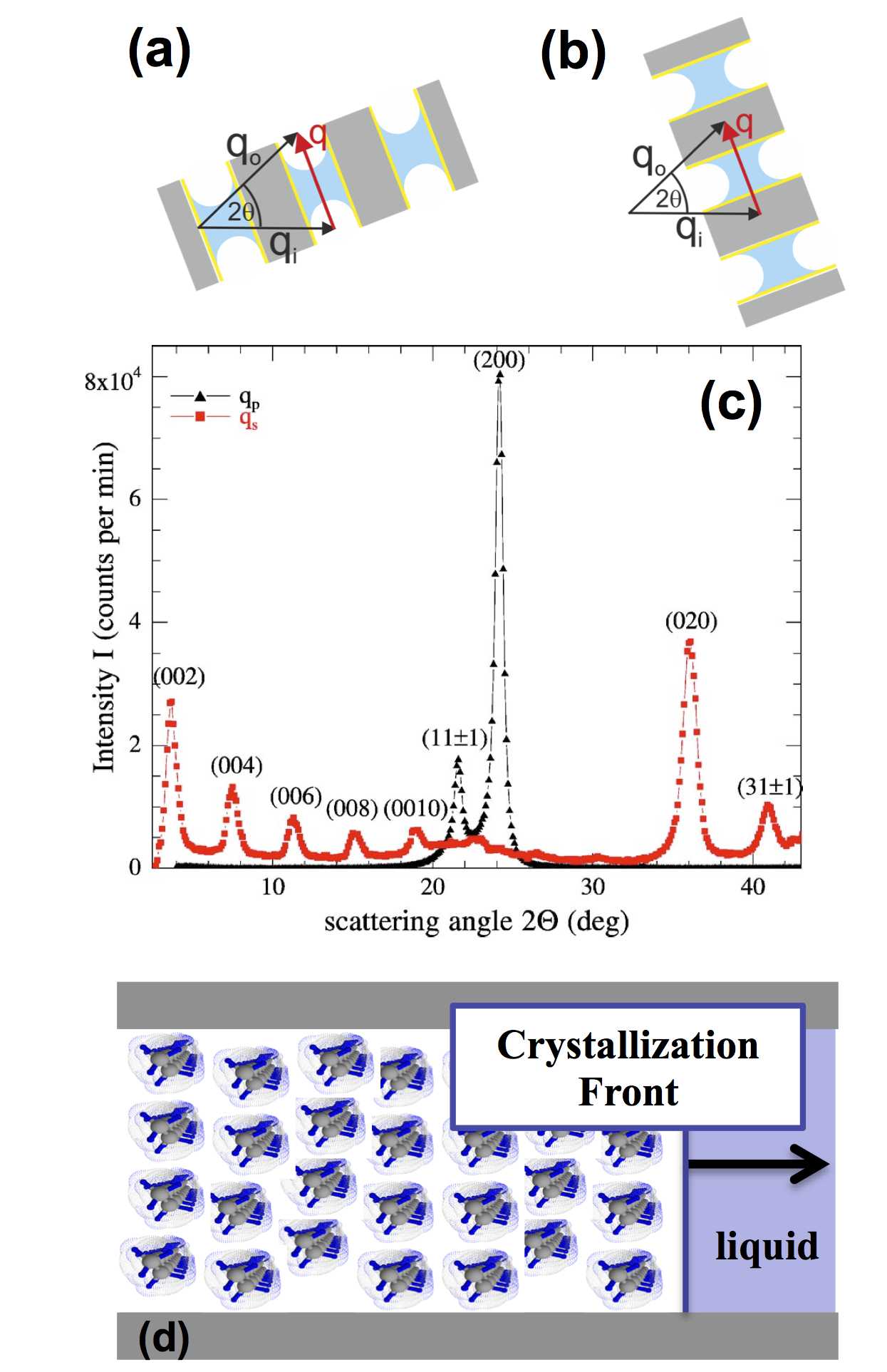}
\caption{X-ray scattering geometry for radial scans with wave vector transfer q$_p$ parallel (a) and q$_s$ perpendicular to the nano channel axes in porous silicon (b). (c) Scattered X-ray intensity for radial scans along  q$_p$ and along q$_s$ of n-heptadecane (n-C$_{17}$H$_{H36}$) confined in nanoporous silicon at 245~K \cite{Henschel2007}. (d) Illustration of the Bridgman crystallisation mechanism for linear hydrocarbons in capillary geometry as discussed in the text. Reprinted (adapted) with permission from Henschel \etalp  \cite{Henschel2007}. Copyright 2007 American Physical Society.} \label{fig:XDiffSiliconC17Texture}
\end{figure}

Upon infiltration in a monolithic porous silicon membrane X-ray diffraction allows one to explore the freezing of these molecular assemblies and to explore the orientation of the crystals with respect to the tubular pores. This is exemplified for C17 in Fig.~\ref{fig:XDiffSiliconC17Texture}: The intensity of the Bragg peaks varies significantly as a function of matrix orientation with regard to the scattering vector. In diffraction scans with wave vector transfer parallel to the long pore axis ($q_p$), the in-plane Bragg peaks dominate the scattering patterns. The layering peaks are absent. Vice versa, in diffraction scans with wave vector transfer perpendicular to the long pore axis ($q_s$)  the layering peaks (00l) have high intensities. This indicates that the stacking direction of the layers is perpendicular to, and the in-plane orientation are parallel to the long axes of the channels. Hence the molecules' long axes are perpendicular to the tubular direction of the pores - see illustration in Fig.~\ref{fig:XDiffSiliconC17Texture}. 

This may seem, at first glance, surprising. As outlined in detail in Ref. \cite{Henschel2007} it can be traced, however, to a crystallisation mechanism first suggested by Percy Bridgman for single-crystal growth in narrow, macroscopic capillaries \cite{Bridgman1925, Palibin1933}. Bridgman found that in such a restricted geometry crystalline directions with high growth rates propagate along the capillary direction, whereas other growth directions die out. A selective influence of a capillary geometry on crystallisation which is routinely employed for single crystal growth. In the case of the hydrocarbons, the fast-growing direction is the $\langle$100$\rangle$ in-plane direction within the molecular layers, where the hydrocarbon backbones attach side-by-side \cite{Esselink1994, Fujiwara1998, Waheed2005, Anwar2013, Luo2015}. Because of the nano-version of the Bridgman mechanism this direction is aligned parallel to the tubular pore axis and the somewhat counter-intuitive perpendicular orientation of the long axes of the alkane molecules emerges - see Fig. \ref{fig:XDiffSiliconC17Texture}. 

Preferred orientations compatible with this selection mechanism for the crystalline orientations have been reported for other simple building blocks confined in nanoporous templates, \textit{e.g.} for argon \cite{Hofmann2005, Huber1999}, nitrogen \cite{ Huber1999, Huber1998a}, oxygen \cite{Wallacher2001, Kojda2014}, other medium length n-alkanes \cite{Henschel2007, Henschel2009Diss}, n-alcohols \cite{Henschel2009, Berwanger2009}, and for the smectic state of liquid crystals \cite{Chahine2010, Chahine2010a}. 

\begin{figure}[htbp]
\center
\includegraphics[width=0.8\columnwidth]{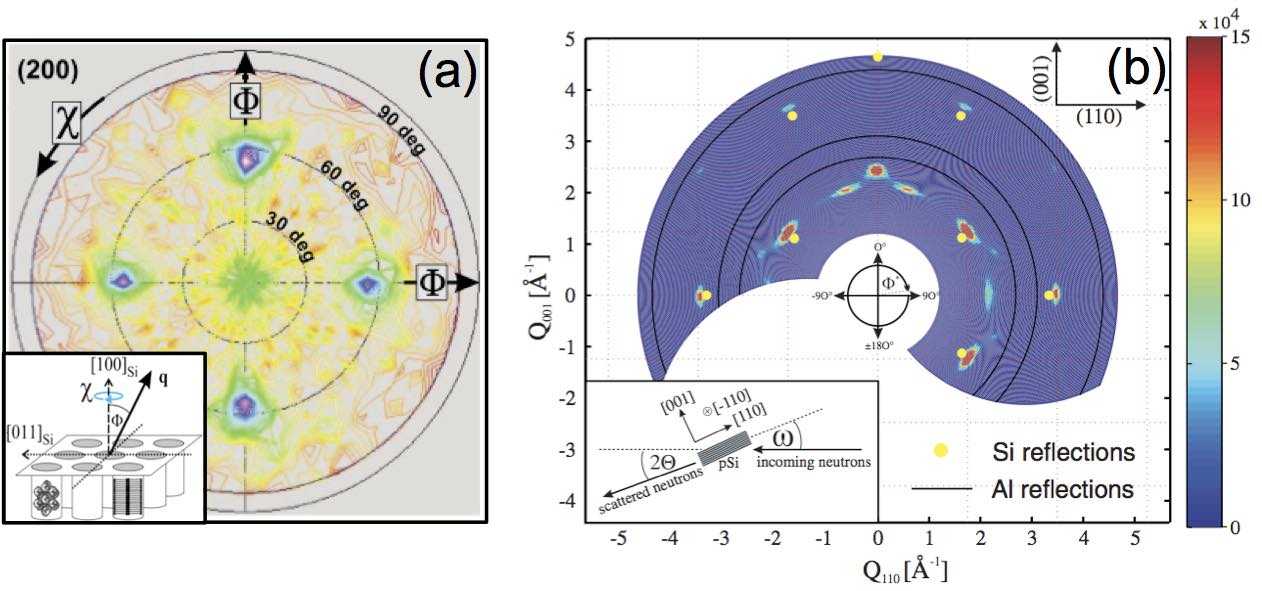}
\caption{(a) X-ray diffraction pole figure of the (200) Bragg reflection of the neat alcohol n-C$_{19}$H$_{39}$OH confined in mesoporous silicon at 298~K. As illustrated in the inset, the orientation of the sample sheet with respect to the scattering vector~\textbf{q} is specified by the polar angle $\Phi$ and the azimuth $\chi$. The two orientational domains of the molecules about the $\chi$ axis are schematically sketched in the figure: The orientation of the long axes of the molecules about the $\chi$-axis differ by 90~deg resulting in a view on the lateral herringbone-type ordered in-plane structure of the alcohols for the first domain (left pore) and a side view on the lamellar order of the chains for the second domain (right pore). Note, the long axes of the molecules in both $\chi$-domains are perpendicular to the long pore axis, which coincides with the [100] direction of the Si host. Reprinted (adapted) with permission from Henschel \etalp  \cite{Henschel2009}. Copyright 2009 American Physical Society. (b) Preferred orientation of deuterium nanocrystals in silicon nanochannels. Scattered neutron intensity in the ($\overline{1}$10)-plane. Q$_{110}$ and Q$_{001}$ depict wave vector transfers in the [110] direction of Si crystals respectively along the [001] surface normal. The inset illustrates the scattering geometry. Reprinted (adapted) with permission from Hofmann \etalp  \cite{Hofmann2013}. Copyright 2013 American Physical Society.} \label{fig:XDiffC19OHDeuteriumSilicon}
\end{figure}

Note, however, that the interaction with the pore wall, in particular in the case of a crystalline substrate, can additionally affect the crystallisation in nanoporous media. Besides the alignment of the fast growing crystalline direction along the pore axis, four-fold orientations in lateral directions, perpendicular to the long channel axes of the pores and coinciding with the four-fold symmetry of the $<100>$ substrate in this plane have been observed in X-ray diffraction texture measurements for n-alkanes and n-alcohols in porous silicon - see the 90$^{o}$-rotation symmetry of the X-ray diffraction pole figure depicted in Fig.~\ref{fig:XDiffC19OHDeuteriumSilicon}(a) \cite{Henschel2009}. Moreover, Hofman \etalp reported a highly textured growth of solid deuterium in porous silicon - see Fig.~\ref{fig:XDiffC19OHDeuteriumSilicon}(b), which could not be traced to the Bridgman mechanism, but rather to an epitaxial growth on the crystalline pore walls \cite{Hofmann2013}. 

Thus, the influence of the structure of the pore wall (amorphous or crystalline, facetted or round) and the intimately related structure of the interfacial layer at the pore wall (solid amorphous, liquid or grown epitaxial) as well as the interaction with the pore wall potential in general may sometimes compete with, but also sometimes enforce the simple geometric Bridgman texture mechanism outlined above. It is interesting to note, however, that the effectiveness of the Bridgman growth mechanism sensitively depends on the filling fraction and thus on the lengths over which the crystallisation front can propagate. A filling fraction dependent study on n-hexane in silicon demonstrated that for complete filling only, hence when the crystallisation front can sweep along the entire nanochannel, the highest degree of texture is observed \cite{Henschel2009}. 

\begin{figure}[htbp]
\center
\includegraphics[width=0.5\columnwidth]{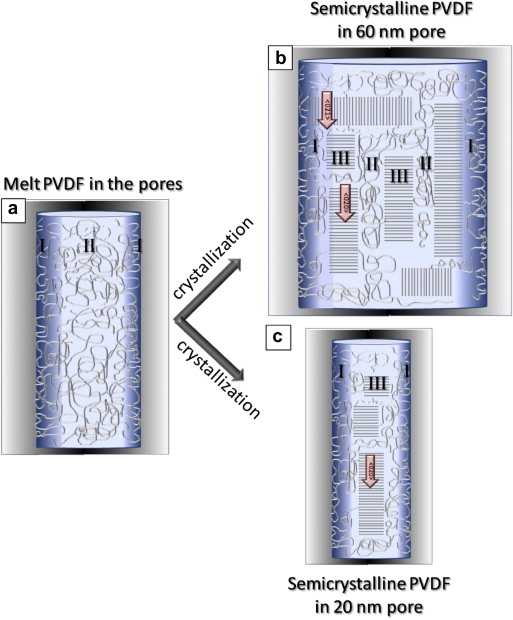}
\caption{Scheme of the proposed model for layers with different mobility and crystallisation. During infiltration of Polyvinylidene fluoride (PVDF) from the melt, an adsorbed layer (I) in contact with the walls is formed (a). Upon cooling, and depending on the pore size, one may find two situations: (b) in large diameter pores, \textit{i.e.}, 60 nm, there is volume enough to accommodate more than a parallel lamella (III). Therefore, there is an amorphous interlamellar region, which relaxes similar to that of the bulk. Moreover, the amorphous region adsorbed to pore wall relaxes in a particular way, as compared to the bulk. On the contrary, the 20 nm pores have volume to accommodate a single lamella oriented flat onto pore walls (III), the amorphous phase is mainly included in the adsorbed layer (I), and therefore, the main relaxation in this sample is the highly constrained one. Reprinted (adapted) from Polymer \cite{Michell2013}, with permission from Elsevier (2013).} \label{fig:SchematicsPolymerAluminaCrystallization}
\end{figure}

Also for more complex molecules, such as polymers, the crystallisation in nanoporous media results in peculiar orientational domain structures and an intimately related partitioning of the pore solid in amorphous, crystalline and semi-crystalline contributions. Steinhart \etalp \cite{Steinhart2006} demonstrated for the crystallisation of polymers in aligned alumina nanochannels distinct texture formation depending on the pore filling and on contact of the pore condensate with bulk material outside the porous medium. Again configurations were observed, where the long polymer chains - somewhat counterintuitively  - arrange perpendicular to the long axes of the channels. One may speculate that this can also be traced to the nano-Bridgman effect outlined above. An illustration of different immobilisation and crystallisation pathways of a polymer and the corresponding partitioning of the pore filling can be found in Fig.~\ref{fig:SchematicsPolymerAluminaCrystallization}. More extensive studies regarding the crystallisation of confined polymers within nanoscale cavities, the relevance of these studies with regard to the ongoing interest in crystallisation of polymers in general \cite{Beiner2008, Strobl2009a} and their nanotechnological importance can be found in the Refs. \cite{Beiner2008, Martin2012, Michell2012, Suzuki2013, Michell2013}. 

\begin{figure}[htbp]
\center
\includegraphics[width=0.8\columnwidth]{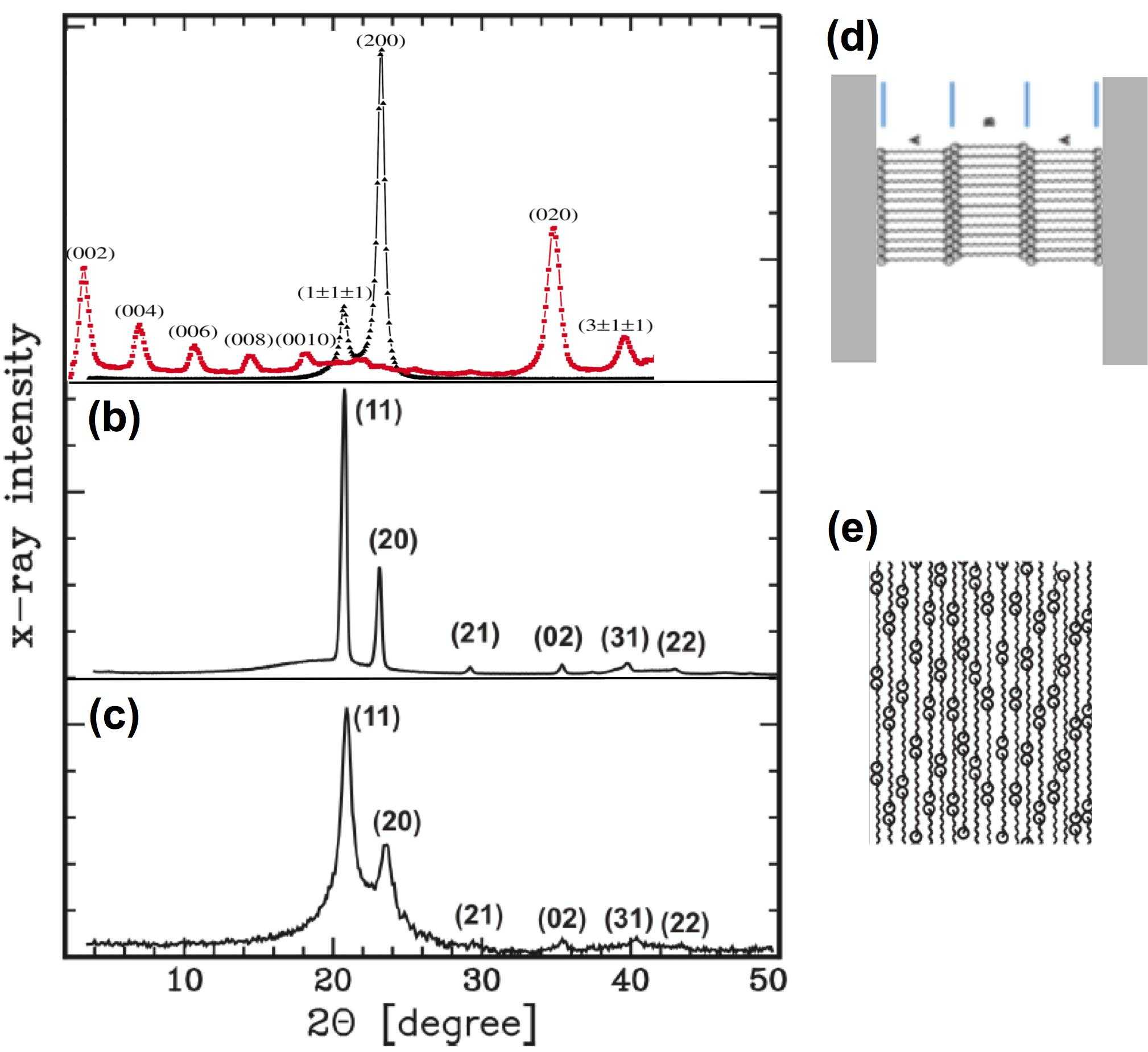}
\caption{X-ray diffraction patterns of highly textured crystalline C17 in tubular channels of silicon \cite{Henschel2007} in comparison with the X-ray profiles typical of C19 confined in tortuous pores of nanoporous Vycor \cite{Huber2004} and of natural bee wax \cite{Huber2008Habil}, both in a nematocrystalline state, as illustrated in the figure. Note the absence of any (00l)-Bragg peaks typical of molecular layering in the nematocrystalline state.} \label{fig:AlkanesNematoCrystallization}
\end{figure}

The influence of the pore morphology on the formation of distinct orientational domain structures and even crystalline phases in nanoporous media can also be inferred from a comparison of the crystallisation of alkanes in two geometries with distinct morphology, but almost identical pore diameter. In sponge-like Vycor glass the peculiar orientation of molecular lamella for medium-length n-alkanes, discussed above, was not observed \cite{Huber2004}. By contrast, the alkanes solidify in this highly tortuous pore space by entirely sacrificing the layering principle, in a ''nematocrystalline'' state. This structure is known from natural waxes with their large chain-length distribution and chemical heterogeneities, most prominently from bee wax - see illustration in Fig.~\ref{fig:AlkanesNematoCrystallization}. In this state, the hydrocarbon backbones are laterally quasi-hexagonal ordered, but not arranged in layers. Thus, by a change of the pore topology it is obviously possible to induce polymorphism and to select different crystalline architectures upon nanopore-confinement.  

This intimate relationship between texture formation, pore morphology, and pore chemistry is not only of academic interest. As exemplified above, this interplay depends also sensitively on the relative ratios of molecular size, molecular-molecular and molecular-wall interaction range to mean pore diameter, which offers a large versatility with regard to possible applications. Given the increasing usage of nanoporous media in the template-assisted preparation of nano objects  \cite{Huczko2000, Yin2001, Steinhart2002, Sander2003, Coakley2003, Ford2005, Hoffmann2006, Jongh2013} (\textit{e.g.}, nanorods or nanotubes) by melt infiltration and electrodeposition, as nano-composites, or even for drug-design \cite{Makila2014, Nadrah2014}, it is therefore also of high practical importance. It allows one to tune the crystalline texture, to affect, and in some cases also to select the phases of the pore-confined solid \cite{Ha2012, Wang2013a,Graubner2014}. The intimately related physicochemical and biological properties of pore solids can thus be tailored by an appropriate choice of the nanoporous template, the filling conditions and the thermal history. 

\begin{figure}[htbp]
\center
\includegraphics[width=0.7\columnwidth]{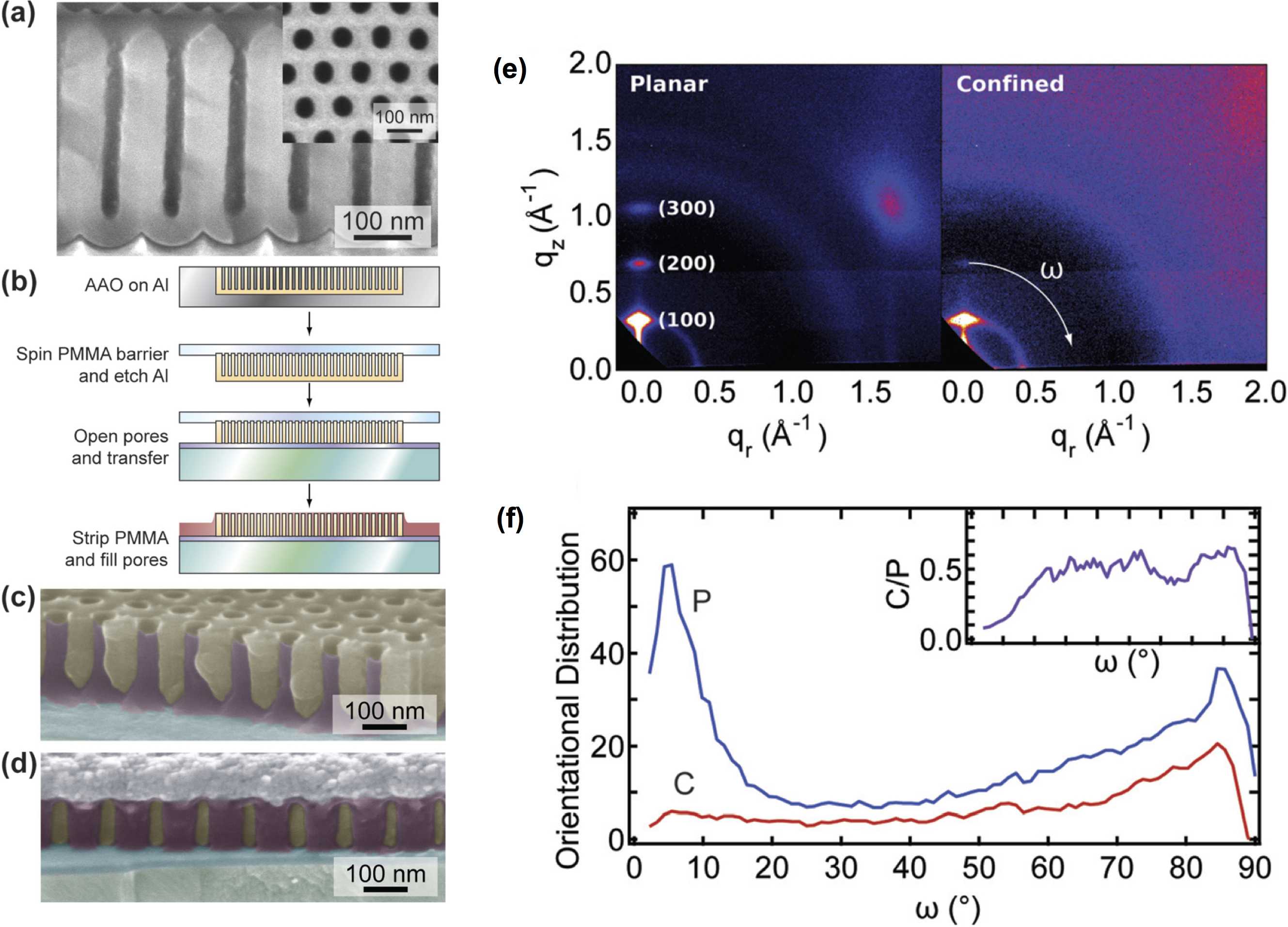}
\caption{Fabrication of templated bulk heterojunction solar cells. (a) Cross sectional view of a 300~nm tall nanoporous alumina template. Inset: top view. (b) Schematic template transfer procedure. (c) SEM image of nanoporous alumina template (beige) filled with an organic semiconductor (purple) on V$_2$O$_5$ (blue)/ITO (green). (d) Electron micrograph of completed templated bulk heterojunction solar cell. (e) X-ray scattering from planar and confined polymer blends. (f) Orientational distribution of the first-order lamellar peak (q=0.38$\AA^{-1}$) for a blend on a planar substrate (blue line, P) and confined in nanometer-scale pores (red line, C). Inset: ratio of C:P as a function of $\omega$ (indicated in the figure). Reprinted (adapted) with permission from Allen \etalp  \cite{Allen2011}. Copyright 2011 AIP Publishing LLC. \cite{Allen2011}} \label{fig:SEMGIXDPolymerblendAlumina}
\end{figure}

As a fine example of this practical impacts may serve the study by Allen \etalp. They show for a polymer blend confined in nanometer-scale cylindrical pores of an organic solar cell a doubling of the supported short-circuit photocurrent density compared to equivalent unconfined volumes of the same blend and an increase of the electron hole mobility in the confined blend by nearly 500 times. These enhanced electrical charge collection properties could be related by grazing incidence X-ray diffraction measurements to the confinement-induced changes in the polymer orientation distribution, suppressing low charge conductivity orientations while simultaneously disrupting polymer ordering - see Fig~\ref{fig:SEMGIXDPolymerblendAlumina}.

\begin{figure}[htbp]
\center
\includegraphics[width=0.4\columnwidth]{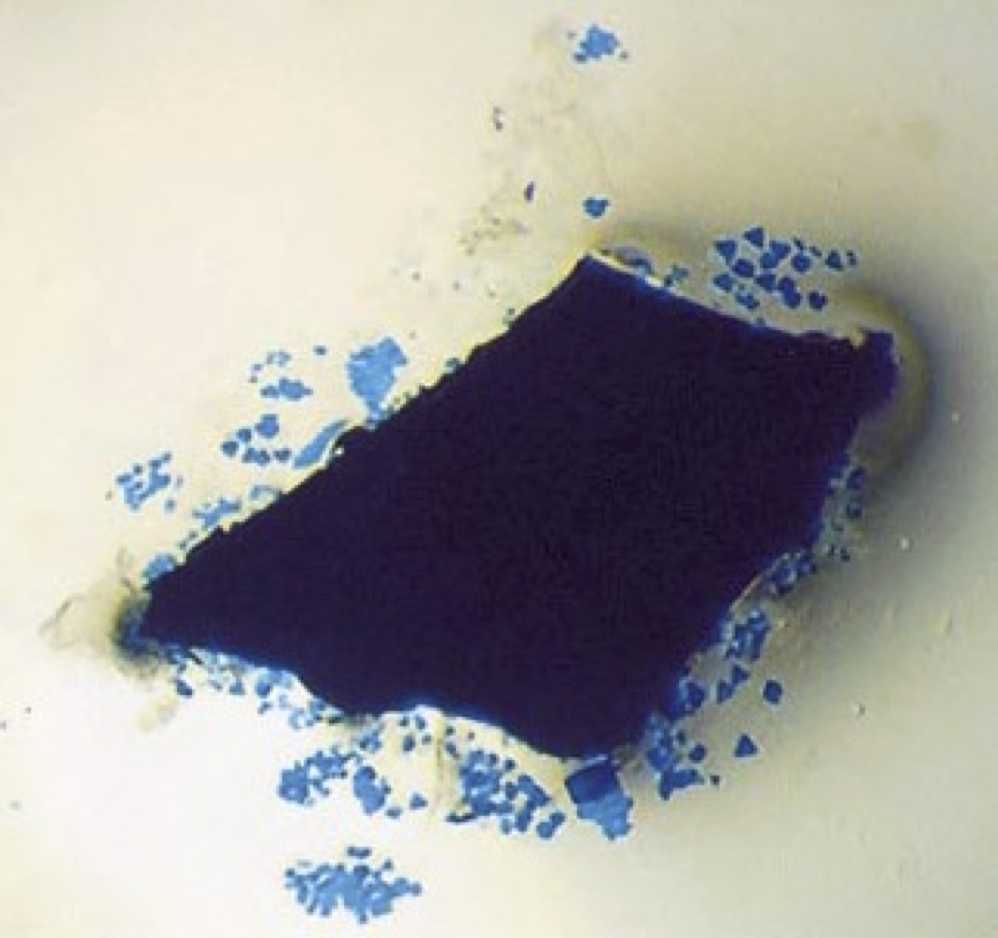}
\caption{Phycobiliprotein crystals growing on and in the proximity of nanoporous silicon fragments. Area shown is 3.0~mm x 2.3~mm. Reprinted (adapted) with permission from Chayen \etalp \cite{Chayen2001}. Copyright 2001 Elsevier.} \label{fig:PictureProteinSiliconCrystallization}
\end{figure}

Finally, it should be mentioned that nanoporous media are also attracting increasing interest with regard to the crystallisation of proteins. Protein crystals play a pivotal part in structural genomics, hence there is an urgent requirement for new and improved methodology to aid crystal growth. One approach is to use mesoporous materials that are likely to constrain protein molecules, immobilise them and thereby encourage them to aggregate in crystalline order \cite{Chayen2001}. In fact, large single crystals were obtained using porous silicon at conditions that are not sufficient for spontaneous crystal nucleation - see Fig.~\ref{fig:PictureProteinSiliconCrystallization}. How the interaction of biopolymers in pore space, their immobilisation as a function of the omnipresent pore size distribution and their crystallisation is related with the nucleation of protein crystals in the surrounding medium is therefore a surprisingly active research field  \cite{Chayen2001, Frenkel2006, Page2006, Meel2010}. The crystallisation peculiarities outlined above for less complex molecules within nanoporous media may help for a better understanding of this phenomenology. For instance, it is conceivable that a possible selection of fast-growing crystallisation modes in nanoporous media, which act not as random, but as fast-growing nuclei for proteins in the surrounding bulk solution may partially explain the success of nanoporous media in protein crystal nucleation. From a deeper understanding of this complex phenomenology may also profit the very active field of ''bioactive glasses", porous media which are used as bone graft implants or as templates for the in vitro synthesis of bone tissue for transplantation \cite{Li1991a, Hench1991, Sepulveda2002}.

\subsection{Structural Solid-Solid Phase Transformations}
Similarly as the liquid-solid transformation also structural solid-solid phase transitions are affected by confinement in nano pores. In particular, for small molecules the modification or suppression of solid-solid phase transitions have been extensively explored - see Ref. \cite{Knorr2008} and references therein. 

\begin{figure}[htbp]
\center
\includegraphics[width=0.5\columnwidth]{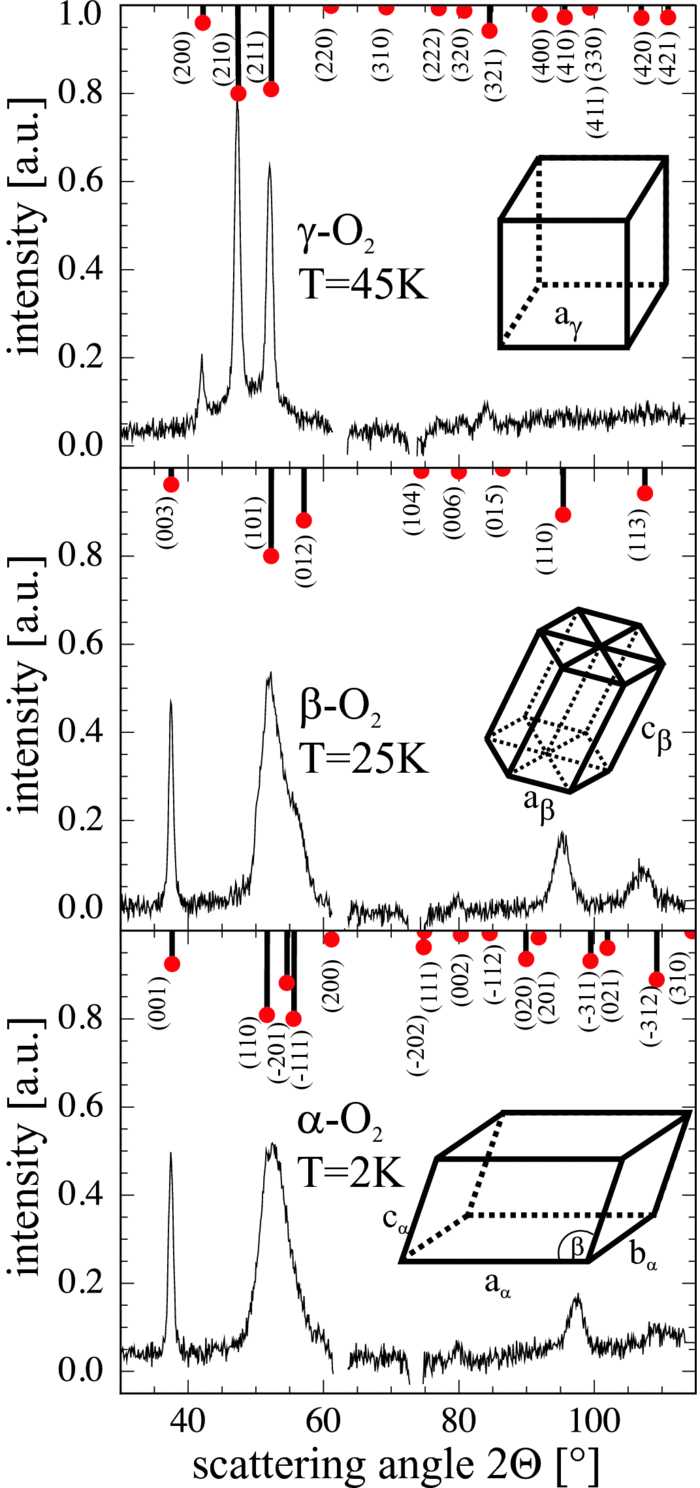}
\caption{Diffractograms of solid oxygen confined in nanoporous alumina: $\gamma$-phase at T = 45 K (top), $\beta$-phase at T = 25 K (middle), $\alpha$-phase at T = 2 K (bottom). Drumsticks mark the position of bulk reflections.Their lengths scale expected powder intensities. Reprinted (adapted) with permission from Kojda \etalp \cite{Kojda2014}. Copyright 2014 AIP Publishing LLC.} \label{fig:NDiffO2AluminaPhases}
\end{figure}

Pioneering studies were performed by Awschalom and Warnock \cite{Awschalom1987}, Molz \etalp \cite{Molz1993} and Schirato \etalp \cite{Schirato1995} on the structural transformations of molecular oxygen. In bulk form, oxygen is stable at moderate temperatures and pressures. It condenses and exhibits an intriguing structural phase sequence \cite{Freiman2004} encompassing a cubic $\gamma$-phase below 54~K, an orthorhombic $\beta$-phase at temperatures lower than 45~K and a monoclinic $\alpha$-phase below 21~K - see illustrations in Fig.~\ref{fig:NDiffO2AluminaPhases}. Structural equilibrium transitions are mainly caused by a complex interplay of isotropic dispersion forces and anisotropic quadrupolar and anisotropic magnetic interactions \cite{Freiman2004}. The latter, strong antiferromagnetic exchange between O$_2$ molecules, is due to unpaired electrons, which combine to a total molecular spin of S=1. Consequently solid phases of bulk O$_2$ differ in nuclear as well as magnetic structure. The monoclinic $\alpha$-phase exhibits long-range antiferromagnetic order. Thus, oxygen is a suitable model system in order to explore confinement effects in a single component system with a remarkable mixture of molecular interactions.

Studies have shown that below a critical pore diameter of 8~nm, phase transitions known from the bulk ($\gamma$-$\beta$, $\beta$-$\alpha$) were incomplete or entirely suppressed \cite{ Wallacher2001, Molz1993, Schirato1995}. SQUID measurements implied not only temperature and pore size dependent magnetic properties of confined O$_2$ but also evidenced a radial anisotropy of the magnetic characteristics inside the pores \cite{Ackermann2003, Ackermann2003a}. Inelastic neutron scattering measurements revealed magnetic fluctuations in confined $\beta$-oxygen similar to corresponding ones in the bulk as precursor of the evolving antiferromagnetic long-range order in the $\alpha$-phase. 

Hofmann $\etalp$ \cite{Kojda2014} showed in a recent neutron diffraction study that the oxygen nanocrystals inside tubular alumina of 12 nm channels do not form an isotropic powder. Rather, they exhibit preferred orientations sensitively depending on the thermal history. Moreover, the very mechanisms and geometric relationships between the solid phases guide also the crystalline texture of the pore solids. In Fig.~\ref{fig:NDiffO2AluminaPhases} neutron diffraction patterns of confined solid oxygen phases are depicted along with the crystalline structures of the $\alpha-$, $\beta-$ and $\gamma$-phase. Upon cooling of the melt the $\langle$111$\rangle$ direction of the $\gamma$-phase is preferentially aligned parallel to the tubular axis. 
However, also the  $\beta-$phase exhibits a pronounced texture, which readily depends on the texture of the high temperature phase. 

\begin{figure}[htbp]
\center
\includegraphics[width=0.5\columnwidth]{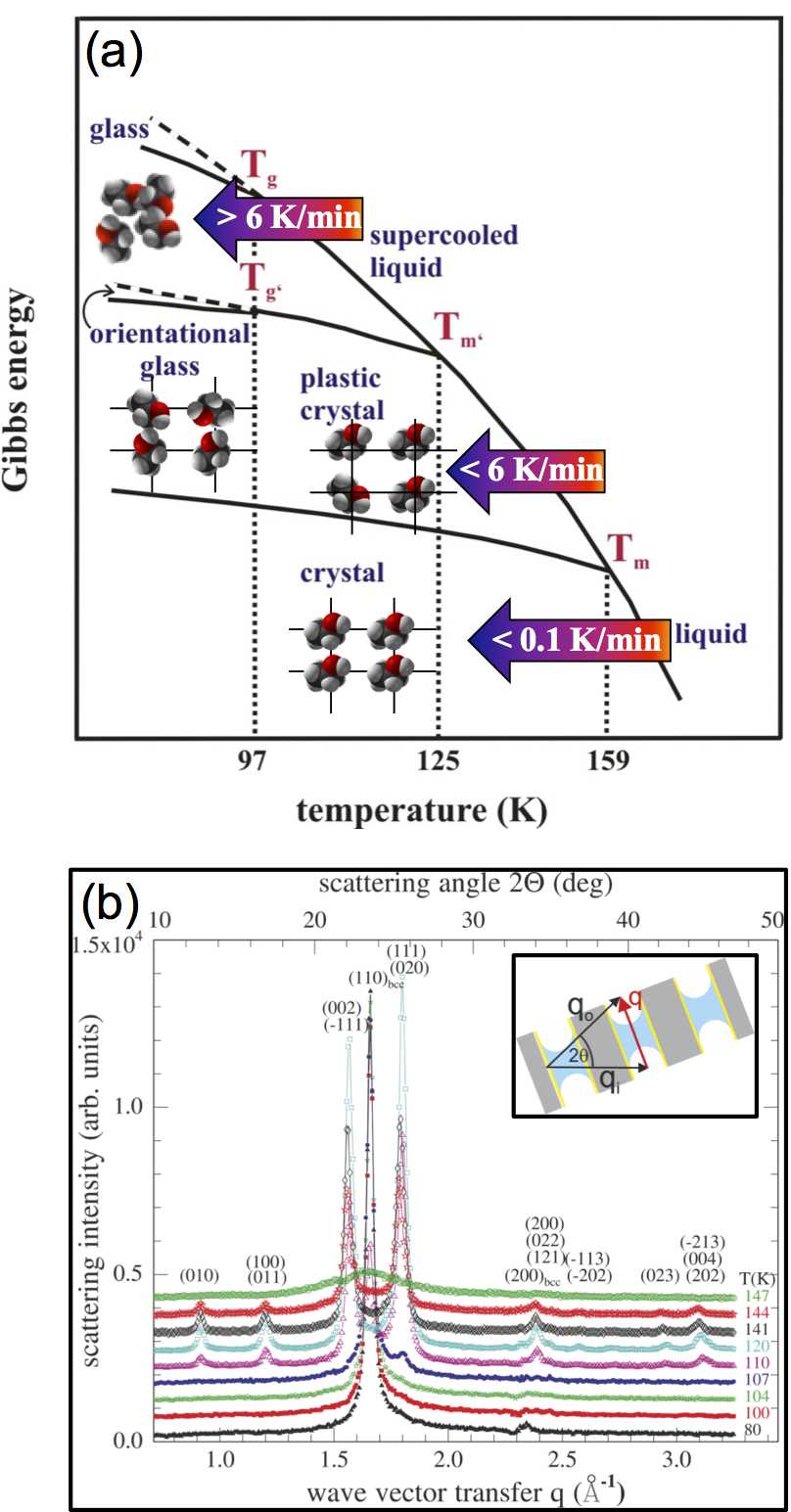}
\caption{(a) Phase diagram of bulk ethanol according to Vieira \etalp \cite{Vieira1997}. Plotted are the Gibbs free energies of the distinct ethanol phases as a function of temperature and cooling rate - see discussion in the text. (b) X-ray diffraction patterns of confined ethanol recorded at selected temperatures for a filling fraction $f=$~0.91 upon heating (lower panel) with 0.08\,K/min, respectively. Plotted is the scattered intensity both as a function of the wave vector transfer (lower abscissa) and scattering angle (upper abscissa) for selected temperatures as indicated in the figure. Reprinted (adapted) with permission from Ref. \cite{Henschel2010}. Copyright 2010 Taylor\&Francis.} \label{fig:XDiffEthanolSilicon}
\end{figure}

Another interesting system exhibiting polymorphism is bulk ethanol \cite{Vieira1997, Ramos1997}, see Fig.~\ref{fig:XDiffEthanolSilicon}(a). When cooled down in a sufficiently slow way, ethanol crystallises in a monoclinic structure. Here half of the molecules are in a {trans-,} the other half in a gauche-conformation. There is residual conformational, but no orientational entropy. On fast cooling (faster than a critical rate $r_{\rm c}$, $r_{\rm c}=6$\,K/min), the liquid-monoclinic transition is bypassed, and ethanol finally forms a glass state, with frozen-in translational, orientational, and perhaps also conformational disorder (a "structural glass"). For intermediate cooling rates and/or special annealing procedures, the liquid crystallises in another modification, with a bcc center-of-mass lattice and disordered orientations of the molecules (a "plastic phase"). At lower $T$, the orientations freeze-in, but the center-of-mass lattice remains bcc. This state is reminiscent of "orientational glasses" \cite{Hochli1990}, known from so-called mixed crystals, and has been named "glassy crystals" or orientationally disordered crystals for one component systems. Thus there are two liquid-to-crystal (at temperatures $T_{\rm m}$ and $T_{\rm m'}$) and two glass transitions at $T_{\rm g}$ and $T_{\rm g'}$ - see Fig.~\ref{fig:XDiffEthanolSilicon}.

What happens with this polymorphism in confinement? Henschel \etalp \cite{Henschel2010} addressed this question by a filling fraction and cooling rate dependent study of ethanol confined in 10~nm diameter channels of porous silicon - see the selected series of X-ray diffraction patterns of Fig.~\ref{fig:XDiffEthanolSilicon}. Whereas a strongly at the pore wall adsorbed ethanol film, corresponding to approximately two monolayers, remains in an amorphous state for the entire temperature range investigated, the capillary condensed molecules in the core of the pores reproduce the polymorphism of bulk solid ethanol. However, the critical cooling rate necessary to achieve a vitrification in the mesopores of silicon is at least two orders of magnitude smaller than in the bulk state. This observation is archetypical of the sensibility of the glass transition to confinement effects in general. For pertinent reviews on this very active and very controversially discussed topic see Refs. \cite{Alcoutlabi2005,McKenna2010}. 

In general, the examples presented here, along with analogous observations for a variety of structural transitions in other molecular \cite{Henschel2007, Berwanger2009, Huber2004, Huber2006a, Guegan2008, Kityk2008d,  Wang2014a, Wang2014b, Su2014} and macromolecular systems \cite{Serghei2013} suggest that confinement in nanoporous media can be employed to fine tune or to affect the polymorphism of solid phases (resulting from solid-solid phase transformations) and thus the intimately related physical properties. Beiner \etalp \cite{Beiner2007} and Graubner et al. \cite{Graubner2014} show that confinement in nanoporous hosts is a versatile way to rationally stabilise otherwise metastable or transient polymorphs of pharmaceuticals, as required for controllable and efficient drug delivery. Moreover, fine tuning of the existence and coexistence of crystalline or glassy phases is also of importance for applications of advanced nanocomposite materials \cite{Farasat2013}. 

\subsection{Thermotropic Liquid Crystalline Order}

Spatial confinement on the micro-, meso-, and nano-scale affects the physics of liquid crystals (LCs) markedly. Modified phase transition behaviour has been found for LCs imbibed into a variety of random porous media, in composite systems consisting of small particles immersed in LCs \cite{Chahine2010, Chahine2010a, Crawford1996, Iannacchione1993, Bellini2001, Kutnjak2003,Kityk2010, Kralj2012, Kralj2013, Busselez2014, Lefort2014}, in semi-confined thin film geometries \cite{Garcia2008, Pizzirusso2012}, and at the free surface of bulk LCs \cite{Ocko1986}.

It has also been demonstrated that there is no ''true'' isotropic-nematic \textit{(I-N)} transition for LCs confined in geometries spatially restricted in at least one direction to a few nanometers \cite{Iannacchione1993}. The anchoring at the confining walls, quantified by a surface field, imposes a partial orientational, that is a partially nematic ordering of the confined LCs, even at temperatures $T$ far above the bulk \textit{I-N} transition temperature $T^{\rm b}_{IN}$. The symmetry breaking does not occur spontaneously, as characteristic of a genuine phase transition, but is enforced over relevant distances by the interaction with the confining walls \cite{Stark2002}. Thus spatial confinement here plays a similar role as an external magnetic field for a spin system: The strong first order, discontinuous \textit{I-N} transition is replaced by a weak first order or continuous paranematic-to-nematic (\textit{P-N}) transition, depending on the strength of the surface orientational field.

For rod-like molecules the degree of orientational molecular order can be quantified by the uniaxial order para{\-}meter $Q= \frac{1}{2}\left \langle 3 \cos^2\phi -1 \right \rangle$, where $\phi$ is the angle between the long axis of a single molecule and a direction of preferred orientation of that axis, the director. The brackets denote an averaging over all molecules under consideration. %The orientation of the director may vary locally. However, it can be dictated by external fields or by surface anchoring conditions over macroscopic distances. Planar silica surfaces enforce planar anchoring of 7CB and 8CB without a preferred lateral orientation \cite{Kumar2001}. Additionally, the director is expected to be oriented parallel to the long axis in a cylindrical silica channel \cite{GrohDietrich1999}. 

The propagation speed of light and thus the refractive index $n$ in an LC sensitively depends on the orientation of the polarisation with respect to the molecular orientation of the anisotropic nematogens. Conversely, the state of molecular order in an LC can be inferred from optical polarisation measurements. To a good approximation, $Q$ is proportional to the optical birefringence $\Delta n=n_{\rm e}-n_{\rm o}$, where $n_{\rm o}$ and $n_{\rm e}$ refer to polarisations perpendicular and parallel to the local optical axis, the so-called ordinary and extraordinary refractive indices, respectively. In a nematic LC the local optical axis agrees with the director. Thus, in principle it is sufficient to determine the experimentally accessible $\Delta n$ in order to determine the molecular arrangement in an LC. 

\begin{figure}[htbp]
\center
\includegraphics[width=0.5\columnwidth]{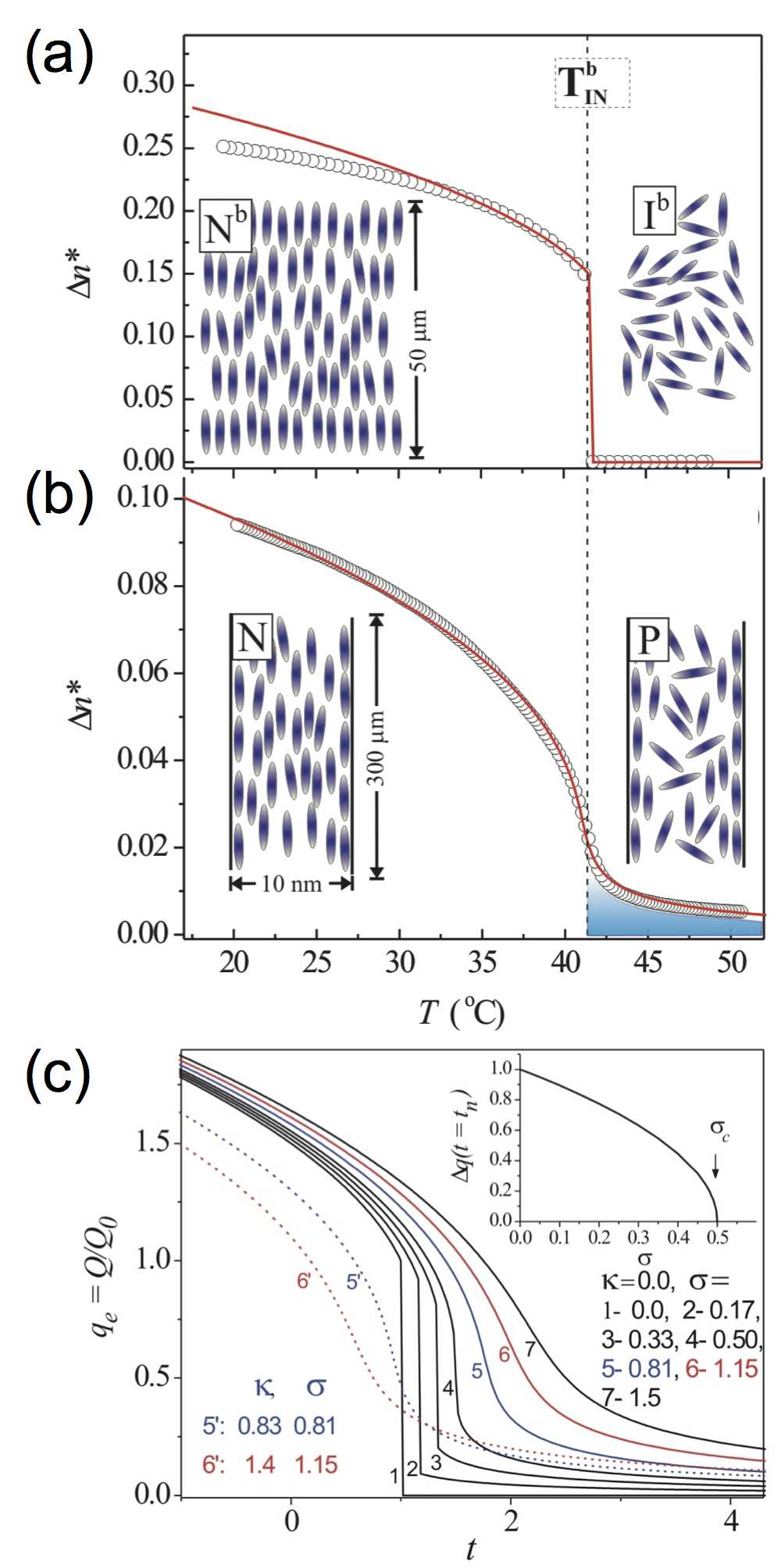}
\caption{Birefringence of the rod-like liquid crystal 7CB measured in the bulk state, panel (a) and in silica nanochannels, panel (b) as a function of temperature in comparison to a fit (solid lines) based on the KKLZ-model discussed in the text. The final birefringence characteristic of the paranematic phases are shaded down to the $P-N$ ''transition'' temperatures, $T_{\rm PN}$. The dashed lines mark the bulk \textit{I-N} and \textit{N-SmA} transition temperatures. As insets in (a) and (b), the bulk isotropic ($I^{\rm b}$) as well as the bulk nematic ($N^{\rm b}$) phases upon homeotropic alignment, and the confined paranematic (P) and nematic (N) phases are illustrated, respectively. (d) Nematic order parameter $q_{\rm e}$ as obtained by minimisation of the free energy of the KKLZ-model as a function of reduced temperature $t$ for selected effective surface fields $\sigma$ and strength of quenched disorder $\kappa$. Inset: Magnitude of the order parameter jump at the \textit{I-N (P-N)} transition as a function of the surface field $\sigma$. Reprinted (adapted) with permission from Kityk \etalp \cite{Kityk2008}. Copyright 2008 American Physical Society.} \label{fig:Birefringence7CBSilica}
\end{figure}

Kityk \etalp \cite{Kityk2008} demonstrate in a study on the thermotropic orientational order of rod-like LCs (7CB and 8CB), that monolithic silica glasses traversed by parallel channels, which can be prepared by thermal oxidation of silicon matrixes (see Fig. \ref{fig:IsothermArC6Silicon}b) is particularly suitable for optical birefringence measurements of confined nematics. 

In Fig.~\ref{fig:Birefringence7CBSilica}a $\Delta n^*$ is plotted for bulk 7CB upon cooling to the solidification temperature. There is a jump in $\Delta n^*(T)$ of bulk 7CB typical of the first-order \textit{I-N} phase transition at $T^{\rm b}_{\rm IN}\approx$~42~\dC. Any pretransitional effects are clearly absent in the bulk isotropic phase, $\Delta n^*(T)=0$ for $T>T^{\rm b}_{\rm IN}$. The nanoconfined LC reveals a considerably different behavior, see Fig.~\ref{fig:Birefringence7CBSilica}b. There exists a residual $\Delta n^*$ characteristic of a paranematic LC state at $T$s far above $T^{\rm b}_{\rm IN}$. Upon further cooling $\Delta n^*$ increases continuously and at the lowest $T$s investigated, the absolute magnitude of $\Delta n^*$ is compatible with an 80\% alignment of the 7CBs' long axes parallel to the channel axes. Hence, the silica nanochannel confinement dictates a substantial molecular alignment, as illustrated in the inset of Fig.~\ref{fig:Birefringence7CBSilica}b. More importantly, it renders the transition continuous.

This peculiar $\Delta n^*$ behavior can be analysed within a Landau-de~Gennes model for the $I-N$ transition in confinement suggested by Kutnjak, Kralj, Lahajnar, and Zumer (KKLZ-model) \cite{Kutnjak2003, Kutnjak2004}. This model is based on the pioneering work by Sheng, Poniewierski, Sluckin \cite{Sheng1976, Poniewierski1987} as well as  experimental work by Yokoyama\cite{Yokoyama1988} on LCs in semi-infinite, planar confinement. The dimensionless free energy density of a nematic phase spatially confined in a cylindrical geometry with planar anchoring conditions reads in the KKLZ-model as:
\begin{equation}
f=tq^2-2q^3+q^4-q\sigma+\kappa q^2 \nonumber \label{eq2}
\end{equation}
where $q=Q/Q(T^{\rm b}_{\rm IN})$ is the scaled nematic order parameter, $t$ is a reduced temperature, and $\sigma$ is the effective surface field. The last term in Eq.~1 describes quenched disordering effects due to surface-induced deformations (wall irregularities). Minimalisation of $f$ yields the equilibrium order parameter $q_{\rm e}$, which is shown in Fig.~\ref{fig:Birefringence7CBSilica}c for selected values of $\sigma$ and $\kappa$ as a function of $t$. In the KKLZ-model, the \textit{I-N} transition is of first order for $\sigma < \sigma_{\rm c}=0.5$. The jump of $q_{\rm e}$ approaches zero while $\sigma \rightarrow 0.5$, see inset in Fig.~\ref{fig:Birefringence7CBSilica}. Thus, $\sigma_{\rm c}$ marks a critical threshold separating first-order, discontinuous from continuous \textit{I-N} behavior.

The solid lines in Figs.~\ref{fig:Birefringence7CBSilica}(a),(b) are the best fits of the dependencies $\Delta n^*(T)$ as obtained by rescaling $q_{\rm e}$ and $t$ while assuming an absence of any surface ordering and quenched disorder fields in the bulk state ($\sigma(bulk)=\sigma(D=\infty)$=0, $\kappa=0$) and final values for this quantities in the confined state ($\sigma(D=10$nm$)$ = 0.81, $\kappa=0.2$). Thereby, an encouraging agreement between the measured $\Delta n^*(T)$ curves in the proximity of $T^{\rm b}_{\rm IN}$ and deep into the nematic phase for both bulk LCs can be achieved. 

In the mean time the KKLZ-model could be successfully applied also for binary liquid crystalline mixtures in confinement \cite{Calus2014}. Optical birefringence measurements revealed that depending on the channel radius the nematic ordering in the cylindrical nanochannels evolves either discontinuously (subcritical regime, nematic ordering field $\sigma<1/2$) or continuously (overcritical regime, $\sigma>1/2$), but in both cases with a characteristic paranematic precursor behavior. The strength of the ordering field, imposed by the channel walls, and the magnitude of quenched disorder varies linearly with the mole fraction $x$ and scales inversely proportional with $R$ for channel radii larger than 4 nm. The critical pore radius, $R_c$, separating a continuous from a discontinuous paranematic-to-nematic evolution, varies linearly with $x$ and differs negligibly between confinement in silica and alumina membranes. 

\begin{figure}[htbp]
\center
\includegraphics[width=0.85\columnwidth]{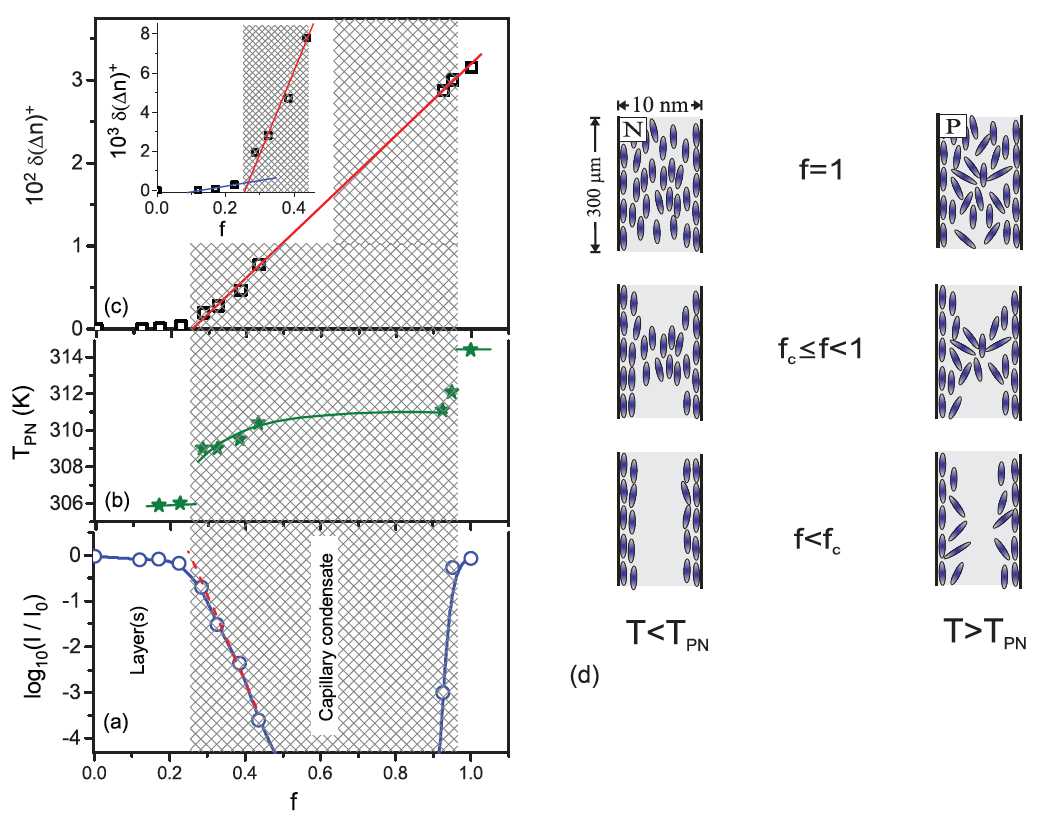}
\caption{Gradual filling of nanoporous silica with the rod-like liquid crystal 7CB. (a) Light transmission $log_{\rm 10}(I/I_{\rm 0})$ vs. filling fraction $f$, (b) Temperature of the paranematic-to-nematic transition $T_{\rm PN}$ vs. filling fraction $f$ as determined from optical birefringence measurements. (c) $f$-dependence of the effective optical birefringence $\delta (\Delta n)^+$ at $T=T_{\rm PN}-10$~K, (d) schematic side views of three characteristic capillary filling regimes: adsorbed monolayer(s), film-condensed regime ($f<f_{\rm c}$), capillary condensate ($f_{\rm c}\le f<1$) and completely filled substrate ($f=1$). The molecular orientations in the different filling regimes are illustrated below (left panel d) and above (right panel d) $T_{\rm PN}$, that is in the nematic and paranematic state of the confined liquid crystal. In the panels (a-c) the $f$-range typical of capillary condensation is shaded. The solid lines in panel (a) and (b) are guide for the eyes, whereas in panel (c) linear fits are presented. Reprinted (adapted) with permission from Kityk \etalp \cite{Kityk2009}. Copyright 2009 American Physical Society.} \label{fig:Optics7CBPartialSilica}
\end{figure}

Whereas most of the studies on LCs in nanoporous media published so far concern completely filled pores, it is also possible to scrutinise the thermotropic orientational behaviour as a function of gradual pore filling \cite{Huber2013, Frunza2006}. In Fig.~\ref{fig:Optics7CBPartialSilica} the results of optical birefringence and light absorption measurements of 7CB in a monolithic silica membrane are depicted \cite{Huber2013}. They reveal four regimes for the thermotropic behavior upon sequential filling of parallel capillaries of 12~nm diameter. No molecular reorientation is observed for the first adsorbed monolayer. In the film-condensed state (up to 1~nm thickness) a weak, continuous paranematic-to-nematic (P-N) transition is found, which is shifted by 10~K below the discontinuous bulk transition at T$_{\rm IN}=$305~K. The capillary-condensed state exhibits a more pronounced, albeit still continuous P-N reordering, located 4~K below T$_{\rm IN}$. This shift vanishes abruptly on complete filling of the capillaries, which could originate in competing anchoring conditions at the free inner surfaces and at the pore walls or result from the 10~MPa tensile pressure release associated with the disappearance of concave menisci in the confined liquid upon complete filling. In general, this study documents that the thermo-optical properties of this nonporous medium can be tailored over a surprisingly wide range simply by variation of the filling fraction with liquid crystals.

Besides rod-like liquid crystals also disc-like liquid crystalline systems have attracted quite some interest both from fundamental and practical point of view. In particular, molecular assemblies consisting of disc-like molecules (DLCs) with an aromatic core and aliphatic side chains can exhibit a rich phase transition behavior \cite{Laschat2007, Sergeyev2007, Bisoyi2010}. Because of the $\pi-\pi$ overlap of their aromatic cores they can stack into columns, which in turn arrange in a two-dimensional crystalline lattice leading to thermotropic discotic columnar crystals. Thermal fluctuations give rise to liquid-like properties \cite{Haverkate2011} and even glassy disorder can occur \cite{Krause2012, Krause2014}. Therefore, DLCs are particularly interesting systems in order to address fundamental questions in soft matter science, such as structure-dynamics and structure-phase transition relationships.

\begin{figure}[htbp]
\center
\includegraphics[width=0.8\columnwidth]{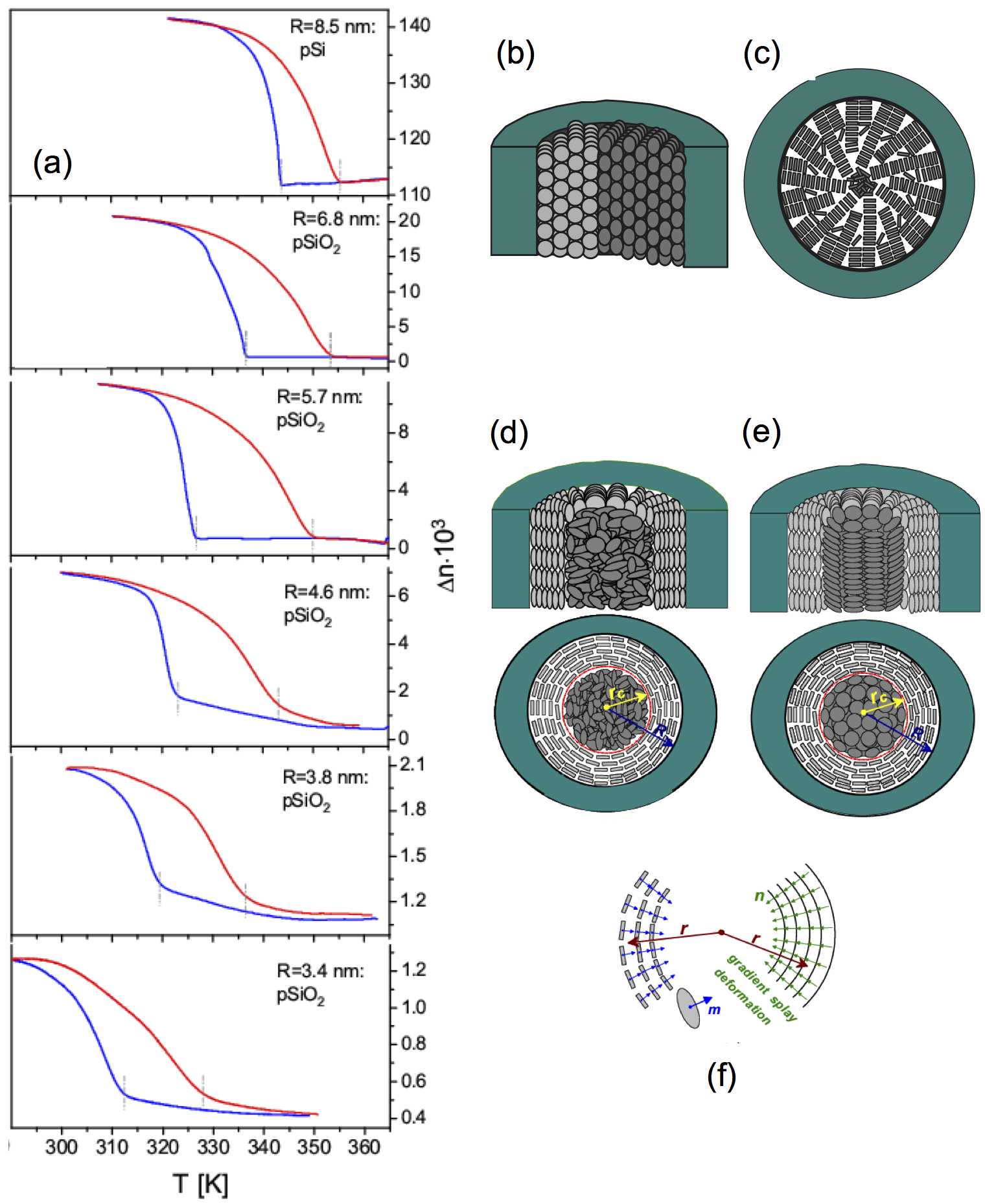}
\caption{(a) The optical birefringence of a discotic liquid crystal vs. temperature during cooling (blue) and subsequent heating (red) as a function of mean channel diameter variation from 8.5 nm to 3.4 nm. (b and c) Schematics of radial molecular ordering of discotics with long-range, hexagonal columnar order confined in cylindrical channels: (b) Sideview on two columnar domains. (c) Topview on a radial structure with dislocation defects and a disordered core.  (d) Radial configuration with isotropic core and (e) escaped radial configuration. Panel (f): The left sectorial fragment represents the molecular ordering, the right sectorial fragment illustrates the splay distortion due to a curvature of molecular layers with the local director, $\vec{n}$, oriented radially. The resulting splay distortion, $\vec{\nabla}\vec{n}$, equals $1/r$, \textit{i.e.}, it is spatially inhomogeneous. The gradient in the splay distortion results in a phase front (see red circle of radius $r_c$), which separates a radially ordered shell from an isotropic core. According to the Landau-de-Gennes calculations presented in Ref. \cite{Kityk2014}, it moves during cooling from the periphery to the channel centre, and vice-versa during heating. Reprinted (adapted) with permission from Kityk2014 \etalp \cite{Kityk2014}. Copyright 2014 Royal Society of Chemistry.} \label{fig:SketchDiscoticsChannel}
\end{figure}

DLCs encompass also advantageous materials properties, including highly anisotropic visible light absorption, long-range self-assembly, self-healing mechanisms, high charge-carrier mobilities along the column axis and a tuneable alignment of the columns \cite{Laschat2007, Sergeyev2007, Grelet2006, Feng2009}. Therefore they represent promising systems for active layers in organic devices, such as field-effect transistors and photovoltaic cells.

The combination of DLCs with nanoporous media offers the opportunity to prepare nanowires by template-assisted melt infiltration and to design hybrid systems \cite{Steinhart2005, Duran2012} with interesting opto- and opto-electronical properties. However, similarly as the thermotropic behaviour of rod-like liquid crystalline systems discussed above, the properties of confined discotic systems has turned out to be particularly susceptible to nano confinement and to be substantially altered in comparison to the bulk systems \cite{Steinhart2005, Kopitzke2000, Stillings2008}. 

\begin{figure}[htbp]
\center
\includegraphics[width=0.4\columnwidth]{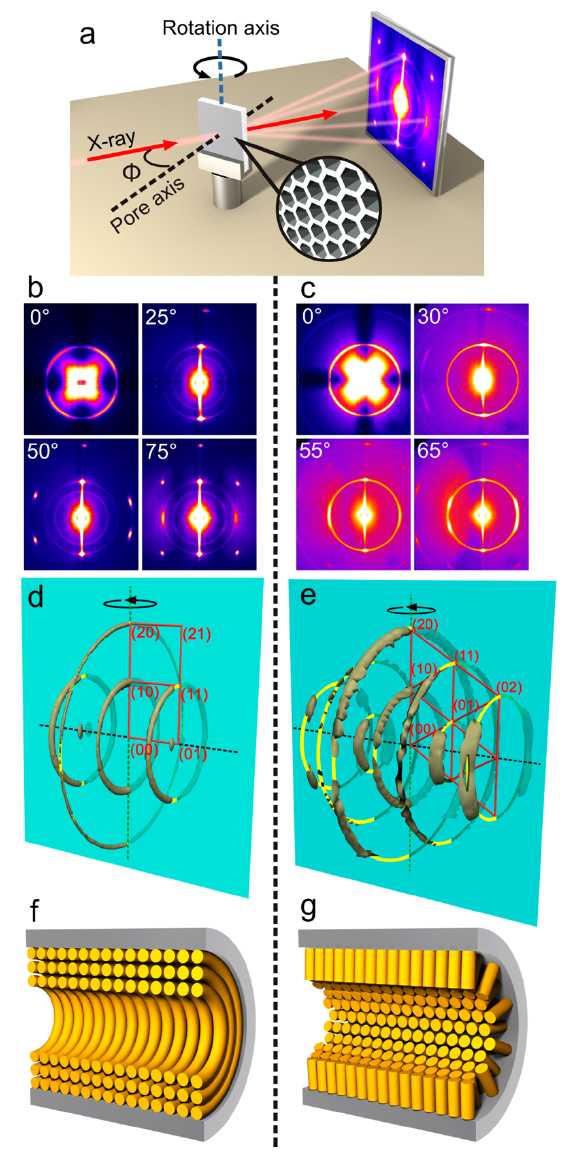}
\caption{X-ray diffraction setup used for the reciprocal space mapping of LC-infiltrated alumina membranes (AAO) in the study by Zhang \etalp \cite{Zhang2014}. The rotation angles $\Phi$ is the angle between the beam and the membrane normal (\textit{i.e.}, the AAO pore axis). (b and c) X-ray diffraction patterns of AAO membranes (pore diameter 400 nm) filled with LC compounds recorded at a series of rotation angles $\Phi$ indicated. Note that the four-quadrant sectorisation in the $\phi$ = 0 patterns is an artefact due to the saturation of the detector with the intense central scatter from the AAO pores. (d
and e) 3-D diffraction patterns of (d) LC compound 1 and (e) compound 2 confined to AAO with a pore diameter of 400 nmin reciprocal space coordinates. The isosurface is set to an intensity level such that the reciprocal space distribution of all three strongest reflection groups, \textit{i.e.}, \{10\}, \{11\}, and \{20\}, could be shown simultaneously. The dashed black line indicates the orientation of the pores in the AAO. The breaks in the rings are mainly due to absorption of X-rays by the AAO. The yellow lines are guides to the eye. (f and g) Two of the possible models of columns in cylindrical confinement that fit the X-ray data: (f) planar circular (for planar, or homogeneous surface anchoring) and (g) planar radial (for homeotropic anchoring); these models assume cylindrical D$_{\inf}$ symmetry within each AAO pore. Reprinted (adapted) with permission from Zhang \etalp \cite{Zhang2014}. Copyright 2014 American Chemical Society.} \label{fig:XDiffDiscoticsAlumina}
\end{figure}

Temperature-dependent optical birefringence experiments on DLCs in nanoporous membranes of silicon, silica and alumina by Kityk \etalp \cite{Kityk2014, Calus2014a} allowed for an exploration of the thermotropic evolution of the microscopic and mesoscopic structures of nanopore-confined DLCs. These measurements revealed a competition of radial and axial columnar order, see Fig.~\ref{fig:SketchDiscoticsChannel}. The evolution of the orientational order parameter of the confined systems is continuous, in contrast to the discontinuous transition in the bulk. For channel radii larger than 10~nm the authors suggest several, alternative defect structures, which are compatible both with the optical experiments on the collective molecular orientation (seen in the birefringence measurements) and with a translational, radial columnar order reported in previous diffraction studies by Cerclier \etalp \cite{Cerclier2012}. For smaller channel radii the observations can semi-quantitatively be described by a Landau-de Gennes model with a nematic shell of radially ordered columns (affected by elastic splay deformations) that coexists with an orientationally disordered, isotropic core. For these structures, the cylindrical phase boundaries are predicted to move from the channel walls to the channel centres upon cooling, and vice-versa upon heating, in accord with the pronounced cooling/heating hystereses observed and the scaling behavior of the transition temperatures with channel diameter. The absence of experimental hints of a paranematic state is consistent with a biquadratic coupling of the splay deformations to the order parameter.

It is also interesting to compare these findings for disc-like systems with melting of spherical building-blocks, \textit{e.g.}, argon in nanochannels. Also for this first-order bulk transition the movement of the ordering/disordering interface has been intensively discussed and experimentally explored in the past \cite{Christenson2001, Alba-Simionesco2006,Knorr2008,Alcoutlabi2005, Webber2010}. Interfacial melting with a radial moving solid/liquid boundary has been proposed based on filling fraction dependent investigations \cite{Alba-Simionesco2006, Wallacher2001a, Schaefer2008, Moerz2012}. Note however, that here the high temperature (liquid) phase is believed to be nucleated at the pore wall and hence the movement of the front boundary is opposite to the scenario outlined in Ref. \cite{Kityk2014, Calus2014a} for the melting of the columnar discotic state.

Moreover, a plethora of experimental and theoretical studies unanimously find that the depression in the melting transition and other structural phase transitions scale with $1/R$ \cite{Christenson2001, Knorr2008, Alba-Simionesco2006, Wallacher2001a, Alcoutlabi2005}. This can be rationalised for "simple" spherical building blocks, independently of the details of the phase transition model, by a competition of volume free energies (scaling with $1/R^3$) and interfacial free energies (scaling with $1/R^2$), leading to the well-established Gibbs-Thomson equation. The validity of the Gibbs-Thomson equation is the base of thermoporometry, that is the determination of pore diameter distributions from the temperature shifts of phase transitions \cite{Mitchell2008, Petrov2009, Riikonen2011, Kondrashova2011}. Therefore confined discotics are important examples, where geometrical constraints render the $1/R$-scaling law inappropriate for a conversion of phase transition shifts into pore radii distributions, if one relies purely on the bare optical birefringence measurements. Rather an appropriate conversion of these data sets to a quantity which corresponds to a specific heat signal, that is the temperature-derivative of the square of the birefringence, follows the simple $1/R$-scaling.

Finally, the experimental advantages of monolithic nanoporous media shall be exemplified by the work of Zhang \etalp \cite{Zhang2014}. Besides optical birefringence also X-ray and neutron diffraction experiments can give valuable information on the molecular arrangement of liquid crystals \cite{Chahine2010, Chahine2010a, Guegan2008,  Cerclier2012, Guegan2006}. In the study by Zhang \etalp small-angle X-ray scattering (SAXS), see Fig.~\ref{fig:XDiffDiscoticsAlumina} and atomic force microscopy (AFM) were used to study orientation patterns of two polyphilic liquid crystals confined to cylindrical pores of alumina. By conducting complete reciprocal space mapping using SAXS, they could conclude that the columns of the two compounds investigated align in planes normal to the AAO pore axis, with a specific crystallographic direction of the LC lattice aligning strictly parallel to the pore axis. AFM of LC-containing AAO fracture surfaces further revealed that the columns of the planar anchoring LC (compound 1) formed concentric circles in the plane normal to the pore axis near the AAO wall. Toward the pore center, the circles become anisometric ÒracetrackÓ loops consisting of two straight segments and two semicircles. This mode compensates for a slight ellipticity of the pore cross section. For the homeotropically anchoring compound 2, the columns are to the most part straight and parallel to each other, arranged in layers normal to the AAO pore axis, like logs in an ordered pile. Only near the pore wall the columns splay somewhat. In both cases, columns are confined to layers strictly perpendicular to the AAO pore axis, and there is no sign of escape to the third dimension or of axial orientation. According to Zhang \etalp the main cause for the two new LC configurations, the ÒracetrackÓ and the Òlog pileÓ, and of their difference from those of confined nematic LC, is the very high splay energy and low bend energy of columnar phases. A conclusion which is in agreement with the observations by Kityk \etalp \cite{Kityk2014, Calus2014} presented above. 

Today, the importance of liquid crystalline systems in opto-electronic applications, particularly in display technologies, can barely be overestimated. It continues to attract considerable interest in many areas of organic electronics \cite{Khoo2009}. Memory effects resulting from frustration, as explored in recent computer simulations by Araki \etalp \cite{Araki2011} for nematic LCs confined in bicontinuous porous structures, may be considered as one prominent example, where orientationally ordered materials are expected to exhibit surprising functionalities because of topological confinement. However, implementation of such ideas in devices based on nanoporous media requires a better understanding of the rather fundamental aspects of geometrical confinement and interface couplings described above. The functionality of these devices also sensitively depends on the molecular mobility, in particular the reorientation dynamics of pore-confined matter, a topic that will be addressed in the following section.

\subsection{Self-Diffusion Dynamics}

The stochastic, thermally excited motions within molecular condensates spatially confined on the nanometer scale is attracting great interest both in fundamental and applied sciences \cite{Drake1990, Keil2000, Alcoutlabi2005, McKenna2010, Zorn2010}. Particularly, self-diffusion of molecular liquids confined in mesoporous hosts plays a dominant role in catalysis and filtering \cite{Schueth2002}. It has also been discussed with regard to drug delivery applications \cite{Bras2011, Ukmar2011a, Tzur-Balter2013, Kurtulus2014} and with respect to the origin of the adsorption-desorption hysteresis found in mesoporous materials \cite{Valiullin2006}. 

In terms of fundamental science, the Brownian motion of molecules in pores a few nanometers across is interesting, since concepts concerning random motions in highly confined, crowded geometries \cite{Burada2009} or for molecules exposed to external fields (\textit{i.e.}, interaction potentials with the confining walls) can be scrutinised \cite{Scheidler2002, Bock2007, Sangthong2008, Zuerner2007, Krishna2009, Ok2010, Raccis2011, Mazza2012, Valiullin2013, Busselez2014,Ndao2014}. Moreover, the direct relation between stochastic motions and viscous properties of a liquid (expressed by the Stokes-Einstein equation) allows one to explore the fluidity of liquids in nano-scale structures and in the proximity of the confining solid walls by studies of their self-diffusion dynamics. This is of obvious importance in nano- and microfluidics, where the exploration of these properties can be experimentally very demanding \cite{Striemer2007, Chan1985, Georges1993, Ruths1999, Becker2003, Eijkel2005, Neto2005, Joly2006, Huber2007, Kusmin2010, Walz2011, Bocquet2010, Gruener2011}.

%Mey2012
\begin{figure}[htbp]
\center
\includegraphics[width=0.5\columnwidth]{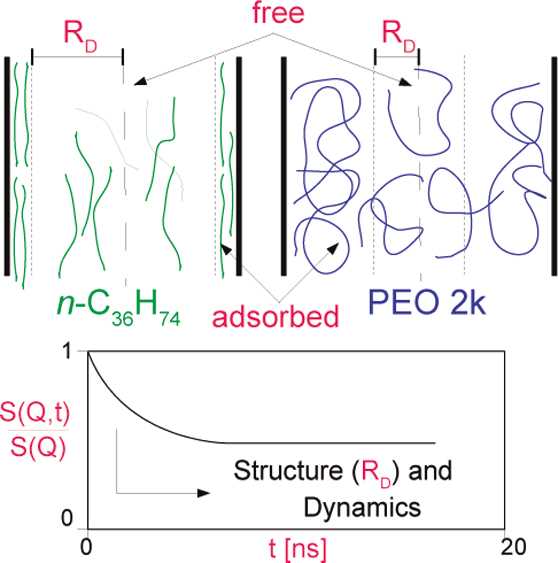}
\caption{Illustration of molecules (left: linear hydrocarbons, right: polymers (Polyethylenoxide PEO)) freely diffusing in the core of a tubular pore in coexistence with a population of immobilised (anchored) molecules at the pore walls, as inferred from wave vector transfer $q$ dependent quasi-elastic neutron measurements of the intermediate scattering function $S(q,t)$ in the time domain $t$ normalised by the static structure $S(q)$, where $q$ is the wave vector transfer in the scattering experiment, see Ref. \cite{Kusmin2010}.} \label{fig:SketchDiffusionMoleculesChannel}
\end{figure}

There is a sizeable number of experimental techniques, most prominently nuclear magnetic resonance (NMR), gravimetric uptake measurements, photon correlation spectroscopy (PCS), dielectric spectroscopy as well as quasi-elastic neutron scattering (QENS), which allow one to study diffusion dynamics in nano- and mesoporous matrices. The advent of monolithic mesoporous hosts with tailorable channels have additionally increased the interest in this field and extended the analytical opportunities, for it allows a better and/or simpler comparison of theory and experiment \cite{Kusmin2010, Gruener2008, Gruener2009, Valiullin2009, Martin2010, Krutyeva2013}.

Many studies of diffusion in restricted geometries are aimed at a comparison of the confined with the unconfined state. Over the years it has turned out, however, that this comparison is often hampered, if one has to solely rely on diffusion data reported in the literature and extracted with different measurement techniques. Even for identical bulk or pore systems alternative experimental methods can result in deviating dynamical properties and upon confinement these differences in the measured quantities can even be more pronounced \cite{Mitra1992, Mitra1993, Chmelik2011, Kimmich2011, Shakirov2012}. Different methods probe different length scales and thus also different time scales of diffusion processes. Translational and rotational self-diffusion in a crowded melt is, however, a highly cooperative phenomenon and can necessitate large scale molecular rearrangements depending on the investigated diffusion length and the molecular species investigated. This translates to a dependency of the diffusion dynamics on the diffusion length, and thus diffusion time investigated. This was particularly convincingly demonstrated by seminal Molecular Dynamics simulations on unconfined, molecular liquids \cite{Alder1970, Alder1970a} and is all the more of importance for spatially nano-confined liquids, where the bare geometrical restriction can hamper both the movements of the diffusing molecule as well as the necessary mesoscopic rearrangement of the surrounding molecules \cite{Cui2005, Devi2010}. 

Given the large interest in confined diffusion there is a huge number of studies of molecular diffusion in nanoporous media. In the following a few, selected studies shall be presented starting from the diffusion of a simple n-alkane.

Hofmann \etalp \cite{Hofmann2012} performed time-of-flight quasi-elastic neutron scattering experiments on liquid n-hexane confined in cylindrical, parallel nanochannels of 6~nm mean diameter and 260~$\mu$m length in monolithic, mesoporous silicon. Those measurements were complemented with, and compared to, measurements on the bulk system in a temperature range from 50~K to 250~K. The measured time-of-flight spectra of the bulk liquid n-hexane can be modelled by microscopic translational as well as fast localised rotational, thermally excited, stochastic motions of the molecules. In the nano-confined state of the liquid, which was prepared by vapour condensation in this study, two molecular populations with distinct dynamics were found: a first fraction which is immobile on the time scale of 1~ps to 100~ps  probed in the experiments and a second component with a self-diffusion dynamics slightly slower than observed for the bulk liquid - see the illustration in Fig. \ref{fig:SketchDiffusionMoleculesChannel}. 

\begin{figure}[htbp]
\center
\includegraphics[width=0.8\columnwidth]{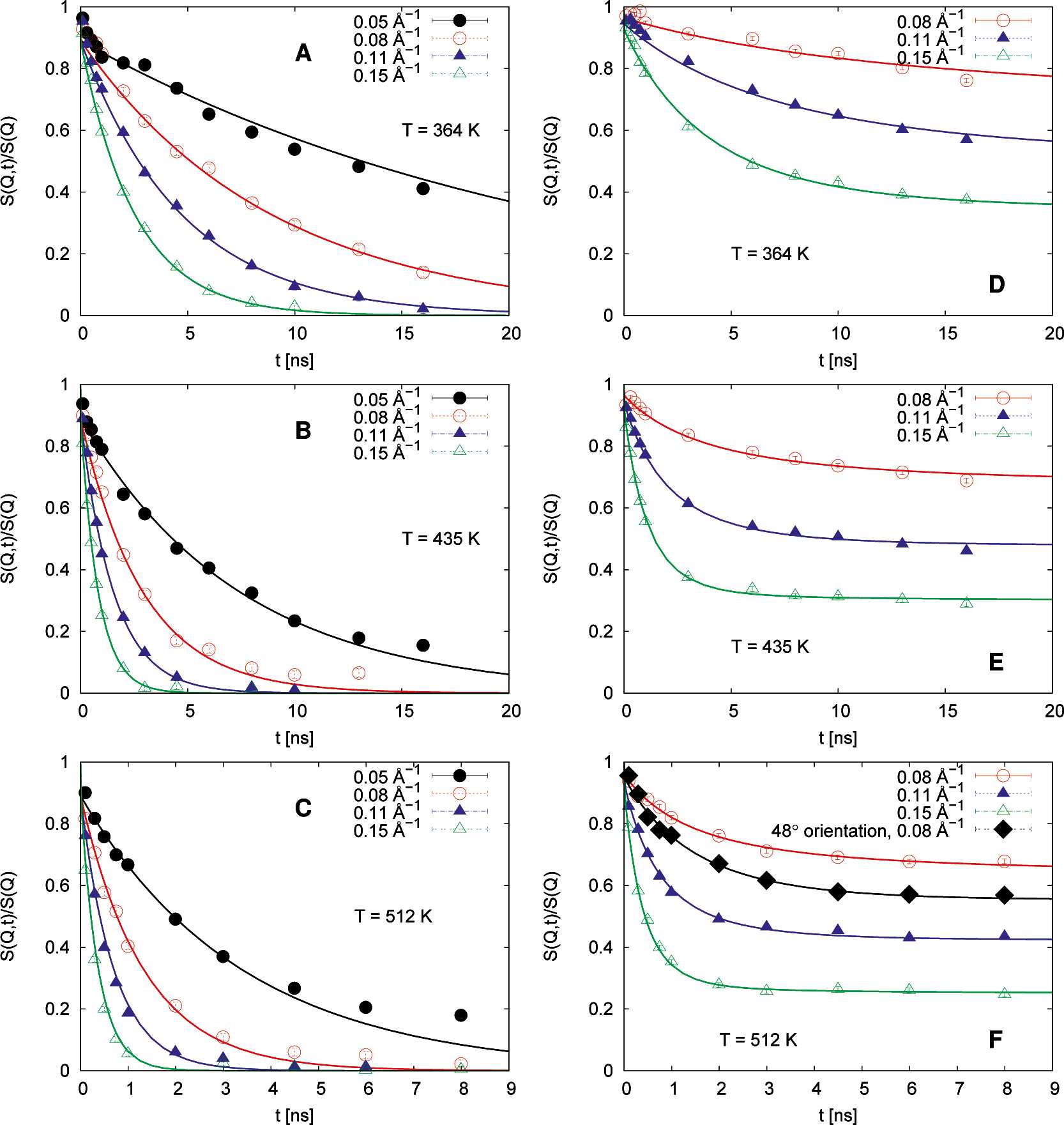}
\caption{(symbols) Measured quasi-elastic neutron spectra of bulk (A, B, C) and nanoporous-silicon-confined (D, E, F) n-C$_{36}$H$_{74}$ for different temperatures and wave vector transfers Q (symbols), as indicated in the figure. In comparison with model calculations (solid lines) of freely diffusing molecules (A, B, C) and of partially immobile, partially freely diffusing molecules (D, E, F), respectively. Reprinted (adapted) with permission from Kusmin \etalp \cite{Kusmin2010}. Copyright 2010 American Chemical Society.} \label{fig:QENSAlkaneSilicon}
\end{figure}

Analogous observations were also made by Kusmin \etalp \cite{Kusmin2010} in spin-echo QENS measurements on a medium-length alkane, C$_{36}$H$_{74}$ confined in porous silicon. With this technique, one has direct access to the intermediate scattering function $S(q,t)$ as a function of wave-vector transfer $q$ in the time domain $t$ and can extract therefrom the translational self-diffusion coefficients. Kusmin \etalp \cite{Kusmin2010} can quantitatively attribute the decays of $S(q,t)$ to the translational self-diffusion coefficient of C36 as calculated with the Stokes-Einstein relation from the viscosity of the melt - see Fig.~\ref{fig:QENSAlkaneSilicon}. For the confined state, two contributions are important, the decays of $S(q,t)$ are again compatible with a bulk-like translational diffusion, whereas the remaining sockets (after the decays) are typical of an immobile fraction of molecules. This sticky molecular fraction, which amounts to about two monolayers at the pore walls at low temperature is decreasing with increasing temperature \cite{Kusmin2010}.
 
The partitioning of the pore confined liquid in a slow component and in one with a dynamics close to the bulk liquid is representative for many other pore condensates and porous media, \textit{e.g.}, polymers \cite{Schoenhals2010} or other small molecules \cite{Kusmin2010, Loughnane2000, Baumert2002, Wallacher2002, Wallacher2004b, Morineau2004} in silica or silicon matrixes. The thermodynamic partitioning of pore-condensates in a film-condensed regime, strongly interacting with the pore walls, and in the capillary condensate in the pore centre, with a thermodynamics close to the bulk system, nicely documented in the discussion on capillary condensation, results obviously in an analogous partitioning of the molecular mobility - see the illustration in Fig.~\ref{fig:SketchDiffusionMoleculesChannel}. Note, however, that the extraction of just two diffusion coefficients is presumably too simplistic in order to describe this dynamical heterogeneity typical of molecular condensates in nanoporous media. Even for small molecules the gradually with the distance from the pore walls decaying interaction potentials suggest rather a smooth variation of the mobilities from the pore wall proximity to the pore centre \cite{Kityk2014a}. A separation of the molecular mobility in a pure core and shell molecular mobility is thus oversimplified \cite{Valiullin1997, Koppensteiner2008, Ji2009, Norton2014}.

\begin{figure}[htbp]
\center
\includegraphics[width=0.5\columnwidth]{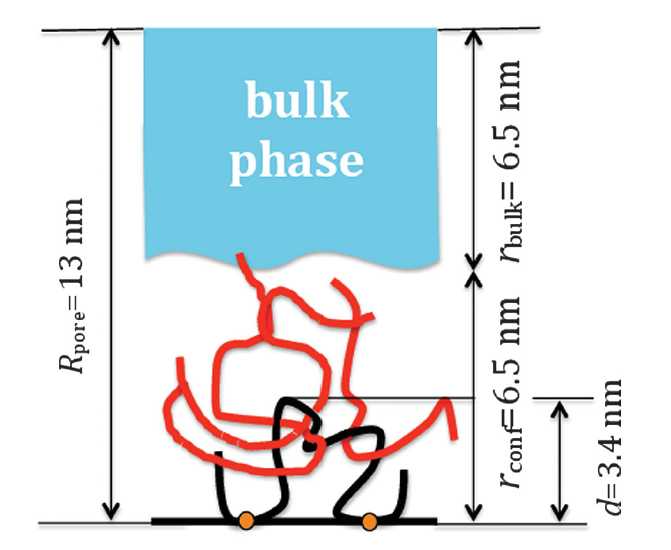}
\caption{Schematic representation of the artificial surface-induced entanglements in the confined polymer melt.
The black line represents the chain adsorbed on the surface of an alumina nanopore, and the red lines show entangled chains in the confined phase \cite{Krutyeva2013}. Reprinted with permission from Krutyeva \etalp \cite{Krutyeva2013}. Copyright 2013 American Physical Society.} \label{fig:InterphasePolymerAlumina}
\end{figure}

Moreover, in the case of macromolecules it can even be anticipated that at the interface between pore wall anchored species and the freely in the pore center diffusing molecules a third, distinct molecular population exists with qualitatively different dynamics: Neutron spin echo experiments by Krutyeva \etalp \cite{Krutyeva2013} document for the polymer dynamics in alumina nano channels the existence of two phases, one fully equal to the bulk polymer and another that is partly anchored at the surface. By strong topological interaction, this phase confines further chains with no direct contact to the surface. These form a \textit{third} population, an interphase, where the full chain relaxation is hampered by the interaction with the anchored chains - see illustration of this dynamical partitioning in Fig.~\ref{fig:InterphasePolymerAlumina}.

Presently, there is an increasing interest in the molecular mobility of room-temperature ionic liquids (RTILs), organic salts with room-temperature melting points, in nanoporous media. RTILs have been intensely studied over the last decade for numerous applications such as green solvents in chemical synthesis, batteries, fuel and solar cells \cite{Wasserscheid2000}. Their great interest to basic science is due to their peculiar molecular structure and composition, a bulky organic cation with a small inorganic anion, which produce unusual molecular packings, and to a complex combination of interactions (van-der-Waals, ionic, dipolar, and hydrogen bonding) seldom occurring together in other materials. 

The self-diffusion dynamics of ionic liquids is of importance for nanotemplate-assisted electrodeposition of ionic liquid structures \cite{Elbasiony2014}. Moreover, a couple of years ago it was demonstrated that ions from RTILs can enter nanoporous carbon, rendering these systems very interesting candidates for supercapacitors \cite{Vix-Guterl2005, Chmiola2006, Singh2014} and thus energy storage. Since these applications rely crucially on diffusion and reversible ion adsorption in the pores, it is particularly important to explore the molecular self-diffusion dynamics of RTILs in such restricted geometries. Reduced mobilities as exemplified above for more simple, uncharged molecules would implicate reduced performances, particularly charging kinetics in the case of super capacitors. 

However, in QENS measurements of a RTIL confined in mesoporous carbon \cite{Chathoth2012, Chathoth2013} by Chathoth \etalp a self-diffusion behaviour faster than in the bulk system was observed. This somewhat surprising finding is corroborated by Iacob \etalp \cite{Iacob2012}, who reported enhanced self-diffusion dynamics in unidirectional nanoporous membranes (porous silicon with pore diameters of 7.5-10.4 nm). By combining broadband dielectric spectroscopy and pulsed field gradient NMR they were able to determine the diffusion coefficient and the diffusion rate over more than 13 decades and to trace its temperature dependence. The enhancement of diffusivities by more than two orders of magnitude attributed the authors to changes in molecular packing and hence in density leading to higher mobility and electrical conductivity. Indeed, Kondrat \etalp \cite{Kondrat2014} found in a molecular dynamics study on nano-confined RTILs hints of pronounced packing effects and, even more importantly, intimately related collective diffusion modes, that lead to diffusion-enhancement and thus rapid charging dynamics for RTILs in nanoconfinement. 

The experimental study of packing effects (and thus the structure factor) of liquids in nanopores is experimentally extremely demanding \cite{Morineau2003}. In this respect it is worth mentioning, that indeed evidence for pronounced molecular layering of RTILs at charged planar surfaces and thus in semi-infinite confinement exists. In high-energy X-ray reflectivity experiments by Mezger \etalp \cite{Mezger2008} strong interfacial layering, starting with a cation layer at the substrate and decaying exponentially into the bulk liquid were observed and the observed decay length and layering period pointed to an interfacial ordering mechanism, akin to the charge inversion effect, which is suggested to originate from strong correlations between the unscreened ions. More importantly, the observed layering (and thus presumably the enhanced diffusion kinetics) is expected to be a generic feature of RTILs at charged interfaces. 
Indeed, a recent combined scanning tunnelling microscopy, atomic force microscopy and density functional calculation study by Carstens \etalp \cite{Carstens2014} corroborate this conclusion. It revealed that multiple RTIL layers are also present at graphite/electrode interfaces and that their rearrangements as a function of applied potential exhibit a large cooperative character. This observation of layering in a molecular system is somewhat analogous to the experimentally and theoretically documented layering (and intimately related non-monotonic oscillatory surface potentials) observed in charged colloidal solutions confined in slit-pore geometry with charged pore walls \cite{Allahyarov1999, Grandner2009}.

The diffusion properties of charged molecules in confinement is also of obvious relevance for the usage of nanoporous media for template-assisted electrochemical deposition \cite{Huczko2000, Yin2001, Sander2003, Kumar2007, Kumar2010}. In an intriguing experimental and theoretical study Shin \etalp \cite{Shin2014} showed that the growth of Cu nanowires within alumina membranes is diffusion-limited, which can result in a morphological instability of the deposited metal driven by a race among the growing nanowires in parallel channels, and eventually even to an incomplete nanowire growth. Interestingly, this growth instability can be markedly reduced by applying a temperature gradient across the porous template. This strategy of manipulating the ion self-diffusion in the pores, increases the length of nanowires.

More generally, most electrochemical processes are limited by diffusion, but in porous media, surface conduction and electro-osmotic flow, an ion transport in the electrical double layer at charged pore walls, can also contribute to ionic flux \cite{Squires2005} and can be significantly affected by pore surface grafting \cite{Campus1999, Calvo2009, Taffa2010}. For example, Deng \etalp \cite{Deng2013} reported experimental evidence for surface-driven over-limiting current (faster than diffusion) and deionization shocks (propagating salt removal) in a mesoporous medium. The results suggest the feasibility of shock electrodialysis as a new approach to water desalination and other electrochemical separations by employing porous membranes \cite{Humplik2011}.

Interestingly, ion transport across mesoporous structures has also been discussed with regard to the observation of a perfectly periodic three-dimensional protein/silica mesoporous structure in siliceous sponges (one of the most primitive and oldest living multicellular animals found in all oceans in the world). Although the functionality of this architecture is not yet completely understood, the authors speculate that the open mesoporous silica structure facilitates ionic transport and, at the same time, provides a rigid mechanical support \cite{Zlotnikov2014}.

Experiments on the self-diffusion dynamics can also give detailed insights on the phase transition behaviour of condensates in nanoporous media, since different phases are characterised by distinct molecular mobilities, most prominently different translation and rotational molecular modes in the case of simple molecules. In the aforementioned QENS study on n-hexane by Hofmann \etalp \cite{Hofmann2012} in pores the immobile fraction amounts to about 5\% at 250~K (corresponding to about one, sticky monolayer adsorbed at the channel walls), gradually increases upon cooling and exhibits an abrupt increase at 160~K (20~K below bulk crystallisation), which indicates pore freezing \cite{Hofmann2012}. Similarly, temperature-dependent NMR measurements on pore condensates gave important insights on phase transformations in nanoporous media \cite{Stapf1997, Kondrashova2010}, most prominently on the liquid-solid transition, see Refs. \cite{Christenson2001, Alba-Simionesco2006, Petrov2009, Webber2010} for pertinent reviews.

\section{Non-equilibrium Phenomena}
 
Transport in nanoporous media is of increasing relevance in the emerging fields of micro- and nanofluidics \cite{Eijkel2005, Gruener2008, Squires2005, Urbakh2004, Stone2004, Majumder2005,  Dittrich2006, Whitby2007, Persson2007, Schoch2008,  Piruska2010, Kirby2010, Koester2012, Bocquet2014, Vincent2014, With2014}. In particular, it is both of fundamental and technological interest, whether macroscopically determined wetting properties or values of fluid parameters, such as the viscosity $\eta$, surface and interfacial tensions $\sigma$, accurately describe a liquid and its transport characteristics down to very small length scales, on the order of the size of its building blocks \cite{Fradin2000, Vinogradova2011}. Moreover, the validity of the continuum approach of classical hydrodynamics is questionable in restricted geometries, given for example the small number of molecules in a crosssection of a nano pore. Also the velocity profile in the proximity of the confining walls plays a crucial role in the determination of the overall transport rates in nanoporous media.

Measurements with the surface-force apparatus (SFA), which allow one to study shear viscosities of thin films with thicknesses down to sub-nanometers, have revealed that depending on the shear rate, the type of molecule and the surface chemistry both viscosity values quite different from the bulk values and viscosities in remarkable agreement with the bulk ones can be found in confinement \cite{Raviv2001}. 

\subsection{Rheology and spontaneous imbibition of liquids}
Pioneering experiments to probe transport behaviour through pores a few nanometers across were performed by Nordberg \cite{Nordberg1944} and Debye and Cleland in the mid of the last century \cite{Debye1959}. Nordberg studied water and acetone flow, whereas Debye and Cleland report on the flow of a series of linear hydrocarbons (n-decane to n-octadecane) through nanoporous silica (Vycor, mean pore diameter $d=7$~nm). Flow rates in agreement with Darcy's law, the generalisation of Hagen-Poiseuille's law for simple capillaries towards porous media \cite{Gruener2009}, were observed. Moreover, they question the validity of the no-slip shear boundary condition at the pore walls - a topic, that has attracted an increasing interest in the field of micro- and nanofluidics \cite{Squires2005, Bocquet2014,Vinogradova2011,Granick2003, Lauga2005,  Jacobs2011, Kriel2014}. Today, it is understood that the core concept of ''no-slip at the wall'' is valid only, provided that certain conditions are met: a single-component fluid, a wetted surface, and low levels of shear stress. In many engineering applications these conditions are not fulfilled and studies sensitive to the near-surface velocity profiles have revealed that slippage, that is a final velocity of the liquid at the wall can occur in systems with surfactants, at high shear rates, and low roughnesses of the confining walls \cite{Bocquet2014, Jacobs2011, Pit2000, Cheikh2003, Schmatko2005, Bocquet2007, Mueller2008, Servantie2008, Sendner2009, Sochi2011}.

As Abeles \etalpÊ\cite{Abeles1991} documented by an experimental study on toluene using again nanoporous Vycor glass, flow in nanoporous media can be through gas (or Knudsen diffusion \cite{Gruener2008}), surface diffusion, and viscous liquid flow driven by capillary forces (termed ''spontaneous imbibition'') or by external hydraulic pressure (called ''forced imbibition''), depending on the size of the pores and on the temperature and pressure of the fluid. In the following we focus mostly on viscous flow in nanoporous media.  

\begin{figure}[!t]
\centering
\includegraphics*[width=.3 \linewidth]{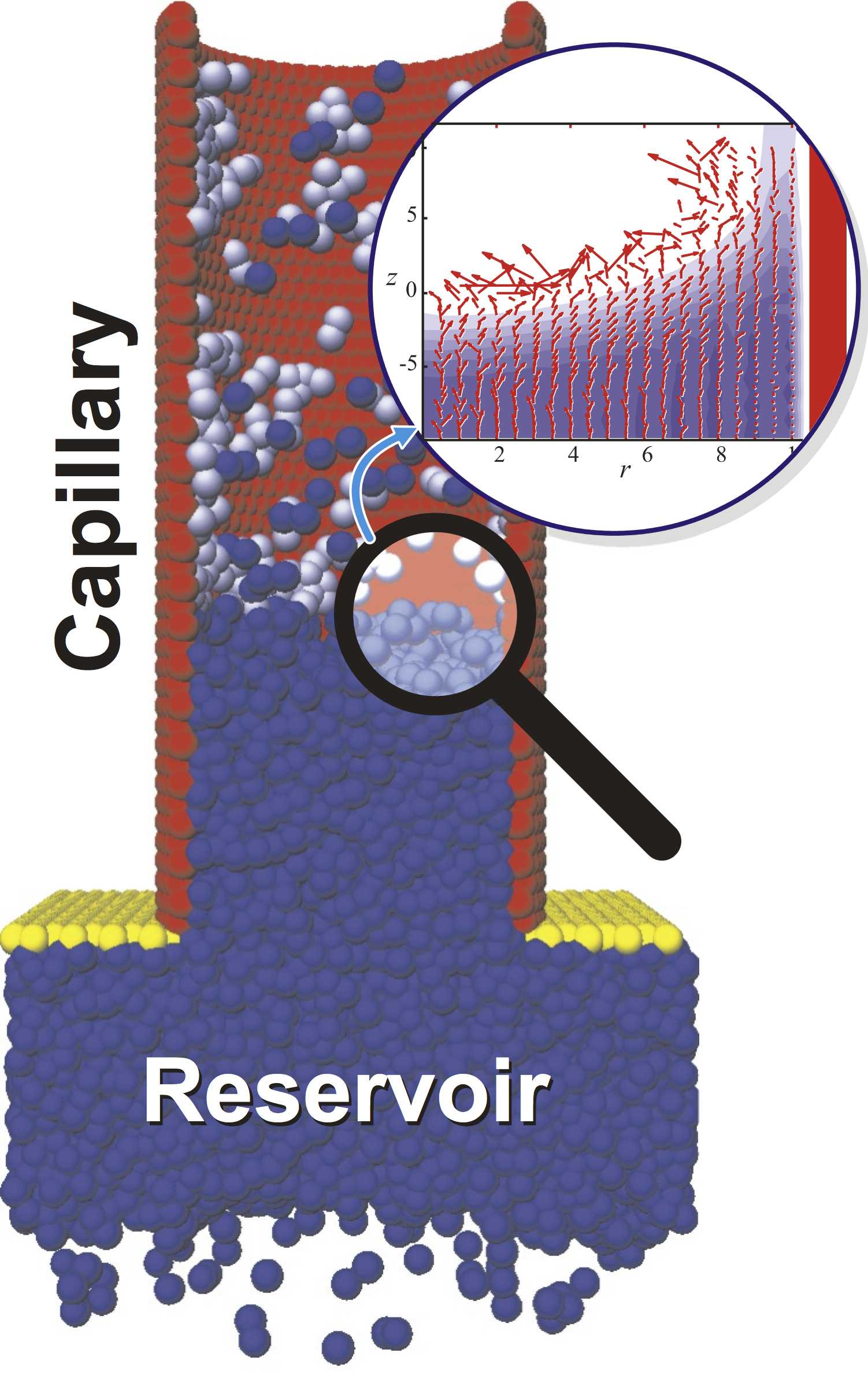}
\caption[CRBinder]{Snapshot of fluid imbibition in a nanocapillary. Fluid atoms are shown in blue. The tube wall is shown in red, and the atoms of the reservoir adhesive wall are in yellow. Reprinted (adapted) with permission from Kurt Binder, Mainz (Germany). Copyright Kurt Binder, Mainz (Germany).}
\label{fig:CRBinder}
\end{figure}

Theoretical studies on nano flows has led to detailed insights with regard to flow across nanoporous media \cite{Dimitrov2007,Gelb2002, Supple2003,  Rauscher2008, Smiatek2008, Caupin2008, Chibbaro2008, Chibbaro2008a, Ahadian2009, Stukan2010, Joly2011, Bakli2012, Chen2014, Schneider2014}: Gelb and Hopkins \cite{Gelb2002} report a simulation study on the capillary rise of xenon into empty cylindrical pores of silica. They found classical Lucas-Washburn (L-W) capillary rise dynamics \cite{Lucas1918, Washburn1921}, that is a dynamics typically observed at the macro scale for spontaneous capillary imbibition (after the initial, inertial capillary entering regime \cite{Quere1997}): The meniscus height increases as a $\sqrt{t}$ with the elapsed capillary rise time $t$. 

This law was first found by Bell and Cameron \cite{Bell1906} and can be phenomenologically quite simply understood: Spontaneous imbibition of a liquid into a porous host is governed by the interplay of capillary pressure, viscous drag, volume conservation and gravity. For nanoscale pores gravity is, however, negligible \cite{Caupin2008}, the sucking pressure at the concave menisci (Young-Laplace pressure) in a liquid wetting a porous matrix with fixed mean pore diameter and porosity along with the linearly with the filling height increasing viscous drag results in this simple scaling law \cite{Huber2007}. The squared pre-factor of the L-W law, the so called invasion speed, $v_{\rm i}$ is given by the ratio of surface tension $\sigma$ and viscosity $\eta$ of the liquid. The liquid-pore wall interaction enters in $v_{\rm i}$ via the cosine of the contact angle $\Theta$ of the liquid at the pore wall $v_{\rm i}=\sigma \cos (\Theta)/\eta$. 

Gelb and Hopkins found in their capillary rise study changes of the fluid properties only for the smallest (1.5~nm diameter) cylinder, which was about four times the particle diameter. Thus, in agreement with the SFA-experiments, probing a plane Couette shear flow geometry, simulation studies on Hagen-Poiseuille capillary flow indicate that the continuum-like fluid behavior is valid down to very small length scales. The macroscopic concepts seem to break down, only, once one reaches spatial confinement on the order of the building blocks of the liquids.

Similarly, a simulation study on spontaneous imbibition in nanopores for a simple Lennard-Jones fluid and a polymer melt by Dimitrov, Milchev and Binder \cite{Dimitrov2007} confirmed the validity of the classic Lucas-Washburn imbibition dynamics. In this study, the authors tracked in detail the flow profile both in the vicinity of the advancing meniscus and at the pore walls - see also Fig.~\ref{fig:CRBinder}. In particular, for polymer melts this group found significant velocity slippage at the walls and had to extend the Lucas-Washburn law accordingly.

In experiments employing monolithic porous glasses (mean pore diameter 10~nm) \cite{Huber2007, Gruener2009, Gruener2009a} and rectangular channels with at least one dimension in the nanometer range \cite{Oh2010} the classical Lucas-Washburn capillary rise dynamics for spontaneous imbibition has been demonstrated for a variety of molecular liquids, most prominently water, linear hydrocarbons and neat alcohols. Moreover, it was found that the relative imbibition speeds $v_{\rm i}$ of short and medium-length n-alkanes and alcohols scale as expected from the macroscopic fluid parameters  \cite{Huber2007}. 

% Remark on shear rates

\begin{figure}[htbp]
\centering
\includegraphics[width=0.6\columnwidth]{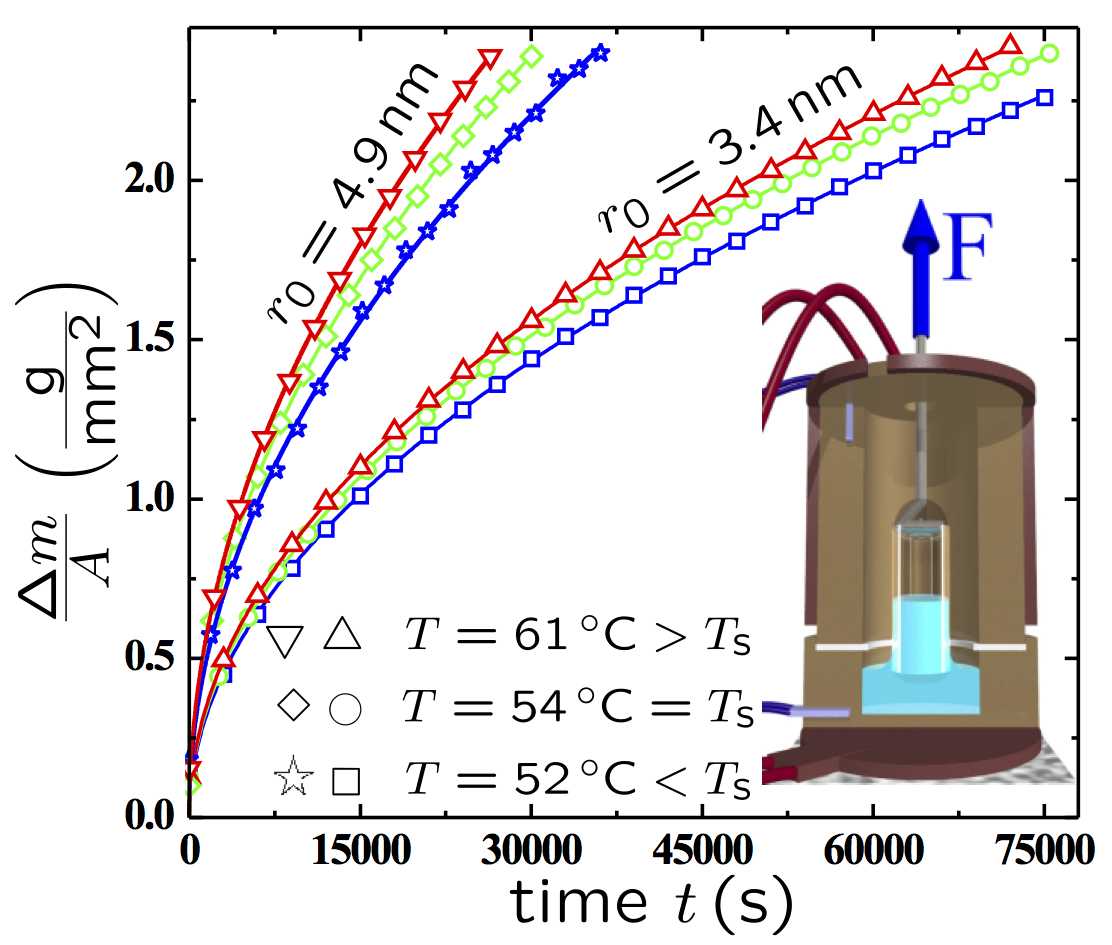}
\caption{\label{fig:HCImbVycor} Specific mass uptake of Vycor for two mean pore radii $r_{\rm 0}$ due to C24 imbibition as a function of time for selected $T$'s close to, but above the bulk freezing temperature. Solid lines correspond to $\sqrt{t}$-fits. Inset: Raytracing illustration of the thermostated imbibition cell employed for the isothermal capillary rise experiments - reproduced from Reference \cite{Gruener2009b}. Reprinted
(adapted) with permission from Gruener \etalp \cite{Gruener2009b}. Copyright
2009 American Physical Society. }
\end{figure}

Whereas the correct scaling of the invasion speeds give only hints for a conserved continuum-like behaviour, but not an unambiguous proof, more recent experimental work corroborates the robustness of macroscopic concepts at the nanoscale \cite{Gruener2009}: Due to the known structure parameters of Vycor \cite{Levitz1991} along with the well established transport physics in porous media, it is possible to determine the absolute invasion speed of water and linear hydrocarbons into the capillary network of this host \cite{Gruener2009b}. 

See Fig.~\ref{fig:HCImbVycor} for gravimetrically determined mass uptakes of Vycor during imbibition of the linear hydrocarbon tetracosane (C24). The dynamics observed can be quantitatively described by the assumption of a sticky boundary layer consisting of a flat-lying, immobile hydrocarbon backbone. This immobile shell is in agreement with the pioneering experiments on forced imbibition of n-alkanes mentioned above by Debye and Cleland \cite{Debye1959} and consistent with experimental and theoretical studies regarding the thinning of n-alkane films in the surface force apparatus \cite{Christenson1982, Chan1985, Heinbuch1989}. Moreover, quasi-elastic neutron scattering measurements are, as discussed in the previous section, sensitive to the center-of-mass self-diffusion of the n-alkanes in the pores and thus the liquid's viscosity. In agreement with the conclusions from the capillary rise experiments they document a partitioning of the diffusion dynamics of the molecules in the pores in two species: One component with a bulk-like self-diffusion dynamics and a second one which is immobile, sticky on the time scale probed in the neutron scattering experiment \cite{Kusmin2010, Kusmin2010a}. 

\begin{figure}[htbp]
\center
\centering
\includegraphics*[width=.6\linewidth]{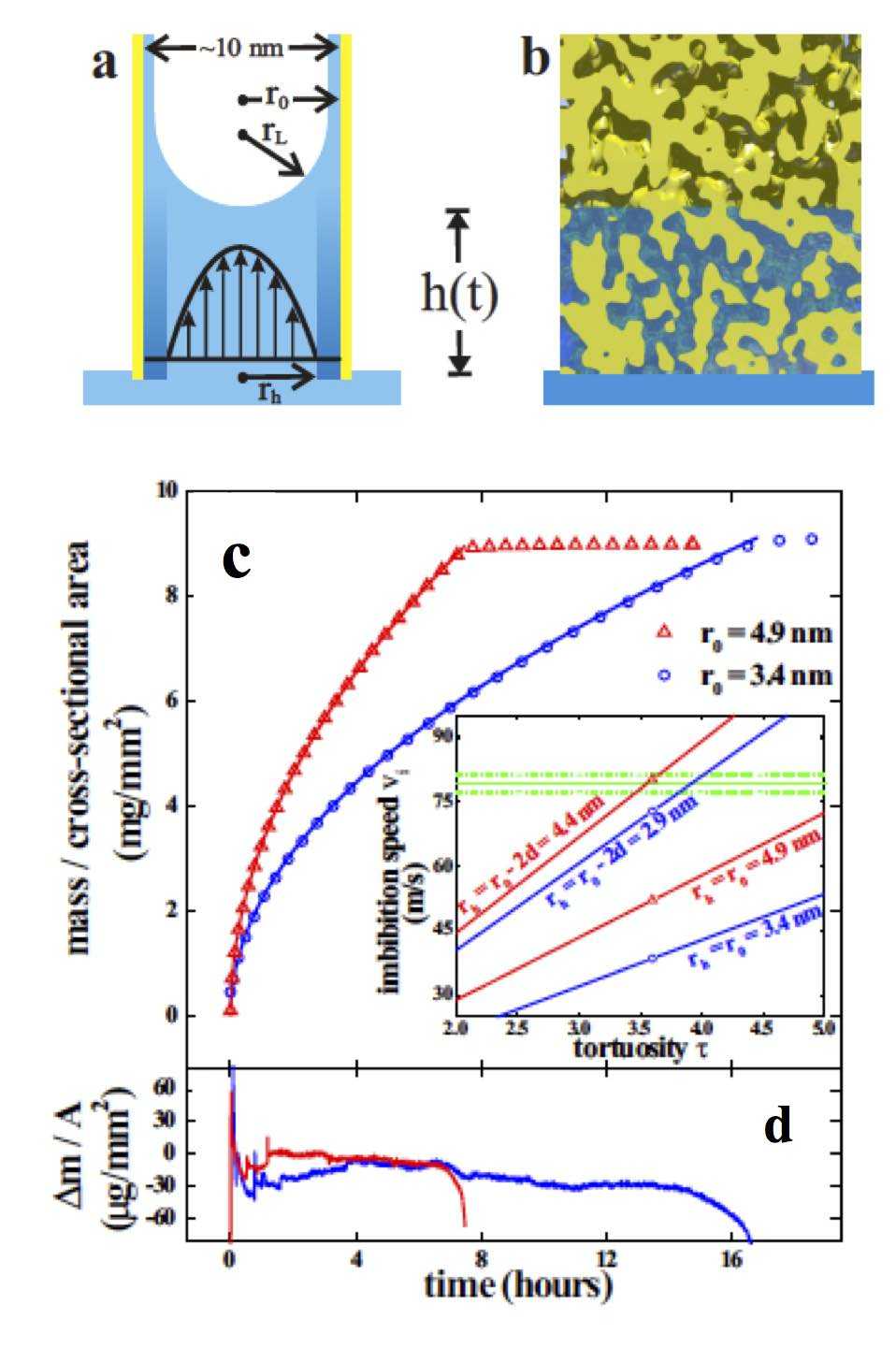}
\caption[]{(a) Schematic side view on the capillary rise of water in comparison with a raytracing illustration of spontaneous water imbibition in Vycor (b). A liquid column has advanced up to the height $h(t)$. A parabolic fluid velocity profile along with pre-adsorbed water layers beyond $h(t)$ and a shaded resting boundary layer are sketched for the nanocapillary in panel (a).(b) Normalized mass uptake $m(t)$ of Vycor with two different pore radii ($r_0=4.9$\,nm: triangles, $r_0=3.4$\,nm: circles) due to water imbibition as a function of time $t$ in comparison with Lucas-Washburn $\sqrt{t}$-fits (solid lines). Inset: Imbibition speed $v_{\rm i}$ - tortuosity $\tau$ map: The lines represent CR speeds from the measured $m(t)$ rate as a function of the tortuosity $\tau$ and for two different hydraulic pore radii ($r_{\rm h}=r_0$ and $r_{\rm h}=r_0-2\cdot d$ resp. where $d=0.25$~nm denotes the diameter of a water molecule) for both types of Vycor as indicated. The predicted $v_{\rm i}$ value for bulk water and its error margins are represented by horizontal lines. (b) Residuals $\Delta m$ between the measured $m(t)$-curve and the $\sqrt{t}$-fits. Reprinted
(adapted) with permission from Gruener \etalp \cite{Gruener2009}. Copyright
2009 American Physical Society.}
\label{fig:SImbibitionWaterVycor}
\end{figure}

The interesting physics of nano confined water \cite{Stanley2007} can also be explored in quite some detail by capillary rise experiments. Such a study performed with hydrophilic Vycor glass \cite{Gruener2009}, see Fig.~\ref{fig:SImbibitionWaterVycor},  also suggests the existence of a sticky boundary layer at silica walls. A conclusion which is supported by Molecular Dynamics studies on the glassy structure of water boundary layers in Vycor \cite{Gallo2000, Bonnaud2010} or more generally hydrophilic surfaces \cite{Sendner2009}, by structural studies documenting a partitioning of water in a core and a surface water contribution in silica pores \cite{Bellissent-Funel2003, Erko2011} and by tip-surface measurements in purified water \cite{Li2007}. These experiments yield additional rheological details: Assuming a parabolic flow velocity profile across the pore cross-section, see Fig.~\ref{fig:SImbibitionWaterVycor}(a), which implies a linear variation of the viscous shear rate, starting with 0 in the pore center to a $t$-dependent maximum at the pore wall $r_{\rm h}$, $\dot{\gamma}_{\rm m} \propto \frac{1}{\sqrt{t}}$.  For the experiment in Fig.~\ref{fig:SImbibitionWaterVycor} with a mean pore diameter of $r_{\rm h}=2.9$\,nm, one can estimate $\dot{\gamma}_{\rm m}$ to decrease from $7\cdot 10^4\,\frac{1}{\rm s}$ after 1\,s to $3\cdot 10^2\,\frac{1}{\rm s}$ at the end of the capillary rise. As no $t$-dependent, and therefore no $\dot{\gamma}$ dependent deviations of $m(t)$ from a single $\sqrt{t}$-fit observable, these measurements also testify the absence of any non-Newtonian behavior of water as well as an unchanged no-slip boundary condition imposed at $r_{\rm h}$, despite the relatively large shear rates probed in this experiment. The latter is not too surprising. The viscous forces of $\mathcal{O}(\eta\,d^2\, \dot{\gamma})$ can only overcome the strong water/silica interactions of $\mathcal{O}(A/d)$ (Hamaker constant $A\sim 10^{-19}$\,J) for $\dot{\gamma} > 10^{12}\,\frac{1}{\rm s}$ \cite{Tabeling2004} -- significantly beyond the $\dot{\gamma}$s probed in this experiment.

The simple, gravimetric liquid uptake measurements presented above necessitate thick, monolithic pieces of nanoporous media. However, in many scientific and technological applications, micrometer thin porous membranes, \textit{e.g.}, mesoporous silicon or mesoporous alumina membranes are of interest, and the handling of these membranes in a gravimetric experiment is challenging. Moreover typical time scales for the filling processes in the gravimetric experiments are several hours. There are, however, many nanoporous matrix/liquid combinations where an exploration of the rheological properties necessitate experiments with a much higher time resolution (on the millisecond timescale).

\begin{figure}[htbp]
\center
\centering
\includegraphics*[width=.6\linewidth]{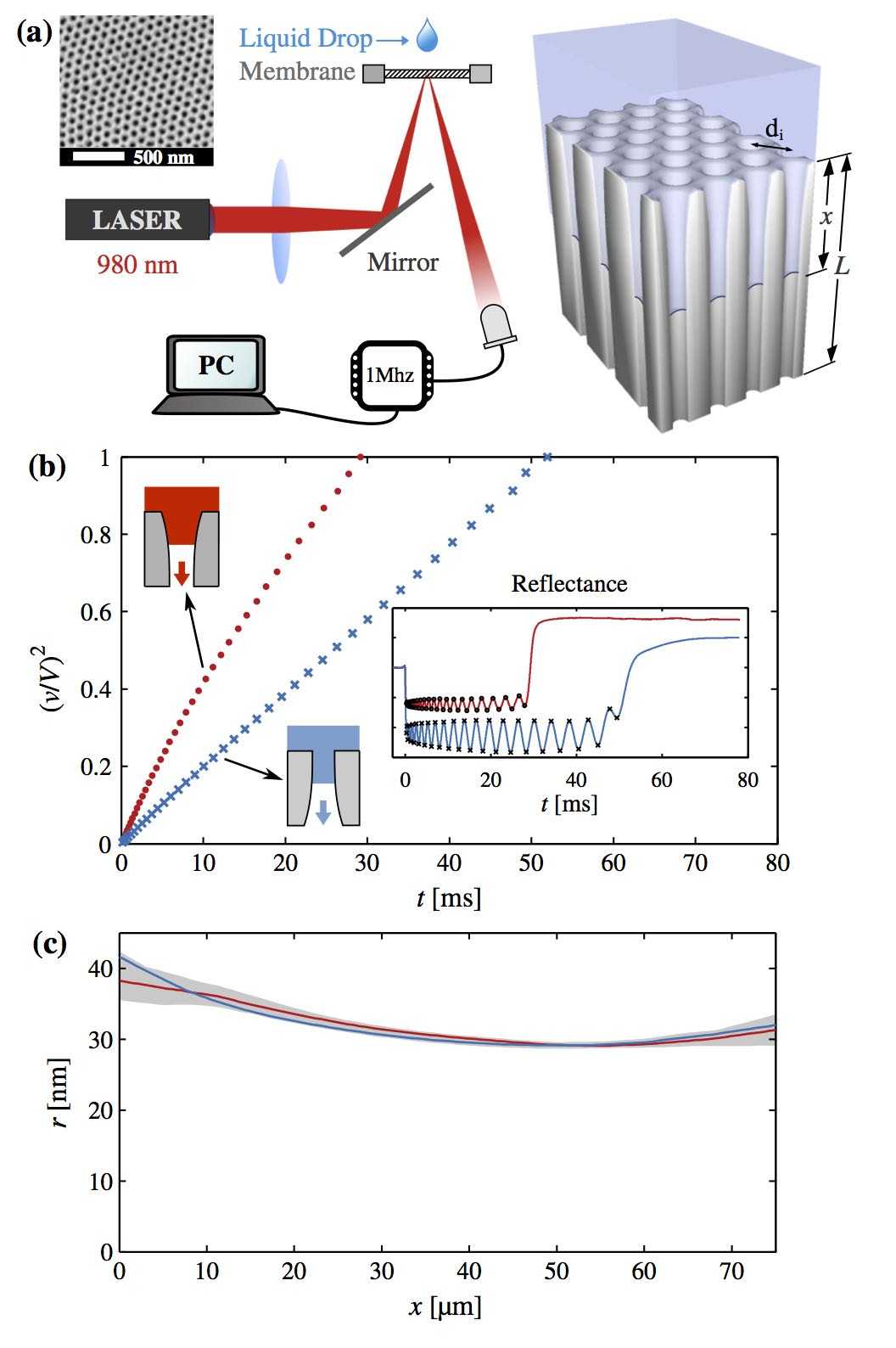}
\caption[Optofluidics]{Optical interferometry experiment in order to address the inverse problem of capillary filling. a) Illustratoin of the experimental setup including a scanning electron microscopy image of the membrane and a schematic representation of the nanochannel array with the imbibing fluid. (b) Square volume fraction of the imbibed fluid as a function of time, obtained from the extremes of the reflectance oscillations (inset), measured from different sides of the membrane. (c) Capillary radius as a function of the axial distance $x$. The continuous red curve and the dashed blue curve are the $r(x)$ functions predicted by solving the inverse problem of capillary filling for the radius variation $r(x)$, as described in Ref. \cite{Elizalde2014}. The shaded area represents the confidence bounds obtained from five different trials in each direction. Reprinted (adapted) with permission from Ref. \cite{Elizalde2014}. Copyright 2014 American Physical Society.}
\label{fig:OpticalLiquidAluminaImbibition}
\end{figure}

An optofluidic method pioneered by Acquaroli \etalp provides an elegant way to overcome these experimental hurdles in order to monitor the filling dynamics of nanoporous media \cite{Acquaroli2011, Elizalde2014} - see Fig. \ref{fig:OpticalLiquidAluminaImbibition}. It employs optical interferometry and the fact that the filled parts of the matrix have a refractive index differing from the one of the empty sample. Hence, upon movement of the imbibition front constructive and destructive interference of reflected laser light occurs which can be related to the progression of the imbibition front in the porous medium. Elizalde \etalp \cite{Elizalde2014} demonstrate by measurements on alumina nanocapillary arrays (a porous alumina membrane) with distinct capillary radius variations along the channel axis the principal suitability of this technique to reconstruct the detailed nano capillary geometry, that is the pore diameter variation $r(x)$ along the capillary axis $x$ by an analysis of the liquid imbibition dynamics in both capillary directions. Hence, they could demonstrate that by addressing the inverse problem of capillary filling dynamics, it is possible to derive otherwise hard accessible geometric quantities of nanoporous membranes.

As briefly introduced in the section on capillary condensation the microscopic and mesoscopic static structure of mesoporous materials have been extensively studied by scattering methods, in particular small angle X-ray and neutron scattering \cite{Schueth2002}. These methods rely on the scattering contrast between the empty pore space and the pore wall material. Upon filling of the matrix this contrast changes and hence these methods can also be employed to analyse the imbibition phenomenology by measuring of time-dependent scattering patterns. Since a recording of a diffraction pattern is time-consuming (from a few minutes up to hours) at conventional X-ray and neutron sources, this technique can usually be employed for high viscous, that are slowly imbibing liquids only, \textit{e.g.}, for polymers.
%Lage s

Intuitively it is clear that the microscopic structure of the fluid/solid interface sensitively depends on the type of fluid/solid interface, on its roughness and the interface chemistry. For chain-like systems, in particular for polymers and long-chain hydrocarbons Molecular Dynamics studies indicate substantial changes in the flow properties across nanoscale channels upon decoration by polymer brushes or by changing the wetting properties of the fluid/wall combination \cite{Mueller2008, Dimitrov2008, Dimitrov2008b}.

In 2007 Shin \etalp \cite{Shin2007} reported a neutron small angle diffraction study aimed at monitoring the filling process of alumina channels by polymers. This study confirmed the classical Lucas-Washburn filling dynamics. However, imbibition speeds enhanced by several orders were found, when compared to the expectations based on the bulk polymer properties. The authors attributed this surprising finding to a reduced viscosity in the channels, which they conjectured to originate in a strongly reduced entanglement density of the polymer chains in the channels. Velocity slippage at the pore walls which could also explain the rapid imbibition process were not discussed. Since such effects are well documented for polymer flow at planar solid substrates \cite{Baeumchen2010, Hatzikiriakos2012} or in simulations on capillary imbibition of polymers \cite{Dimitrov2007}, one may speculate that a changed slip-boundary condition in the nanochannels is responsible or at least sizeably contributed to the rapid polymer imbibition observed in the SANS experiment. At least, quasi-elastic neutron scattering measurements found only a marginal reduction of the entanglement density in alumina channels \cite{Martin2010}. 

\begin{figure}[htbp]
\center
\includegraphics[width=0.5\columnwidth]{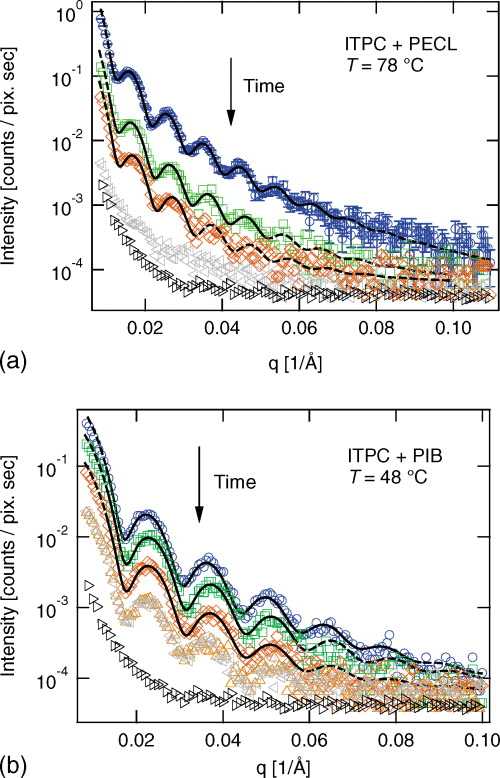}
\caption{X-ray scattering intensity curves recorded at different, selected times for spontaneous imbibition of two polymers in an ion-track-etched polycarbonate membrane. top: filling with hydrophilic poly-$\epsilon$-caprolactone at t=0, 21, 240, 1095~min (from top to bottom, pore radius $\sim$ 30 nm) and bottom: filling with hydrophobic polyisobutylene at t= 0, 90, 238, 735, 915 (on top of 735)~min (pore radius $\sim$ 23 nm). The lowest scattering curve in each figure belongs to the scattering of pure polycarbonate and the lines through the data points are fits of model calculations discussed in Ref. \cite{Engel2010}. As time continues, the scattered intensity decreases due to contrast reduction. A slight shift in the oscillation maxima to higher $q$-values can bee seen, meaning a decrease in the pore radius with the start of or during the filling process. Reprinted (adapted) with permission from Engel \etalp \cite{Engel2010}. Copyright 2010 AIP Publishing LLC.} \label{fig:SAXSPolymerImbibition}
\end{figure}

Small angle diffraction can provide also mesoscopic details of the infiltration process. This has been demonstrated by Engel et al. \cite{Engel2010}: Their time-dependent SAXS measurements on the capillary-filling of polymers in parallel-aligned ion track etched polycarbonate pores \cite{Engel2010}, see Fig. \ref{fig:SAXSPolymerImbibition} and aligned carbon nanotube arrays \cite{Khaneft2013} confirmed the classic L-W law for the infiltration dynamics. More importantly, they found indications of film spreading at the pore wall in front of the main imbibition front, the concave menisci in the pore centres. Thereby, they could confirm simulations by Chibarro \etalp \cite{Chibbaro2008, Chibbaro2008a} which stressed the importance of precursor film spreading for nano capillary infiltration. 

The progress in electron microscopy techniques allows nowadays also to observe liquid flow in nanotubes \textit{\textit{in-situ}}. This was recently demonstrated for liquid lead in a zinc oxide nanotube \cite{Lorenz2014}. Single-shot images elucidate not only the filling mechanisms, but stroboscopic electron diffraction patterns recorded simultaneously provide also the heating and cooling rates of the confining single nanotubes.The temporal changes of the images enable studies of the viscous friction involved in the flow of liquid within the nanotube, as well as studies of mechanical processes such as those that result in the formation of extrusions with nanoscale resolution.

\begin{figure}[!t]
\centering
\includegraphics*[width=.5 \linewidth]{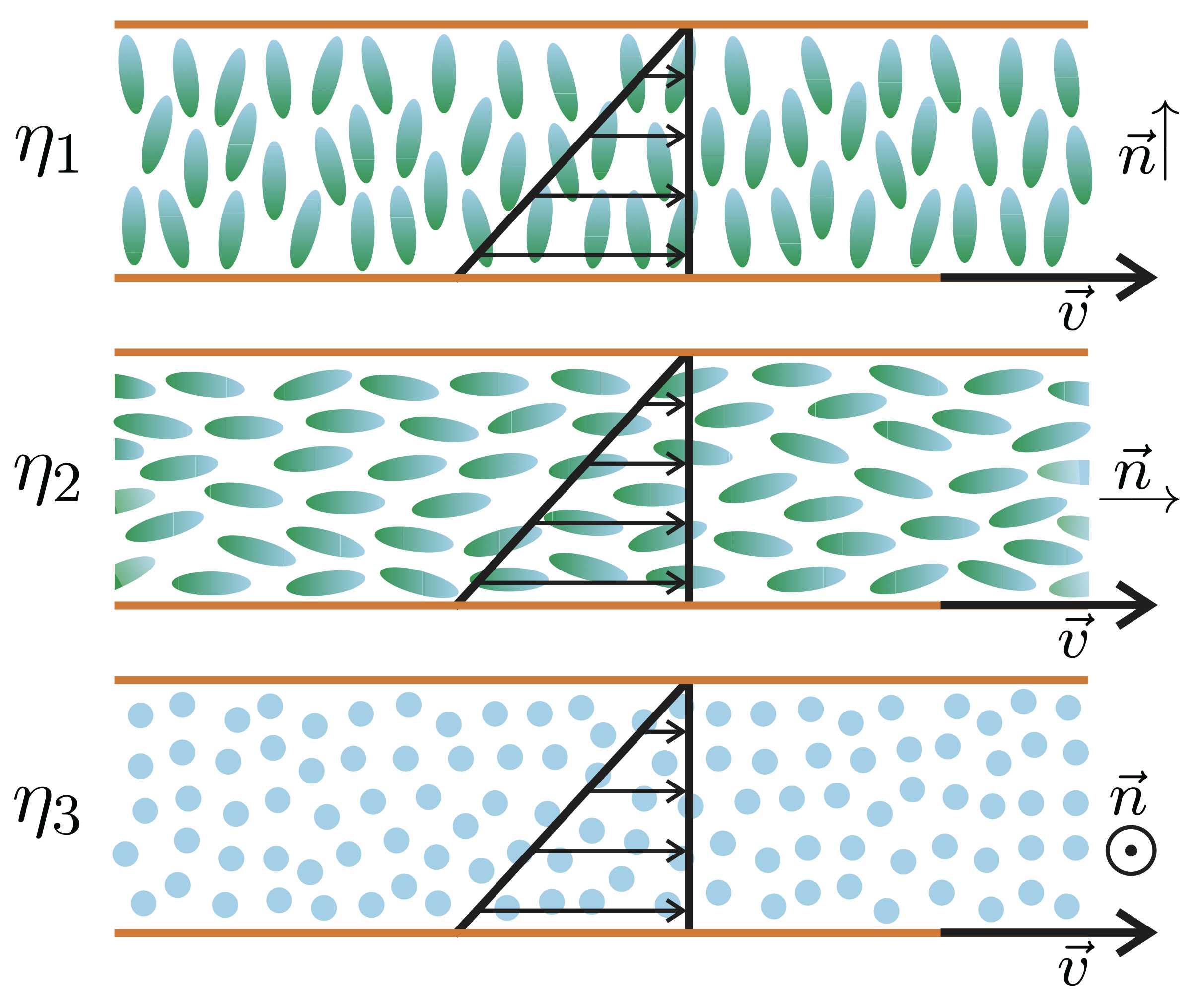}
\caption[Illustration of the three Miesowicz shear viscosity coefficients of nematic liquid crystals]{The experimental Couette flow conditions for measurements of the three Miesowicz shear viscosity coefficients of nematic liquid crystals: $\eta_1$ for $\vec{n}\,\bot\, \vec{v}$ and $\vec{n} \,||\, \nabla \,v$, $\eta_2$ for $\vec{n}\,||\, \vec{v}$ and $\vec{n} \,\bot\, \nabla \,v$, $\eta_3$ for $\vec{n}\,\bot\, \vec{v}$ and $\vec{n}\, \bot\, \nabla \,v$.}
\label{fig:LC_visc}
\end{figure}

The fluidity of more complex liquids, such as liquid crystals, in confinement of the order of a couple of nanometers has so far been investigated almost solely in the surface-force apparatus geometry \cite{Idziak1996, Janik1997, Ruths2000}. Geometric and shear-alignment effects have been found, which also depend on the exposure periods. As illustrated in Fig. \ref{fig:LC_visc} the shear viscosity of a rod-like liquid crystal depends sensitively on the collective arrangement of the molecules with respect to the shear field. It has been found that upon entering of the nematic phase (from the isotropic state) a systematic reduction of the observed shear resistance can be observed. This rheological peculiarity can be traced to a collective rearrangement of the rods in order to minimise the viscous dissipation \cite{Jadzyn2001}. In a recent study on several rod-like systems, it was demonstrated that this reorganisation is suppressed upon flow in the narrow pores of Vycor \cite{Gruener2011}. The extreme spatial confinement hampers orientational rearrangements and changes similarly as the equilibrium phase behavior (discussed above) markedly.

\begin{figure}[!t]
\centering
\includegraphics*[width=.5 \linewidth]{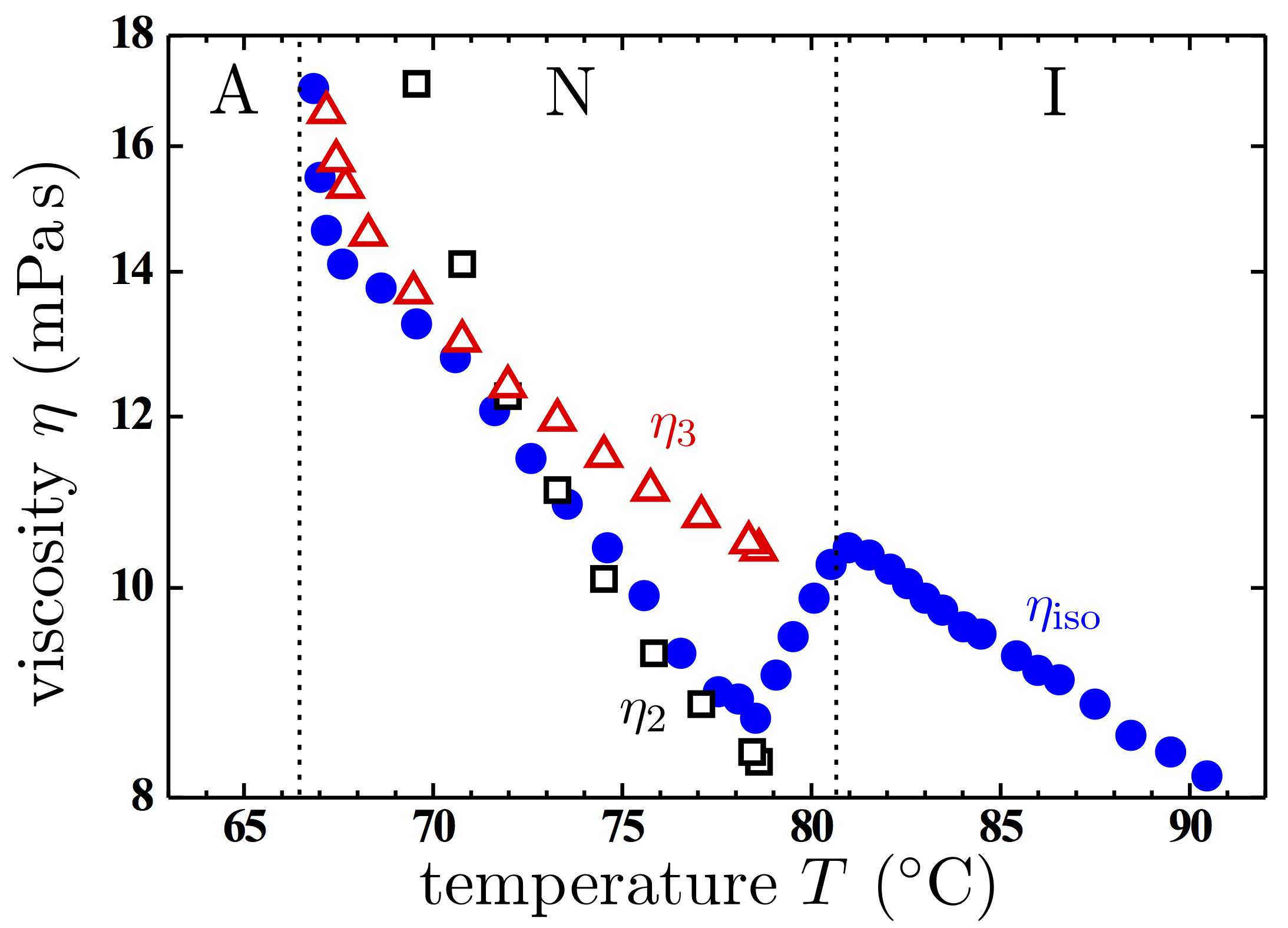}
\caption[Miesowicz and free flow viscosities of 8OCB]{Miesowicz shear viscosities $\eta_2$ (squares) and $\eta_3$ (red triangles) of the liquid crystal 8OCB compared to its free flow viscosity $\eta_{iso}$ (circles) according to Ref. \cite{Jadzyn2001}.}
\label{LC_visc_8OCB}
\end{figure}

\subsection{Switchable and forced liquid imbibition}
As exemplified for a couple of systems above the imbibition dynamics are determined by the geometry of the porous host, the fluid-wall interaction, the fluidity and capillarity of the liquid imbibed. Except for temperature-induced wetting transitions \cite{Zhang2006} or hydrodynamic instabilities \cite{Chen2007} those properties are static or hardly externally changeable during the transport process. This renders a flexible, active control of the fluid flow, as it is desirable in view of possible applications of imbibition in nano- or microfluidic systems \cite{Whitby2007,Wheeler2008} or in the template-assisted formation of nano structures by melt infiltration \cite{Jongh2013}, very challenging.

\begin{figure}[htbp]
\center
\includegraphics[width=0.7\columnwidth]{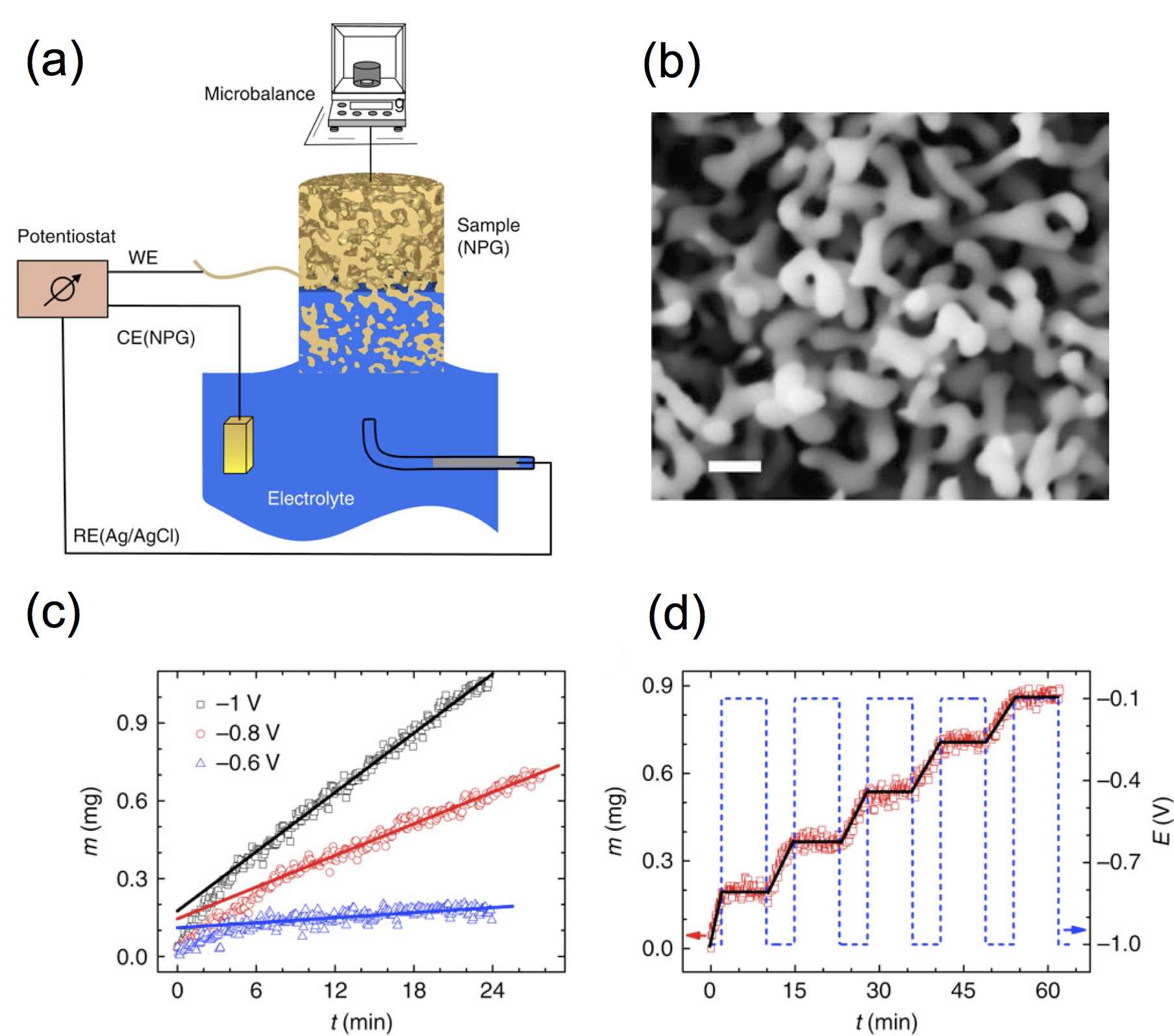}
\caption{(a) Schematics of the experimental setup for imbibition of aqueous electrolytes into nanoporous gold under controlled electrode potential. WE, working electrode; RE, reference electrode; CE, counter electrode. (b) Scanning electron micrograph showing the microstructure of nanoporous gold. Scale bar, 100 nm. (c) Mass, m, versus time, t, for experiments at various constant values of the electrode potential, as indicated by labels. (d) Response of imbibition to step potentials. Symbols and left ordinate: mass, m, evolution with time, t, during imbibition of KOH into cyclohexane-saturated nanoporous gold. The potential was stepped as indicated by the dashed line and right ordinate. The black solid line is a guide to the eye. Reprinted (adapted) with permission from Xue \etalp \cite{Xue2014}. Copyright 2014 Nature Publishing Group.} \label{fig:ImbElectrolyteGold}
\end{figure}

Xue \etalp showed that this shortcoming of nanoporous media can be overcome by employing a nanoporous metal, \textit{i.e.}, nanoporous gold \cite{Xue2014}. They demonstrate for spontaneous aqueous electrolyte imbibition in nanoporous gold that the fluid flow can be reversibly switched on and off through electric potential control of the solid-liquid interfacial tension (direct electrowetting \cite{Mugele2005}) and thus by an external control of the menisci shape in the porous medium. They can accelerate the imbibition front (increase of the concave menisci curvature), stop it (flat menisci) and have it proceed at will - see Fig.~\ref{fig:ImbElectrolyteGold}. 

%Note that the authors perform two types of experiments. In one sort of experiments a wetting electrolyte displaces air, whereas in the second experiment one liquid displaces a second immiscible liquid in the pore space. In the particular case of cyclohexane and water the shear viscosities are practically equal, which results in a simple linear movement of the imbibition front.  

Macroscopic bodies of nanoporous metal, and specifically nanoporous gold are readily fabricated by the controlled electrochemical corrosion (``dealloying") of Ag-Au alloy\cite{Erlebacher2001,Li1992,Parida2006,Hodge2007,Jin2011}. It has been demonstrated that the high electric conductivity along with the pathways for fluid/ionic transport in nanoporous gold provides various opportunities for creating novel functional materials in which external strain, electric resistance, or mechanical strength are controlled through electric or chemical signals \cite{Jin2011, Kramer2004,Biener2009,Wahl2010,Detsi2012,Weissmueller2010, Lang2011}. The additional control of fluid transport opens up additional nanofluidic applications, \textit{e.g.}, by a combination with existing lab-on-a-chip technologies \cite{Mognetti2010}. It is conceivable that nanoporous gold can be employed as an integrated nano-filter, where the matrix is both the filter and an electro-capillary pump without any mechanically moving parts for the pumping process. 

A disadvantage of many nanoporous media with respect to both fundamental fluid dynamics studies as well as applications is a variation of the pore diameter, even in the case of a membrane with parallel-aligned nano channels as in porous silicon or alumina. Siria \etalp \cite{Siria2013} reported recently on the fabrication and use of a hierarchical nanofluidic device made of a single boron nitride nanotube that pierces an ultrathin membrane and connects two fluid reservoirs. Such a transmembrane geometry allows them to study fluidic transport through a single nanotube under diverse forces, including electric fields, pressure drops and chemical gradients. They discovered very large, osmotically induced electric currents generated by salinity gradients, exceeding by two orders of magnitude their pressure-driven counterpart. A finding which could be traced to an anomalously high surface charge carried by the nanotube's internal surface in water at large pH. 

In general, nanoscale carbon tubes and pipes can be readily fabricated using self-assembly techniques and they have useful electrical, optical and mechanical properties. The transport of liquids along their central pores is now of considerable interest both for testing classical theories of fluid flow at the nanoscale and for potential nanofluidic device applications \cite{Holt2006,  Whitby2007,Sholl2006, Noy2007, Falk2012}. In particular, nanoporous graphene \cite{Yuan2014}, that is single-layer freestanding graphene with nanometer-scale pores is considered as a versatile membrane to effectively filter NaCl salt from water. Classical molecular dynamics simulation by Cohen-Tanugi and Jeffrey C. Grossman \cite{Cohen-Tanugi2012} indicate that the membraneÕs ability to prevent the salt passage depends critically on pore diameter with adequately sized pores allowing for water flow while blocking ions. Moreover, their simulations suggest that the water permeability of this material is several orders of magnitude higher than conventional reverse osmosis membranes, and that nanoporous graphene may have a valuable role to play for water purification in the future. In this respect, nanoporous graphene may compete with nanoporous titania membranes in the future, which are also intensively explored for water purification, because of their documented photocatalytic oxidation of organics dissolved in water \cite{Zhang2008, Chong2010}.

\subsection{Front Broadening in Capillarity-Driven Imbibition}

From a practical point of view, imbibition of fluids in bodies with nanoscale dimensions provides an elegant and effective way to propel nano flows. It can also be exploited to explore the rheology of liquids in spatial confinement and this under very high shear rates \cite{Gruener2011, Gruener2009, Shin2007, Serghei2010a,  Spathis2013} or characteristics of the imbibition geometry on the nanometer scale \cite{Elizalde2014,Haidara2008, Grzelakowski2009, Oko2014}. Moreover, it is now frequently employed for the synthesis of novel hybrid materials \cite{ Wang2013, Lang2011,Yuan2008} or used in the functionality of nano-devices \cite{Schoch2008, Piruska2010}. 

In particular for the latter applications, it is important that a nanoporous medium is infiltrated homogeneously, without empty pore segments created during the imbibition process. However, many nonporous media have a complex geometry (variation of the mean pore diameter within isolated channels, meandering of the pores, varying pore connectivity) \cite{Sahimi1993, Halpinhealy1995, Hinrichsen2000, Alava2004}. The inhomogeneities result in variations in the local bulk hydraulic permeability and in the capillary pressure at the moving interface resulting in broadening of the imbibition front and thus the coexistence of empty and filled pore segments during the imbibition process. 
As explored for macro porous systems, this broadening displays universal scaling features on large length and time scales, which are independent of the microscopic details of the fluid and matrix \cite{Spathis2013, Planet2007, Dube2007, Buldyrev1992, Horvath1995, Miranda2010, Hernandez-Machado2001, Geromichalos2002, Leoni2011}. This parallels the elegance of critical phenomena \cite{Alava2004}. In the past most imbibition front broadening studies focused on sand and paper. In these systems, pore space is laterally highly interconnected. This results in a continuous liquid-gas interface, whose advancement is spatially correlated due to an effective surface tension \cite{Dube2000}. Consequently menisci advancement beyond the average front position is slowed down whilst menisci lagging behind are drawn forward. 

\begin{figure}[htbp]
\center
\includegraphics[width=0.7\columnwidth]{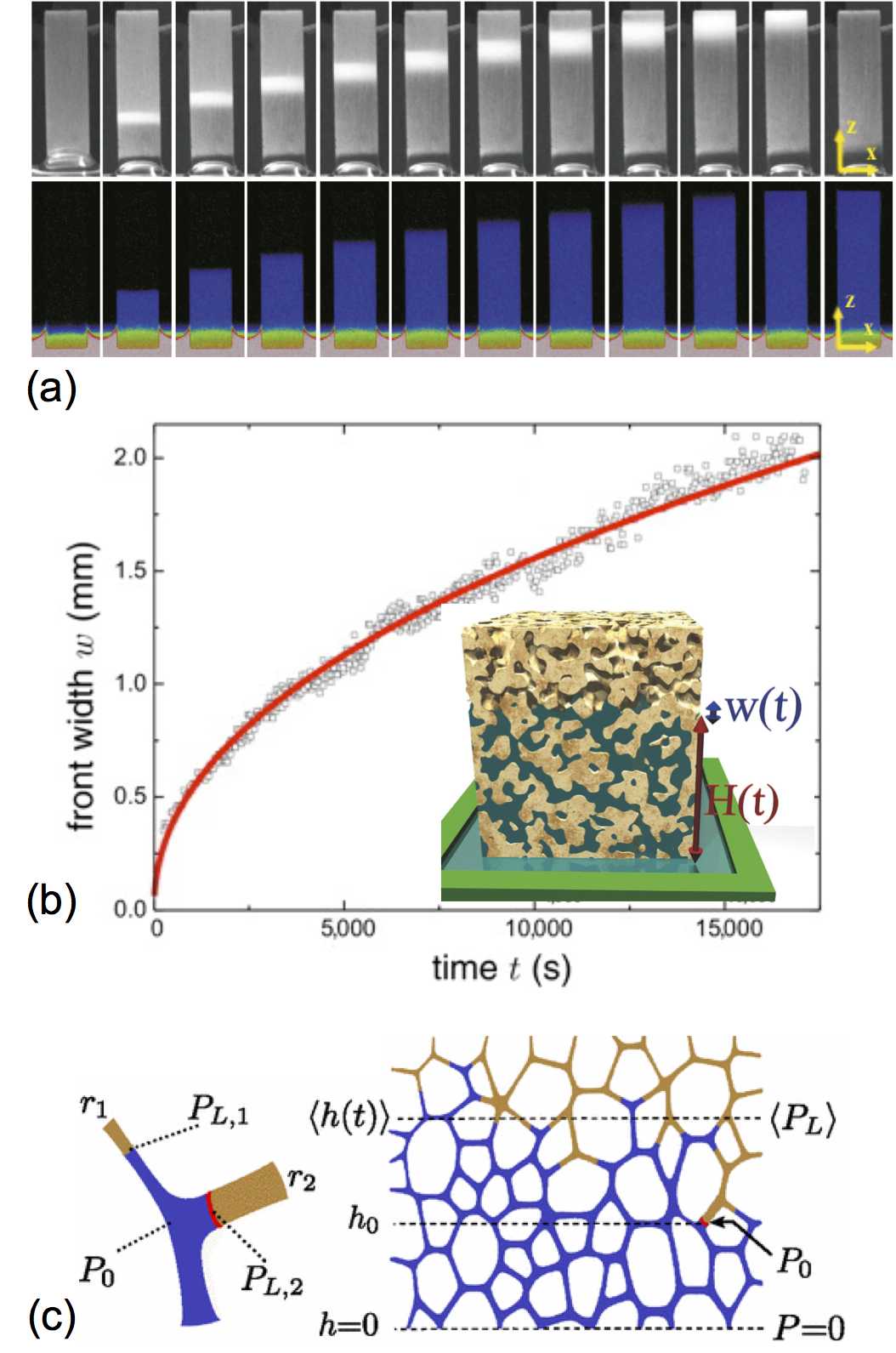}
\caption{(a) Direct observation of spontaneous imbibition of water into nanoporous Vycor glass using visible light (top row) and neutrons (lower row). The reflected light intensity and the local liquid concentration, \textit{i.e.}~the filling degree $f(x,z,t)$, are shown in grayscale and pseudocolors, respectively. Snapshots are recorded for 0.1~s (light) and 30~s (neutrons) about every 15~min. The lateral direction $x$ and the height $z$, \textit{i.e.}~the direction of capillary rise, are indicated. The width and height of the sample are 4.6~mm and 20~mm, respectively. (b) Evolution of the front width $w(t)$ along with a fit of $w\propto t^\beta$ (solid line). The inset shows the same data in a log-log representation. Axes units of the inset agree with the ones of the main plot. Reprinted (adapted) with permission from Gruener \etalp \cite{Gruener2012}. Copyright 2012 National Academy of Sciences of the United States of America. (c) Sketch of a junction in a pore network with elongated pores. r$_i$ and P$_{L,i}$ denote the radius and Young-Laplace pressure, respectively, in pore $i$, and $P_0$ denotes the hydrostatic pressure in the junction (left panel). In the right panel , $<h(t)>$ and $<P_L>$ denote the average height at time $t$ and the average Young-Laplace pressure. Reprinted (adapted) with permission from Sadjadi and Rieger \cite{Sadjadi2013}. Copyright 2013 American Physical Society.} \label{fig:NImagingWaterVycor}
\end{figure}

Gruener \etalp \cite{Gruener2012} reported recently on the broadening of the imbibition front of water in nanoporous Vycor glass. For this porous medium imbibition front roughening manifests itself quite impressively by a white, diffuse light scattering at the advancing capillary rise front - see Fig.~\ref{fig:NImagingWaterVycor} or watch the movie on imbibition of water in this hydrophilic nanoporous silica medium stored as supplementary information \cite{ImbVycorWaterMovie2005}. This can be traced to the coexistence of vapour- and liquid-filled pore segments on visible light length scales. Its detailed study is difficult since the liquid/vapour interface is deeply buried inside the matrix \cite{Callaghan1991, Howle1993}. Neutron radiography \cite{Perfect2014}, however, allowed Gruener \etalp to image this process and to document that the interface width $w(t)$ increases much faster than observed previously for imbibition front broadening in other porous materials, most prominently paper and sand, namely $w(t)\propto t^\beta$ with $\beta\approx0.5$. Moreover, the neutron radiography measurements revealed that lateral correlations of the invasion front are short-ranged and independent of time. This indicated that, for water invasion in a nanoporous silica glass, surface tension is irrelevant in a coarse grained description of interface broadening on macroscopic length scales. 

More precisely, Sadjadi and Rieger revealed in pore network simulations that spontaneous imbibition crucially depends on the pore aspect ratio $a$ \cite{Sadjadi2013}. For short pores (small $a$), neighbouring menisci coalesce and form a continuous imbibition front. Thus the smoothening effect of an effective surface tension within the interface leads to a slow broadening of the front. By contrast, for a large $a$ (like in Vycor) individual menisci form in pore space and the resulting strong broadening mechanism can be phenomenological understood in the following manner - see Ref. \cite{Sadjadi2013}: At pore junctions where menisci with different Young-Laplace and thus sucking pressures compete, see Fig. \ref{fig:NImagingWaterVycor}(c) the meniscus propagation in one or more branches can come to a halt, when the negative Young-Laplace pressure of one of the menisci (here $P_{L,2}$) exceeds the hydrostatic pressure within the junction at height $h_0$. These menisci arrests last until the hydrostatic pressure in the junction is larger than the Young-Laplace pressure of the halted menisci. Caused by the viscous-drag in the liquid column behind the advancing menisci, the junction pressure increases linearly with the distance of the moving menisci from the junction and this distance follows the Lucas-Washburn dynamics. Hence, the arrested menisci start to move, when the moving menisci have advanced by a distance, which is exactly proportional to $\sqrt{t}$ \cite{Sadjadi2013}. Identifying in this simple picture the width of the front with the maximal distance between arrested and moving menisci, the broadening of the front scales also exactly according to a $\sqrt{t}$-law. 

Extensive simulations \cite{Gruener2012, Sadjadi2013, Lee2014} of spontaneous imbibition in networks of elongated pores with random radii show indeed that the scaling relations for the emerging arrest time distribution of the halted menisci and the average front width resulting therefrom is proportional to the height, yielding a roughness exponent of exactly $\beta=1/2$, in good agreement with the neutron imaging experiments and the simple consideration above. This demonstrates, that analogous to the analysis of the pure front propagation \cite{Elizalde2014}, also an analysis of the broadening of the imbibition front, can yield important, otherwise hard accessible information on the pore geometry of the considered porous medium. In fact, depending on the degree of pore variation and the aspect ratio $a$, Lee, Sadjadi, and Rieger found in their pore-network simulations also significant deviations of the L-W behaviour for the front propagation and even situations, where the menisci movements are collectively arrested and thus the imbibition process stops entirely \cite{Lee2014}. 

In this respect, it is worth mentioning that in the experiment by Gruener \etalp all but the bottom facet of the porous matrix were sealed, in order to avoid water evaporation into the surrounding. Otherwise, the increasing evaporation rate from the liquid-filled side facets results in an exponential relaxation to a final capillary rise height, where the front movement also stops. This vivid manifestation of a dynamic equilibrium of capillarity-driven water uptake and evaporation-induced water loss by the nanoporous medium can be watched in a movie available for water imbibition in a Vycor sample with open side facets in the supplemental information \cite{ImbVycorWaterMoviePinning2005}. Whereas the capillary rise height limit determined by Jurin's law \cite{Caupin2008}, which considers a balance of gravity and capillarity (hydrostatic pressure = Young-Laplace pressure), would be at 3.6~km for this Vycor sample ($h= 2\gamma/(r \rho g)$, where $\rho$ is the mass density of water, $g$ gravitational acceleration and $r=4$~nm the mean pore radius). This happens with open side facets for this nanoporous monolith under ambient conditions (35\% humidity) after $\sim 2~cm$ front movement.  

It is important to stress that the strong imbibition front broadening found in the study on liquid imbibition in Vycor is not linked to the nanometer size of the pores. However, its experimental observation over large length and time scales significantly benefits from the dominance of capillary forces over gravitational forces, which results from the nanometer-sized pores. Therefore, strong interfacial broadening is a consequence of any spontaneous imbibition process in porous structures with interconnected elongated capillaries independent of their macroscopic extension and mean pore diameter.  Hence, it is not only important to nanofluidics, but for liquid transport in pore networks in general and the study by Gruener \etalp is a fine example that nanoporous media can be particularly suitable to study fundamental aspects of transport in porous media. 

%It should be mentioned that in the case of experiments where wetting liquids displace air from nanoporous media, e.g. water infiltration into nanoporous silica, the huge capillary pressure highly compresses entrapped air which is subsequently dissolved in water and hence does not affect experiments. In the case of an incompressible second fluid this situation is of course much more complex.

\section{Conclusions}
As exemplified above, the large range of pore morphologies and nanoporous materials readily available nowadays offer a huge playground for the condensed matter physicist to explore the influence of different types of spatial confinement and pore wall interactions on the properties of molecular soft matter. As all nanoporous media are interfacial-dominated, this research field profits a lot from the established surface sciences, with respect to theoretical concepts, experimental techniques as well as known structures and thermodynamics at planar surfaces, and thus in semi-infinite geometric confinement. However, we saw also many examples where the pore wall curvature, the tubular geometric confinement and the pore topology really matters, most prominently with respect to crystallisation and the intimately related texture formation, thermotropic orientational phase transitions, capillary-filling kinetics and roughening of imbibition fronts.

For most of the simply van-der-Waals interacting condensates the equilibrium and non-equilibrium physical behaviour can be understood by the interplay of a radial arrangement of the condensate and a clear partitioning in two components, i.e. the material in the proximity of the wall and the material in the pore centre. This partitioning is reflected in the thermodynamics as well as in the structure and manifests itself also in a pronounced dynamical heterogeneity. Quenched disorder, in particular the pore size distribution and surface roughness, also affects both the phase behaviour and the transport characteristics.

Whereas for many thermodynamic equilibrium phenomena, such as capillary condensation and the liquid-solid transitions, the research field of molecular condensates in nanoporous media can be considered as quite mature, the non-equilibrium properties are largely unexplored and new effects can be expected. Moreover, single component condensates were mostly in the focus in the past and only recently nanopore-confined binary liquids and solids increasingly attract interest. A preferred interaction of one species in a multicomponent condensate and thus a preferred adsorption at the pore wall in a porous medium, can lead to peculiar new phase and flow behaviour compared to the unconfined multicomponent systems. Also the influence of external electric and magnetic fields has been largely unexplored for condensates in nanoporous media and provides a promising perspective for active control of nano structured soft condensed matter systems. 

The phenomenologies discussed in this review have already resulted in a broad interest in nanoporous media in a remarkably large area of fundamental and applied sciences ranging from basic physics and chemistry via biology and geology to materials science and medicine. In this sense, the exploration of soft condensed matter systems in nanoporous media is of very interdisciplinary and versatile character and one can anticipate that many interesting fundamental and technological aspects remain to be uncovered in the future.
\\

%Finally, the field of active soft matter, i.e. self-propelled molecules or particles, may also be one of the new by the same time very challenging but also very interesting topics, to be addressed in the rapidly evolving research on soft matter confined in nanoporous media in the future.

%\end{counted}

%\textit{The text contains \thewords\ words.}
\ack
I would like to thank many people who stimulated over the years my own interest in the physics of molecular condensates confined in nanoporous media, Klaus Knorr, in particular, who introduced me to this research field. Moreover, I enjoyed collaborations with Ralf Ackermann, Johann Albers, Luca Bertinetti, Mark Busch, Moshe Deutsch, Gennady Gor, Stefan Egelhaaf, Mechthild Enderle, Gerhard Findenegg, Eric Grelet, Simon Gruener, Bernhard Frick, Peter Fratzl, Yuri Gogotsi, Anke Henschel, Tommy Hofmann, Andriy Kityk, Alexei Kornyshev, Tobias Kraus, Jan Krueger, Pusphendra Kumar, Andre Kusmin, Ronan Lefort, Denis Morineau, Sebastian Moerz, Ronan Lefort, Ben Ocko, Oskar Paris, Peter Pershan, Rolf Pelster, Dieter Richter, Heiko Rieger, Christof Schaefer, Andreas Schoenhals, Volker Schoen, Wilfried Schranz, Oleg Shpyrko, Viktor Soprunyuk, Martin Steinhart, Yunlan Su, Howard Stone, Rustem Valiullin, Ulrich Volkmann, Dirk Wallacher, Joerg Weissmueller, Matthias Wolff, Yahui Xue and Reiner Zorn. They contributed to the research efforts as students, colleagues or guest scientists. Support by the German Research Foundation (DFG) within the graduate school 1276, ÔStructure formation and transport in complex systemsÕ (Saarbruecken, Germany), the research project HU850/3 and the collaborative research initiative ÔTailor-Made Multi-Scale Materials Systems M3Õ (SFB 986, Hamburg, Germany) are gratefully acknowledged.\\
%We appreciated the financial support of the Deutsche Forschungsgemeinschaft and by the German Academic Exchange Service (DAAD).
%\\

\bibliographystyle{unsrt.bst}

\end{document}